\documentclass{pasa}%

\usepackage{graphicx}
\usepackage{amsmath}
\usepackage{flushend}
\usepackage{subcaption}
\usepackage{booktabs}
\usepackage{ulem}
\usepackage{hyperref}

\title[A Minimal Space Interferometer]{A Minimal Space Interferometer Configuration\\ for Imaging at Low Radio Frequencies}

\author[Akhil Jaini et al.]{Akhil Jaini$^{1,2,3,4,}$\thanks{\textbf{Author for correspondence:} Akhil Jaini, E-mail: \href{mailto:work.jainiakhil@gmail.com}{work.jainiakhil@gmail.com}}, Avinash A. Deshpande$^{1,5,6}$ and Sainath Bitragunta$^{2}$
    
	\affil{$^1$Raman Research Institute, Bangalore, India.}
	\affil{$^2$Birla Institute of Technology and Science-Pilani, Pilani, India.}
	\affil{$^3$Indian Institute of Astrophysics, Bangalore, India.}
	\affil{$^4$University of Calcutta, Kolkata, India.}
	\affil{$^5$Inter-University Centre for Astronomy and Astrophysics, Pune, India.}
	\affil{$^6$Indian Institute of Technology, Kanpur, India.}
}

\jid{PASA}
\doi{10.1017/pas.\the\year.xxx}
\jyear{\the\year}

\usepackage{aas_macros}
\usepackage{hyperref} 
\hypersetup{colorlinks,citecolor=blue,linkcolor=blue,urlcolor=blue}

\begin{document}
	
	\begin{frontmatter}
		\maketitle
		
		\begin{abstract}
			\\
			The radio sky at lower frequencies, particularly below 20 MHz, is expected to be 
			a combination of increasingly bright non-thermal emission and significant 
			absorption from intervening thermal plasma. 
			The sky maps at these frequencies cannot therefore be obtained by 
			simple extrapolation of those at higher frequencies. 
			However, due to severe constraints in ground-based observations, 
			this spectral window still remains greatly unexplored. 
			In this paper, we propose and study, through simulations, a novel minimal configuration for a space 
			interferometer system which would enable imaging of the radio sky 
			at frequencies well below 20 MHz with angular resolutions comparable 
			to those achieved at higher radio frequencies in ground-based observations 
			by using the aperture-synthesis technique. 
            The minimal configuration consists of three apertures aboard 
			Low Earth Orbit (LEO) satellites orbiting the Earth in mutually 
			orthogonal orbits. Orbital periods for the satellites are deliberately 
			chosen to differ from each other so as to obtain maximum $(u,v)$ 
			coverage in short time spans with baselines greater than 15000 km, 
			thus, giving us angular resolutions finer than 10 arcsec even 
			at these low frequencies. 
			The sensitivity of the $(u,v)$ coverage is assessed by varying 
			the orbit and the initial phase of the satellites. 
			We discuss the results obtained from these simulations and 
			highlight the advantages of such a system.
			
		\end{abstract}
		
	\end{frontmatter}

	\section{Introduction} \label{secintro}
	
	In the present era of multi-wavelength (in fact, multi-messenger) astronomy, the sky
	remains poorly explored at the very low radio frequencies, particularly below 20 MHz. 
	The radio sky at such low frequencies would be expected to be very bright due 
	to the dominance of non-thermal emission, however there would also be rapidly increasing 
	free-free absorption from intervening thermal plasma. The sky images at these 
	frequencies would, therefore, be expected to offer an unprecedented opportunity 
	to probe both of these contributors, through the spectral evolution of their 
	combined effect, which is not possible to infer from trivial extrapolation of available 
	measurements at the higher frequencies. However, ground-based observations of 
	astronomical sources in this spectral band are severely limited owing to 
	the spectral cutoff due to the Earth’s ionosphere, and also suffer from
	spectral contamination by man-made Radio Frequency Interference (RFI). 

	To overcome these constraints, such radio observations have already been attempted 
	from above the ionosphere, but were initially
    limited to using a single antenna setup, with 
	no significant angular resolution. 
	To image the radio sky from space with angular resolutions comparable 
	to those routinely achieved in ground-based observations at higher radio frequencies, 
	we would necessarily need to appeal to the aperture-synthesis technique, 
	using multi-element space interferometers. Even if aperture sizes of the individual
	elements may be relatively small, due to practical considerations, implying 
	limited instantaneous sensitivity, significantly finer resolution angular 
	resolution can be achieved, equivalent to that of an Earth-size aperture.
	
	\begin{equation} \label{angres}
	\hspace{1cm} \theta_{arcsec} \approx 0.6\frac{(\lambda_{m}/30 m)}{(D_{km}/10000 km)}\\ 
	\end{equation}
	
	where, $\theta_{arcsec}$ is the angular resolution of the telescope, 
	expressed in arcsecond, $\lambda_{m}$ is the wavelength in meters, 
	and $D_{km}$ is the size of the effective aperture synthesized 
	(or the maximum baseline length in case of interferometry) in km.
	
	Orbiting Very Long Baseline Interferometry (OVLBI) is not a new concept 
	to radio astronomers, and has been employed multiple times in the past 
	(see for example, \cite{gurvits2018radio}). 
	Multi-element space-based radio interferometers have been used, 
	but only in conjunction with ground-based telescopes, 
	therefore, rendering them unusable for observations in the lower frequency window. 
	For example, the first ever mission to have operated in space providing 
	us the proof of concept of OVLBI was the Tracking and 
	Data Relay Satellite System (TDRSS), from 1986 to 1988 
	(see \cite{teles1995overview}). 
	Following that came the Space VLBI Satellite: MUSES-B 
	(see \cite{hirabayashi1998overview}) 
	under the Japanese VLBI Space Observatory Program (VSOP) 
	and RadioAstron (Spektr-R, see \cite{kardashev2012radioastron}) 
	operated by the Russian Astro Space Center. 
	Similarly, the Chinese Cosmic Microscope
	is another proposed Earth-space VLBI mission with two 
	satellites, though at high frequencies (see \cite{an2020space}).
    
    The increasing interest in radio sky at very low frequencies has, 
    in the past few decades, resulted in several space-interferometric 
     missions being proposed (such as SURO, OLFAR; see 
    \cite{baan2012suro}, \cite{engelen2010olfar}, 
    \cite{bentum2020roadmap}, and references therein). These proposed swarms 
    of small satellites (close to the Earth or at other special locations 
    within our solar system, such as L2 or lunar orbit) would have a large number of 
    apertures working together in aperture synthesis mode,
    but with baseline lengths limited to 100 km or so.

	It is clear that sensitive imaging observations
    with very high angular resolution at such low frequencies possess
    immense potential to reveal the yet unexplored view of the entire 
    radio sky. Motivated by these prospects, a suitable 
    space interferometer system to enable such high resolution imaging
    is considered.

    Our exploration and identification of a novel minimal configuration
    consisting of three apertures in LEO, which is the focus of the 
    present study, is only a first step towards the design and development 
    of such a system. We consider a radio astronomy payload on each satellite 
	as consisting of a suitable antenna providing a very wide field-of-view (FoV). 
	Such a configuration would allow us to sample baselines of over 15000 km, 
	which is larger than the diameter of the Earth, thus enabling angular 
	resolution as fine as 1 arcsec even at 4 MHz. 
    
    Here, it is worth noting the difference between (usually stationary)
    interferometers on the Earth and those formed by moving apertures in LEO,
    in terms of the speeds of both, the $(u,v)$ coverages and access to sky area.
    For interferometers on the Earth, the rate of change of projected
    baselines (i.e. the $(u,v)$ spacing) is dictated by the rotation rate of the Earth,
    and the chosen set of aperture locations, where the typical duration required to
    sample the available spacings is about half a day, for most sky directions
    in the view.
    Similarly, unless the Earth-based interferometers are spread over the entire globe,
    access to different areas of the sky, bringing them within the field-of-view
    of the apertures, is again dictated by the Earth rotation, where mere {\it viewing}
    of the entire sky, if at all possible, will require at least half a day,
    even with a very wide field-of-view.
    Both these aspects together imply a one-day cycle for relevant measurements
    across the entire accessible sky.
    In contrast, each of the apertures in LEO, with adequately wide
    (say, hemispheric) field-of view, can access effectively the entire sky
    in the duration of a single orbit, which is an order of magnitude
    shorter than a day.
    
    As for interferometers using apertures in LEO, the preliminary
    implications of orbital periods are also worth noting.
    Let us consider both cases, short and long period orbits.
    Shorter the orbital periods, faster will be the realization
    of the sequence of baselines possible with a given pair of orbits,
    implying in turn, a faster $(u,v)$ coverage.
    However, the available integration time for visibility measurement
    at a given $(u,v)$ spacing is correspondingly reduced. To ensure the
    amount of integration in a given $(u,v)$ cell to be same in both cases,
    correspondingly more cycles of revisit to the cell are required
    in case of faster orbits, and as would be expected, the total time
    span requirement tends to be independent of the orbital periods.
	
	In this paper, we describe our simulation of a system consisting 
	of small apertures aboard multiple LEO satellites for observations 
	of the entire sky at low radio frequencies. 
	We begin with a minimal system consisting of three satellites 
	in mutually orthogonal orbits, and show that this configuration 
	gives desired spatial-frequency $(u,v)$ coverage of the entire sky. 
	We perform multiple tests by varying the (relative) phase and period of the orbits
	of the satellites to assess improvements in the performance 
	in terms of $(u,v)$ coverage. 
	Our system consists of one satellite orbiting the Earth in the equatorial 
	orbit, while the other two satellites are in mutually perpendicular polar orbits. 
	Each of the satellites has an orbital height which is slightly different 
	than the other two, which results in a different (non-redundant) baseline 
	at different points of time, thus greatly improving sampling in $(u,v)$ plane. 
	We quantitatively evaluate the upper and lower limits of the percentage coverage,
	or the filling factor, in the $(u,v)$ plane achievable for different
	possible source directions, by considering a few extreme cases of relative 
	source direction. The sensitivity of the $(u,v)$ coverage is also assessed 
	by varying the orbit and the initial phase of the satellites. 
	We discuss the results obtained from these simulations, and highlight 
	the advantages of such a system.
	
	Section~\ref{secm} describes the model, including the parameter definition,  
	the governing equations and the main assumptions. Section~\ref{secrad} 
	presents the analytical results, with separate subsections for results 
	related to different objectives and different special cases.
	In Section~\ref{seccfs}, we first discuss some of the assumptions
    relating to our model, their validity and implications.
    Before concluding, we touch upon some of the challenges 
    relating to RFI, synchronization, and data rates, as relevant to interferometry.

	
	\section{A space interferometer configuration: Our model} \label{secm}
	Here, we describe our model configuration for a space interferometer,
	presenting the governing equations and defining the various relevant parameters 
	to make the model as realistic as possible. We also state the assumptions made,
	and related justifications, including how the $(u,v)$ plane has been defined 
	for our fully space-based configuration. 
	
\begin{figure}[t]
		\begin{center}
			\includegraphics[scale=0.55]{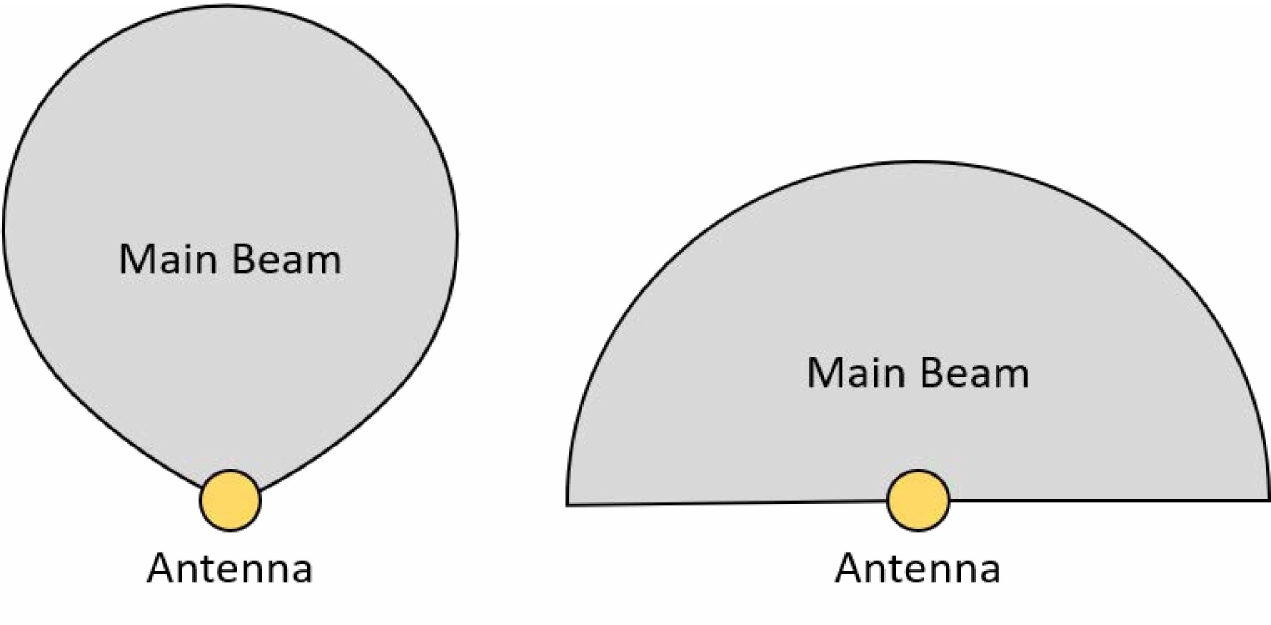}
			\caption{An illustration contrasting the typical main beam of 
				a usual widefield radio astronomical antenna (left) with the semi-isotropic beam (right) 
				of a special antenna (system) considered in our model.} \label{figure1}
		\end{center}
	\end{figure}

	Our model configuration is defined in a reference coordinate system with its
	origin at the center of the Earth\footnote{We assume the Earth radius of $6371$ km, 
	and the current obliquity of $23.4\,^{\circ}$, and the
	Earth-Sun (center to center) distance of $146*10^6$ km.}.
	As can be readily visualized, employing only two satellites with fixed 
	geocentric orbits, regardless of their orbital parameters, would not 
	be sufficient to map the entire sky, even when each of their radio 
	astronomical antennas have a field of view of as much as $2\pi$ steradians 
	(refer to Figure~\ref{figure1}). Hence, our starting model consists of 
	three satellites in mutually orthogonal orbits; one in equatorial orbit 
    and the other two in mutually orthogonal polar  orbits. 
     These orbits have their axes along the x, y and z axes 
     of the reference coordinate system.
  
	The number of interferometric baselines is given by, 
	
	\begin{equation} \label{baselines}
	\hspace{1cm} n_b = \frac{n(n-1)}{2}
	\end{equation}
	
	where n is the number of antennas in the system. 
	Thus, in our model with three antennas, the number of possible 
	baselines would also be three.

	We consider all three orbits to be different from each other,
	with different orbital heights 
	(measured from the surface of the Earth), 
	making their orbital periods distinctly different, and the
	heights are deliberately chosen such that the orbits are mutually {\it asynchronous}.
	The latter ensures that the apparent baselines are different 
	during each orbital cycle, thus giving us significantly 
	filled $(u,v)$ coverage in a relatively short time span.

	In general, the satellites would orbit around the Earth in elliptical orbits, 
	with the Earth center being at one of the two foci. 
	The magnitude of the velocity (or the orbital speed) 
	of a satellite in this general case is given by:
	
	\begin{equation} \label{velocitygen}
	\hspace{1cm} s = \sqrt{GM_e(\frac{2}{r}-\frac{1}{a})}
	\end{equation}
	
	where, $G$ (= $6.67 * 10^{-11} N m^2 / kg^2$) is the 
	universal gravitational constant, $M_e$ (= $5.982*10^{24} kg$) is 
	the mass of the Earth, $r$ is the instantaneous radial distance of the 
	orbiting satellite
	from the Earth center,
	and $a$ is the semi-major axis of the elliptical orbit. 
	In our model, we have assumed, for simplicity, the LEO satellites to be in 
	a nearly circular orbit (i.e. $a \approx r$), so as to improve the 
	{\it uniformity} in $(u,v)$ coverage in both dimensions. Therefore, the average magnitude, $s$
	of the satellite velocity would be,
	
	\begin{equation} \label{velocity}
	\hspace{1cm} s \approx \sqrt{\frac{GM_e}{r}}
	\end{equation}
	
	and the corresponding orbital period T, 
	
	\begin{equation} \label{timeperiod}
	\hspace{1cm} T \approx \frac{2 \pi r}{s}
	\end{equation}
	
	where $r = R + h$, $R$ (= $6371 km$) is the radius of the Earth, 
	and $h$ is the height of the orbit above the Earth surface.

	The Earth gravity dominates, as expected, the considerations that 
	dictate the geocentric motion of the satellites,
	and in comparison, any effects of, particularly, the Sun and the Moon and 
	other solar system bodies can be ignored.
	Noting the long time scales associated with nutation, and even longer for precession
	(see, \cite{balmino1974coriolis}),
	compared to the time scales relevant to obtaining desired $(u,v)$ coverage,
	these two effects are not included in our model. Of course, the apparent astronomical
	source coordinates (such as RA, Dec) which are defined with reference to Earth rotation,
	do routinely need to take in to account the evolution of the Earth's spin axis.
	Also, the effects due to the apparent forces, such as the Coriolis force 
	(arising as a result of the Earth's rotation), those related to atmospheric drag, 
	tidal effects, solar wind pressure, etc., are relatively small
	(see for more details, \cite{balmino1974coriolis}), 
	and are therefore not considered.
	Even though some of these effects might be significant in the long-term operability 
	of the satellites, and the apparent directions of the source may be redefined
	in the equatorial coordinate system as the spin axis of the Earth on relevant
	long time-scales, these aspects have little effect on how the mutual separation 
	of the satellites would vary and the resultant spatial frequency coverage they
	provide.
	
	\begin{table*}[ht]
		\centering
		\caption{Defined parameters of the model.}\label{table1}
		\begin{tabular*}{0.7\textwidth}{@{}c\x c\x c\x c@{}}
			\toprule
			\textit{Index} & Orbital Height & Orbital Speed & Time Period \\
			~ & Above Earth's Surface & ~ & ~ \\ \midrule
			\textit{Satellite 1:} & 770 km & 7.48 km/s & 100.01 min \\ 
			\textit{Satellite 2:} & 1085 km & 7.32 km/s & 106.70 min \\
			\textit{Satellite 3:} & 1400 km & 7.17 km/s & 113.53 min \\ \bottomrule
		\end{tabular*}
	\end{table*}
	
	\begin{figure}[t]
		\begin{center}
			\includegraphics[width = 1\linewidth]{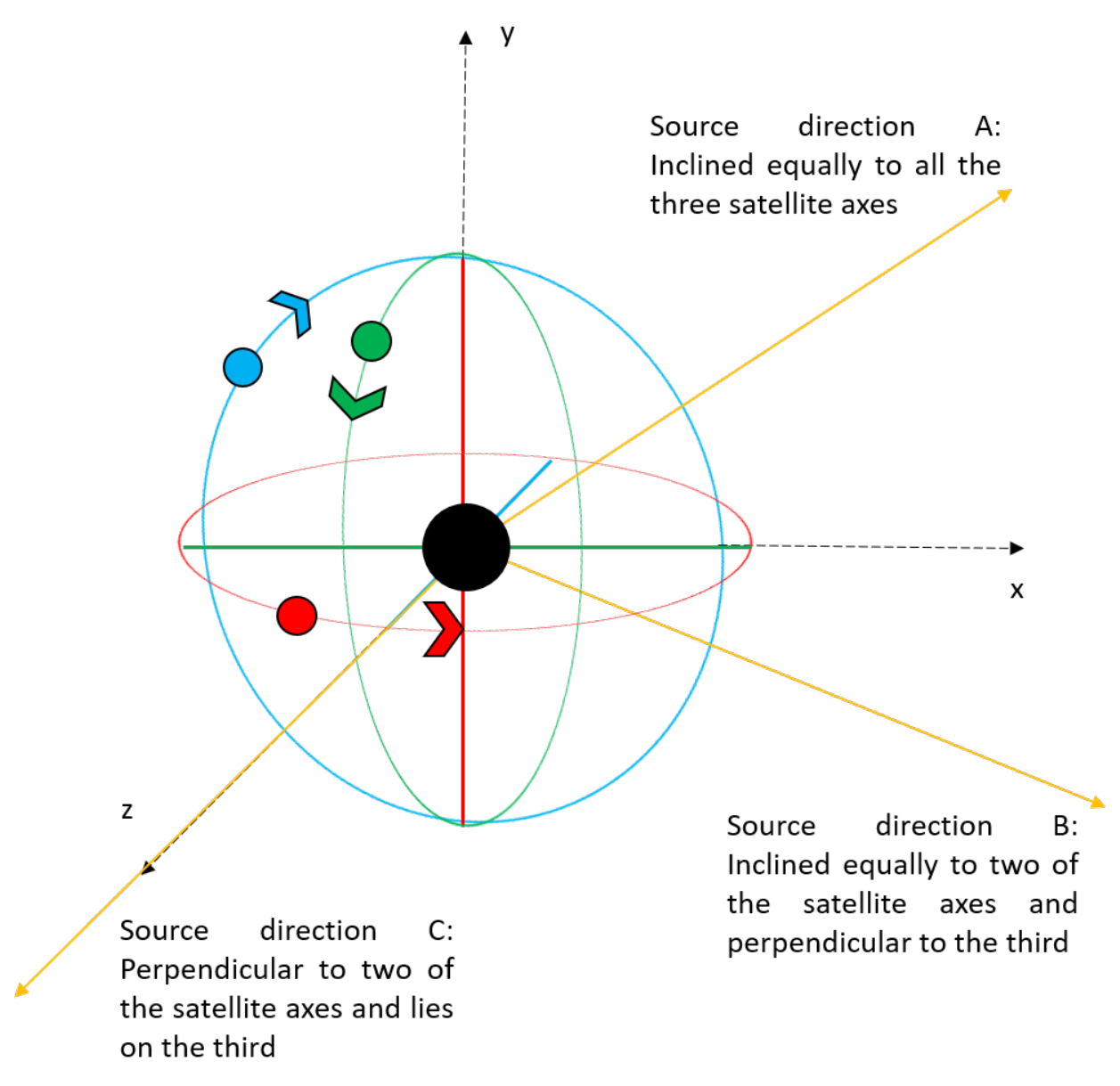}
			\caption{An illustration of our model configuration: 
				The near circular orbits (not to scale, and appearing in projection as ellipses) 
        for the three satellites are shown. The first satellite has an equatorial orbit 
        (red), and the other two (satellite 2 and 3) have polar orbits (blue and green, 
        respectively). The axes associated with the orbits coinciding with our 3-D 
        coordinate frame with its origin at the center of the Earth (denoted by the 
        black circle). The yellow rays indicate the three exclusive sky directions 
        we consider (to be discussed in Section~\ref{secrad}).} \label{figure2}
		\end{center}
	\end{figure}

	Unless otherwise mentioned, the beam width or the FoV of each antenna is assumed, for
	simplicity, to be 180$^{\circ}$, or $2\pi$ steradians, respectively, even though
	in reality it is impossible to have such a sharp truncation in antenna response 
    (refer to Figure~\ref{figure1}).
	Usually, beam widths of typical antennas would be expected to be narrower. 
	However, a near semi-isotropic beam can be achieved even with a short dipole 
	with a reflector, or using a suitably arranged array of aperture elements, discussion
	of which is beyond the scope of the present theme. 
	Our model does have a provision to consider narrower beam widths, 
	but most of our simulations assume the default beam width, 
	for the sake of computational efficiency. 
	
	Figure~\ref{figure2} shows the essentials of our model configuration, 
	with the Earth at the center, along with the assumed coordinate system 
	denoted by the x, y and z axes. 
	The satellites are shown in their orbits, with their orbit axes 
	along the x, y and z axes.
	Also indicated are the three distinct specific source directions 
	for which the $(u,v)$ coverage are assessed 
	(as detailed in the beginning of Section~\ref{secrad}). 
	
	Our model parameters, which can be varied, include 
	the orbital period (which in turn would change the orbital radius and 
	the orbital velocity),
	as well as the starting orbital phase, of each of the satellites.
	
	Although most of the distant astronomical sources have 
	a unique direction, commonly defined
	as the Right Ascension (RA), Declination (Dec), 
	our simulations allow computation 
	of spatial frequency $(u,v)$ coverage even 
	in the case of a relatively nearby source. Thus, in general, 
	the source distance can be provided, in addition to 
	the apparent direction (i.e. RA and Dec).
	By default, the source distance is assumed to 
	be very large (say, $10^9$ AU or $\approx$15800 light
	years), unless specified otherwise. 
	
	For a given direction RA and Dec, represented by $\alpha$, $\delta$ in radians,
	the 3-D position of a source at finite distance, $d_{source}$, can be expressed as,
	
	\begin{align} \label{sourcex}
	x_{source} &= (d_{source}*\cos(\delta))*\sin(\alpha) - x_{earth}\\
	y_{source} &= (d_{source}*\sin(\delta)) - y_{earth}\\
    z_{source} &= (d_{source}*\cos(\delta))*\cos(\alpha) - z_{earth}
	\end{align}
	
	where, $(x_{earth}, y_{earth}, z_{earth})$ define the Earth location w.r.t. 
	the Barycenter of the Solar system, and will vary depending on the phase of 
	the Earth orbit around the Sun.
	For the distant sources ($d_{source} \> \> AU$), the above relations
	will reduce to the more familiar description of a unit vector 
	($\hat{x}$,$\hat{y}$,$\hat{z}$), in the direction of the source, 
	neglecting the terms relating to the Earth's orbital position,
	and normalization by the distance $d_{source}$.

	\begin{figure*}[ht]
		\centering
		
		\begin{subfigure}{\linewidth}
			\includegraphics[width=0.33\linewidth]{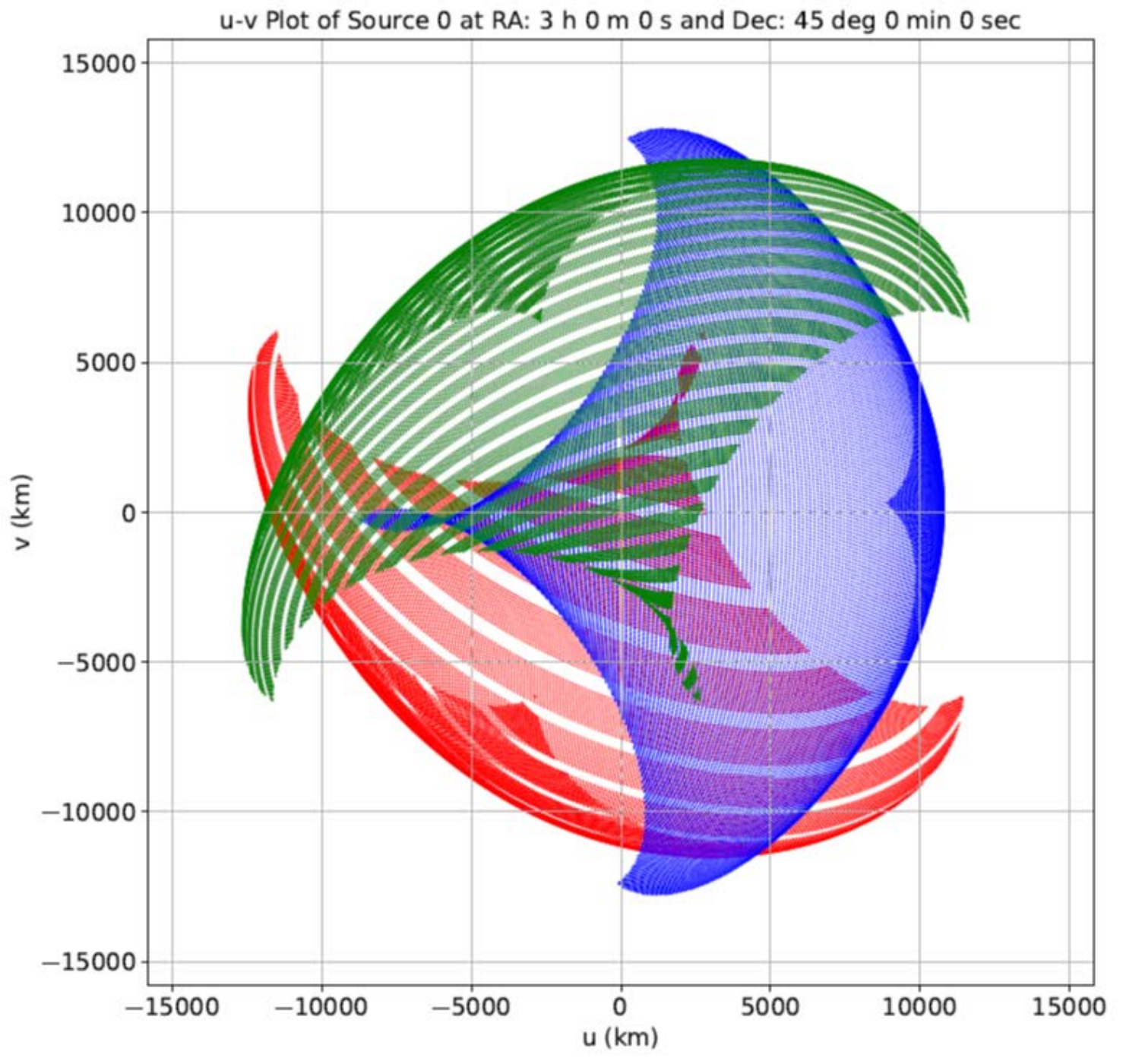}\hfill
			\includegraphics[width=0.33\linewidth]{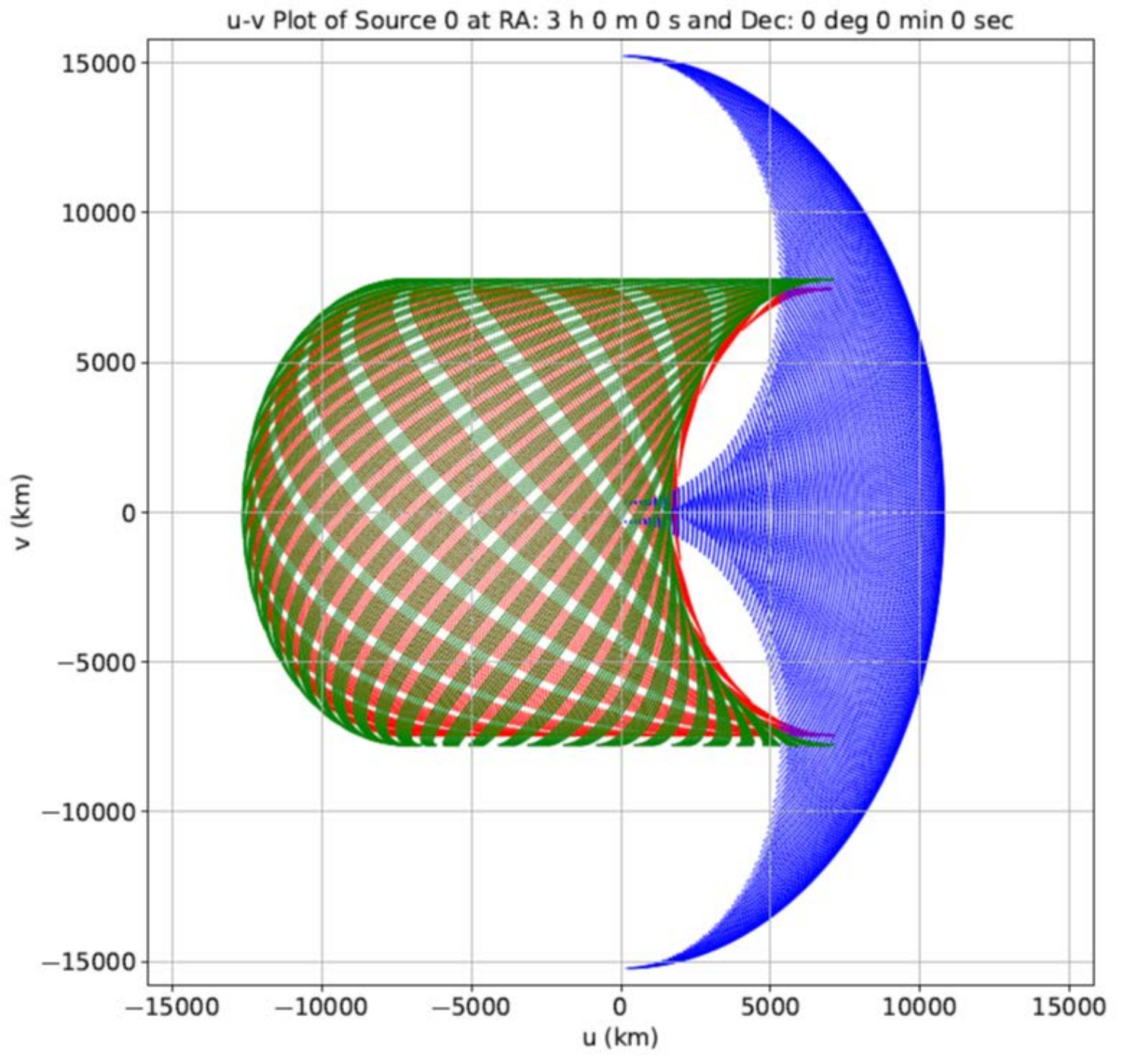}\hfill
			\includegraphics[width=0.33\linewidth]{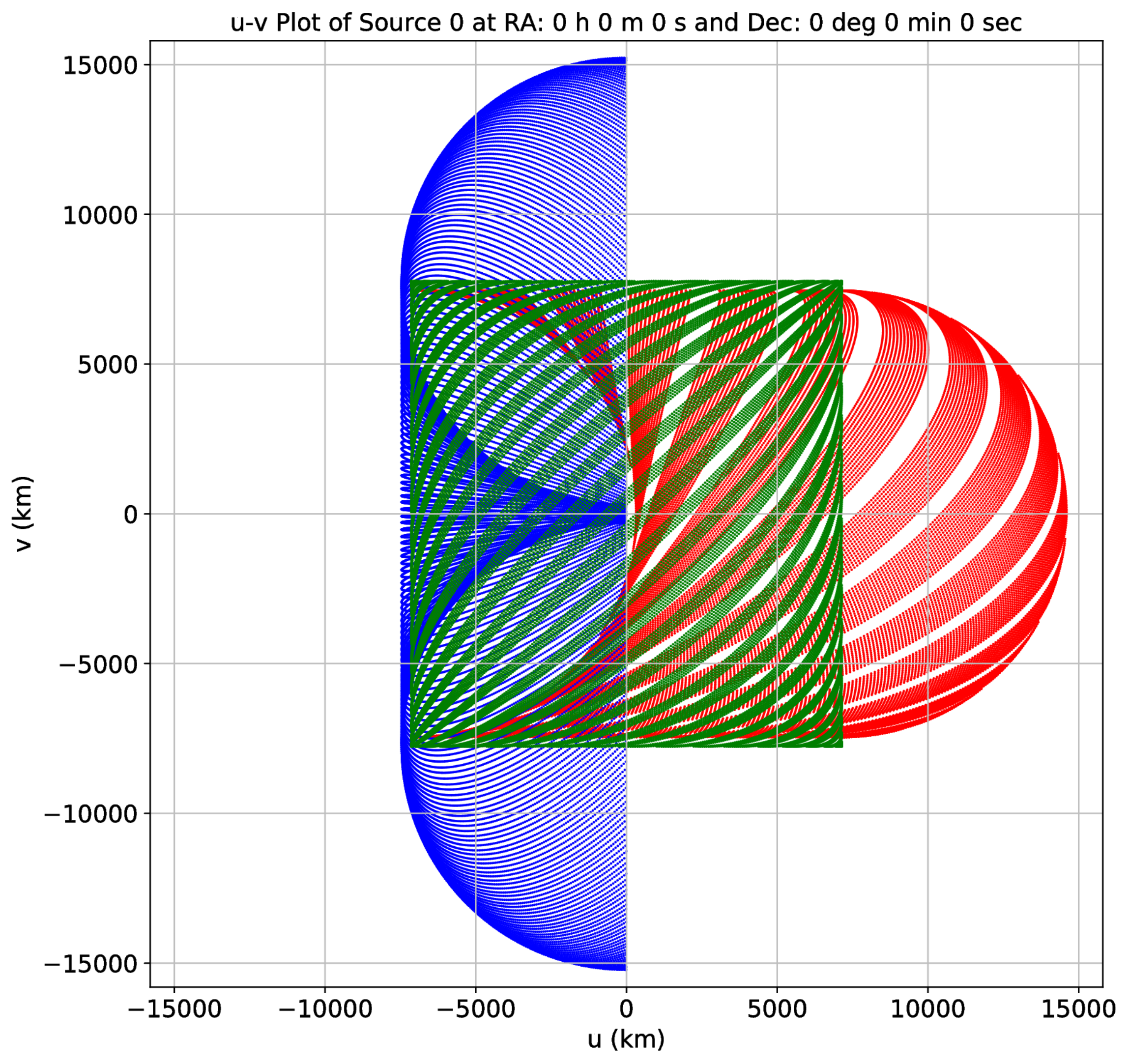}
		\end{subfigure}\par\medskip
		
		\begin{subfigure}{\linewidth}
			\includegraphics[width=0.33\linewidth]{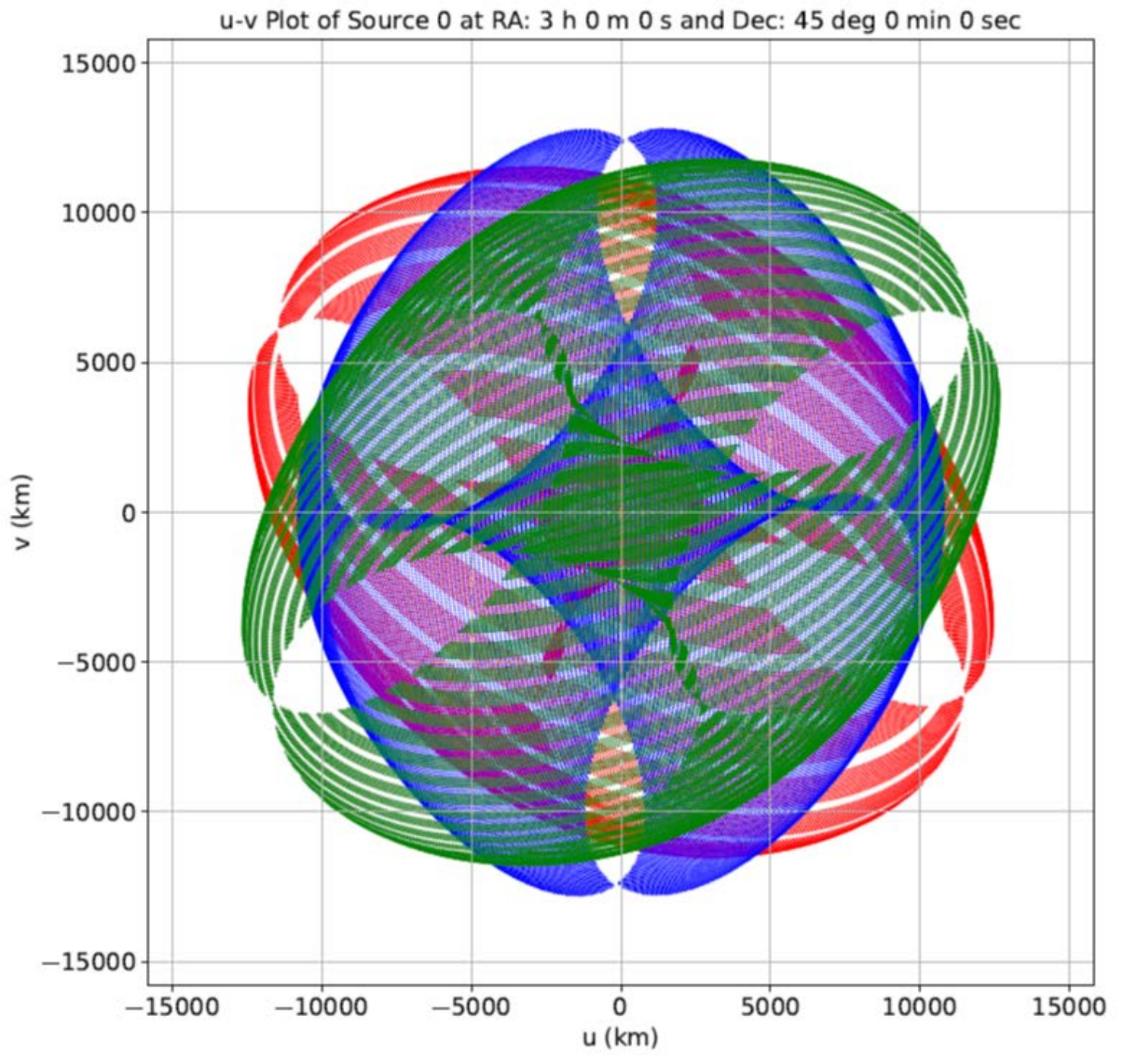}\hfill
			\includegraphics[width=0.33\linewidth]{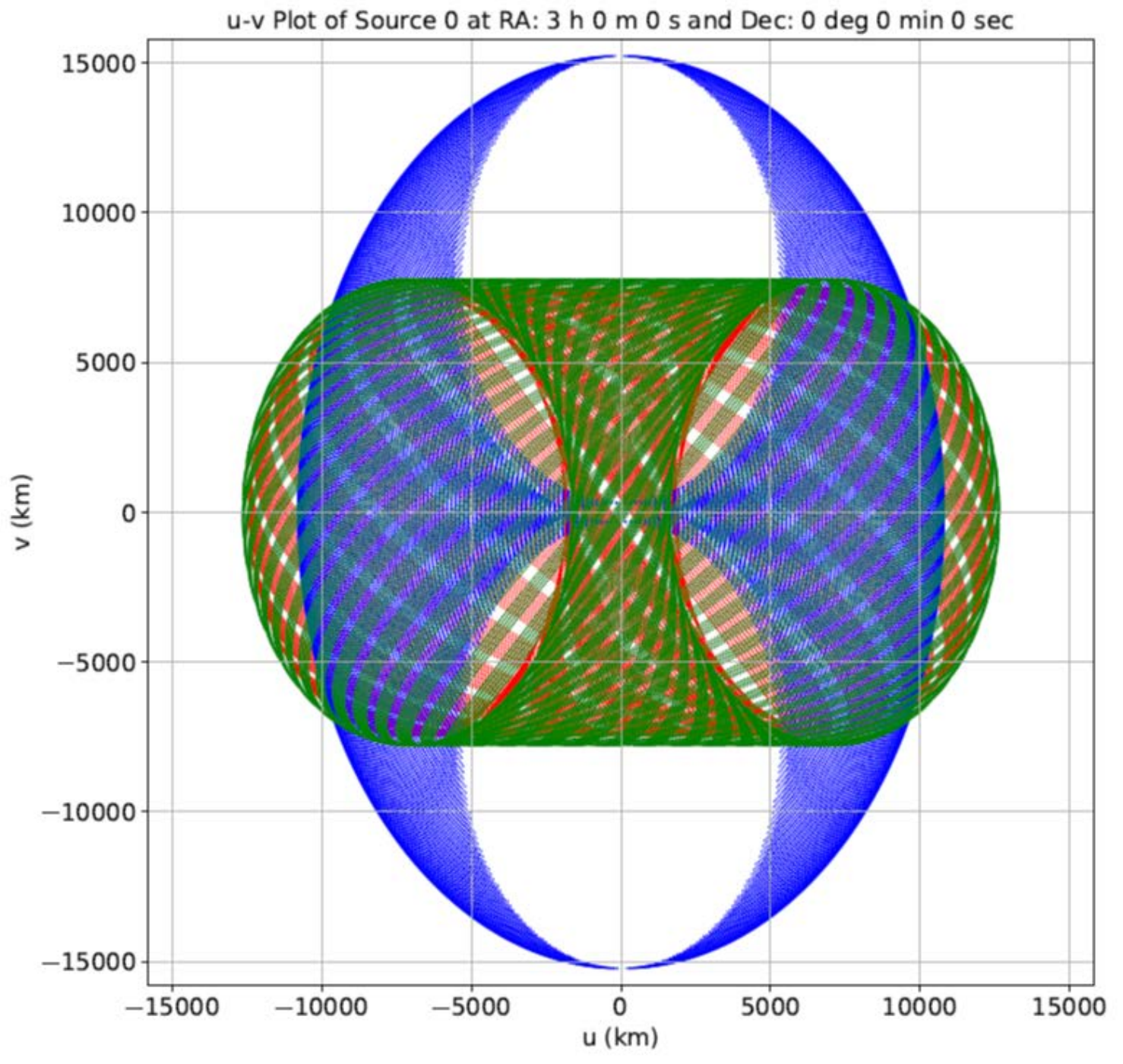}\hfill
			\includegraphics[width=0.33\linewidth]{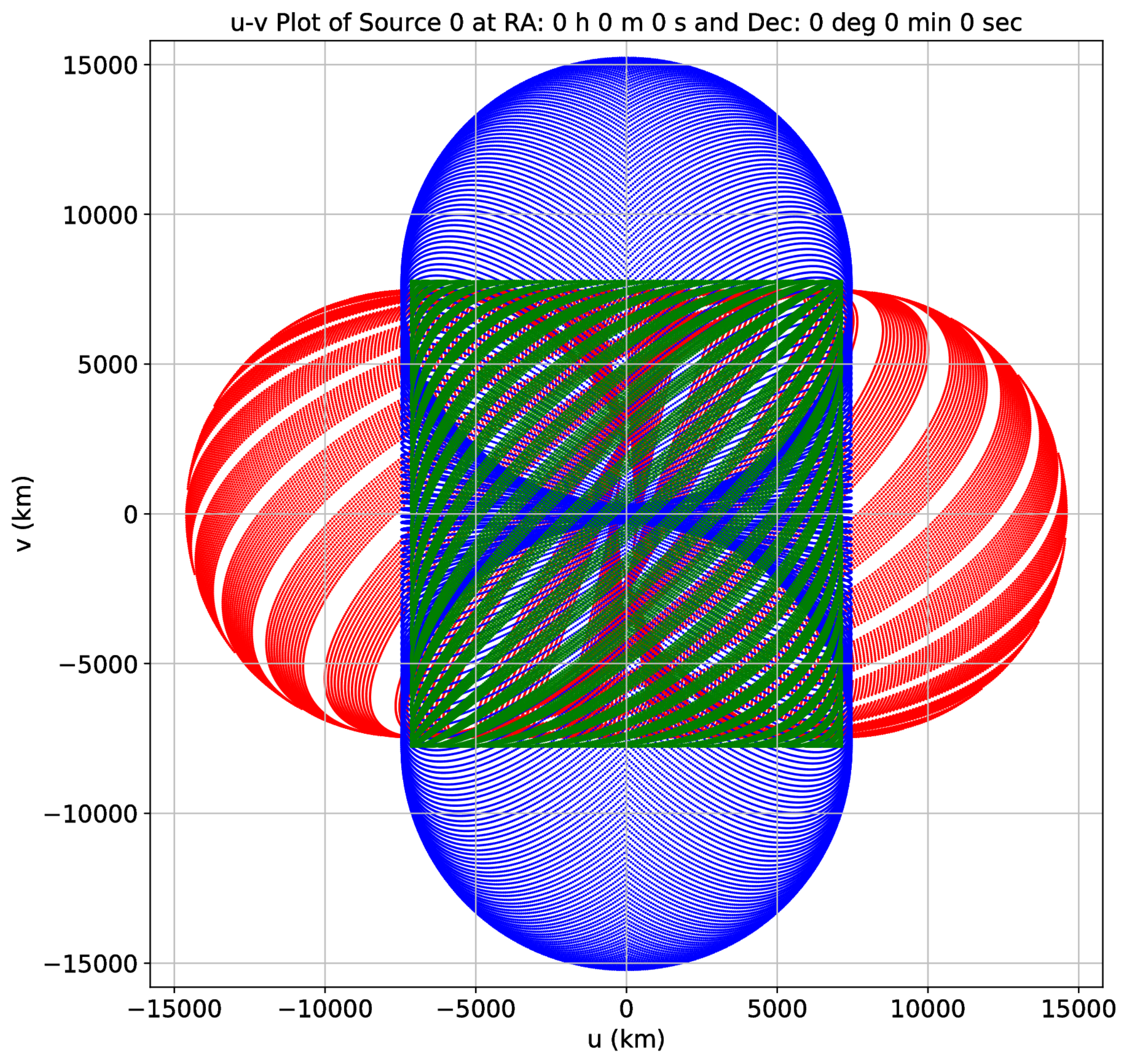}
		\end{subfigure}
		
		\caption{The $(u,v)$ coverages by the system of 3 satellites over the duration 
     of 16 days, for the three pre-defined source directions. The top and the bottom 
     rows of plots correspond to the $(u,v)$ coverage without and with the inclusion 
     of the sampling symmetry. The pair of plots in the three columns (left to right) 
     refer to the three special source directions, namely the sources A, B and C, 
     respectively, as defined in the main text. The red, blue and green tracks represent
     the baseline corresponding to the satellite pairs 1-2, 2-3 and 3-1 respectively. 
     The u, v axes are indicated in km.}
		\label{figure3}
	\end{figure*}

	The $(u,v)$ coverage obtainable in a time span, 
	by default, of typically 16 days is simulated, 
	unless a different span is specified. The observing time span 
	of 16 days is chosen after extensive testing because having longer time periods 
	does not provide any significant improvement in coverage. 
	
	The various parameter values defined for each satellite are given 
	in Table~\ref{table1}. Sensitivity of the $(u,v)$ coverages to 
	values of the listed (interdependent) parameters was examined by
	varying the relevant parameters.
	
	The orbital height of satellite 1 (in the closest orbit)
	is chosen to be greater than 700 km
	(in our simulations it is about 770 km)
	so as to be able to neglect the effects of atmospheric drag which 
	maybe substantial at orbital heights below 400 km and to also 
	avoid attenuation of the radio waves due to the ionospheric plasma which 
	is significant at orbital heights below 700 km 
	(more about this is discussed in Section~\ref{seccfs}). 
	
	The orbital radii of satellites 2 and 3 are larger by 315 and 630 km, respectively, 
	than that of the satellite 1,
	so as to realize as distinct baselines from the different pairs as possible,
	while still keeping the satellites well within the Low Earth Orbit limit,
	and having comparable time periods. 
	The present values of the orbital radii are seen to 
	give the best coverage, as evident from the tests 
	described in the later sections, 
	regardless of their starting orbital phases. 
	Many other combinations of orbits at larger radii
	can also potentially provide similar coverages, but 
	would necessarily require correspondingly longer time spans.
	
	Since our model is different from the usual Earth-based and 
	Earth-and-space-based interferometer setups, 
	we need to define the $(u,v)$ plane explicitly in a suitable, 
	though different, manner.
	
	In order to do that, we have defined the $(u,v)$ plane with its origin
	at the center of the Earth, and the 
	mutually orthogonal (unit) vectors, $\hat{u}$ and $\hat{v}$ are both also 
	perpendicular to the chosen source direction ($d_{source}$; same as that of the
	unit vector $\hat{w}$).
	
	Following the usual convention, $\hat{v}$ always lies in the plane containing
	the source direction and the Earth rotation axis (i.e., both the north and 
	the south poles). 
	
	Note that the $u$ and $v$ values are presently expressed in km 
	and not in the conventional units of wavelength. The Hermitian symmetric 
	nature of the visibilities and the consequent symmetric sampling/coverage in the
	$(u,v)$ plane is shown only in some of the figures. In other cases, only 
	one sample point in $(u,v)$ plane is counted/shown per baseline per time 
	(instead of two due), to avoid cluttering in the display of coverage. 

	\begin{figure*}[ht]
		\centering
		
		\begin{subfigure}{\linewidth}
			\includegraphics[width=0.33\linewidth]{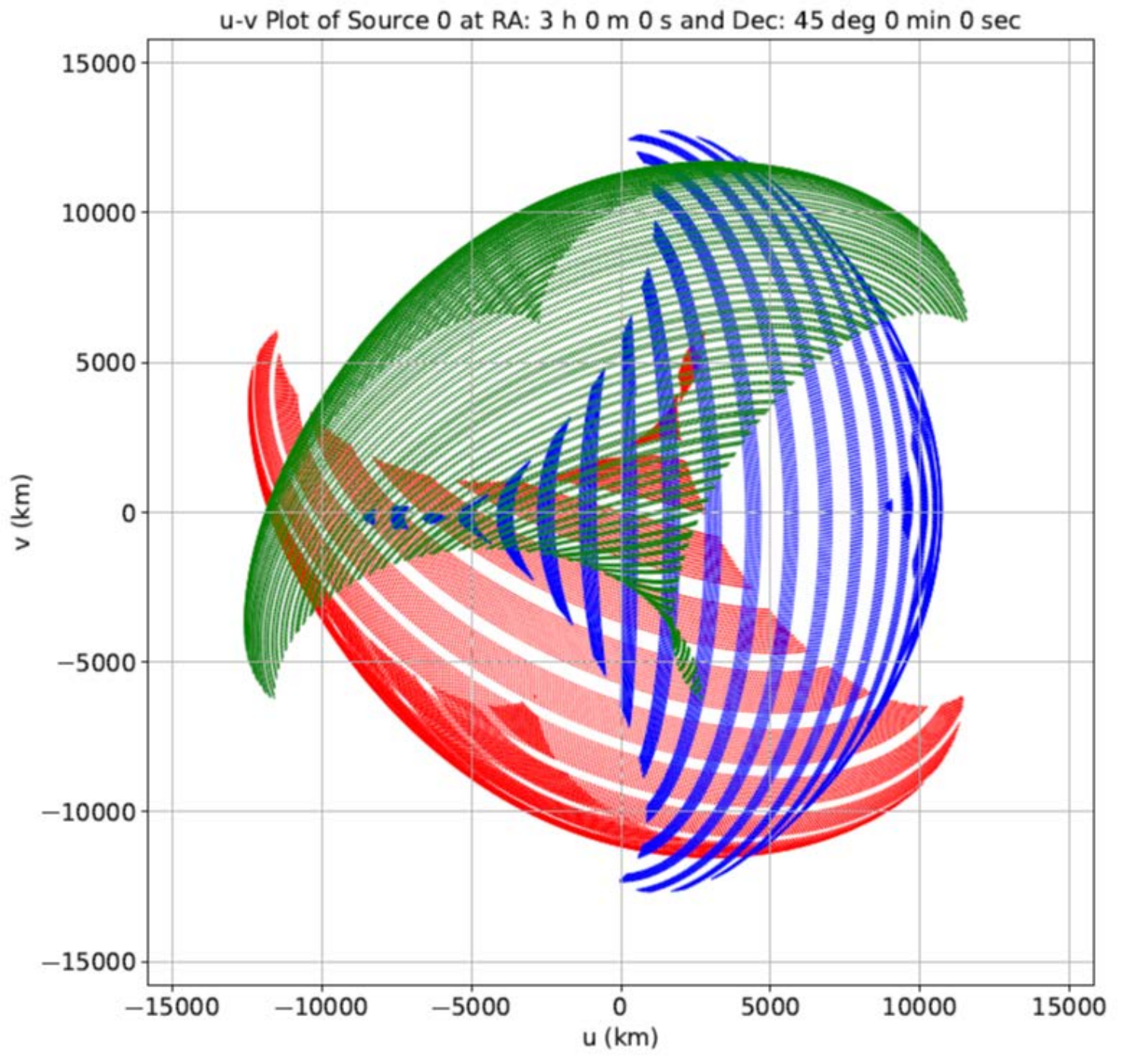}\hfill
			\includegraphics[width=0.33\linewidth]{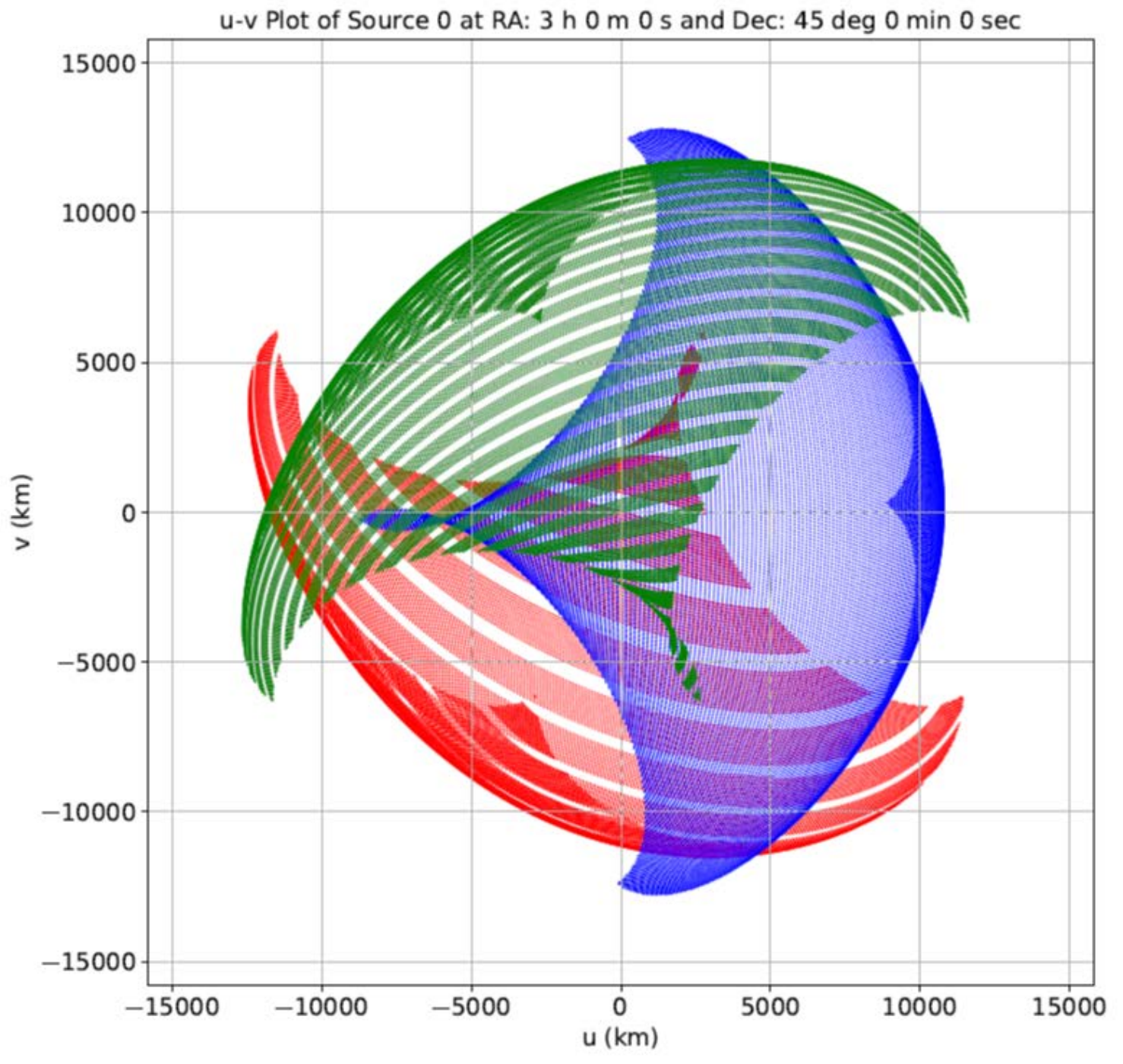}\hfill
			\includegraphics[width=0.33\linewidth]{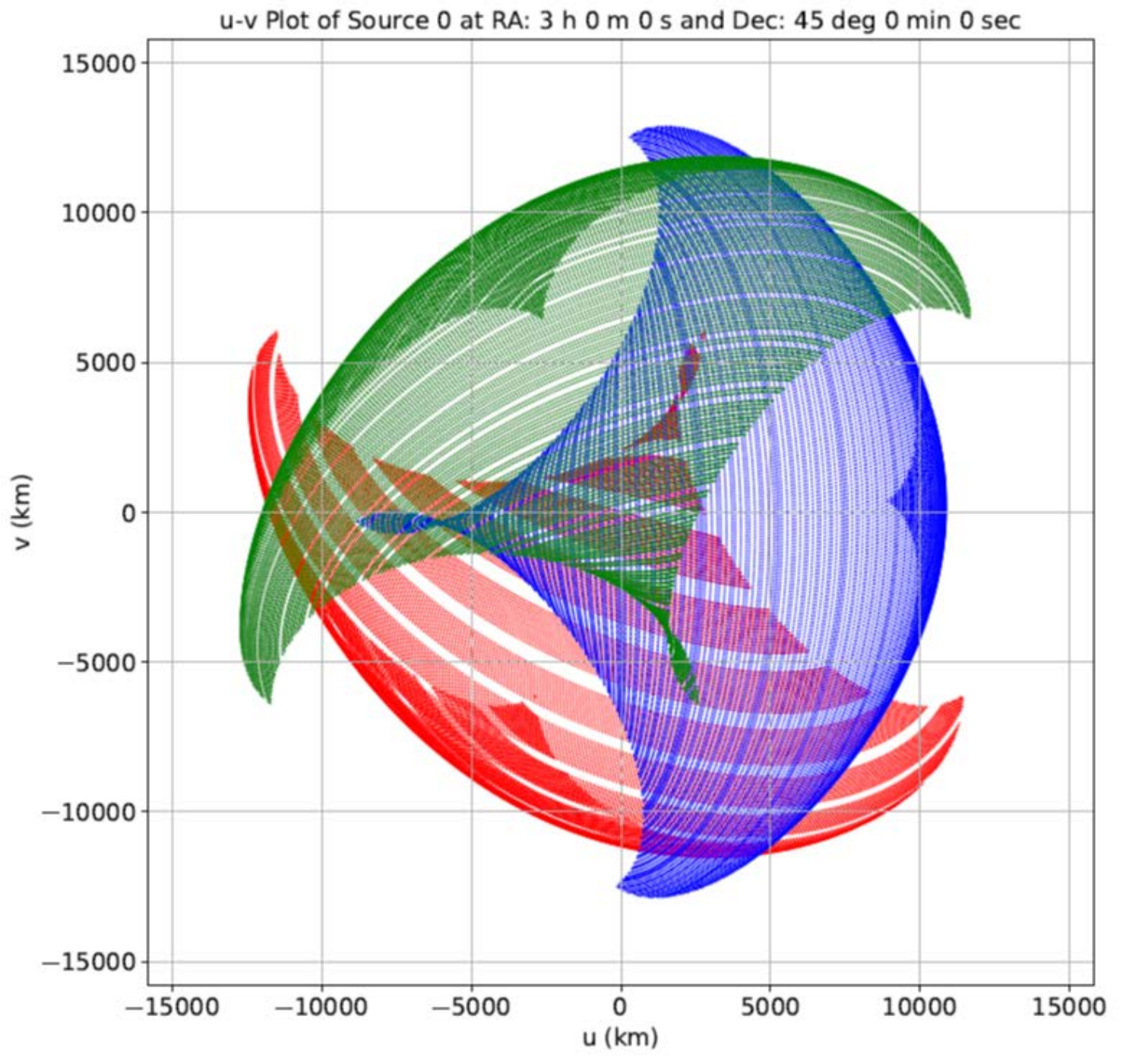}
		\end{subfigure}\par\medskip
		
		\begin{subfigure}{\linewidth}
			\centering
			\includegraphics[width=0.33\linewidth]{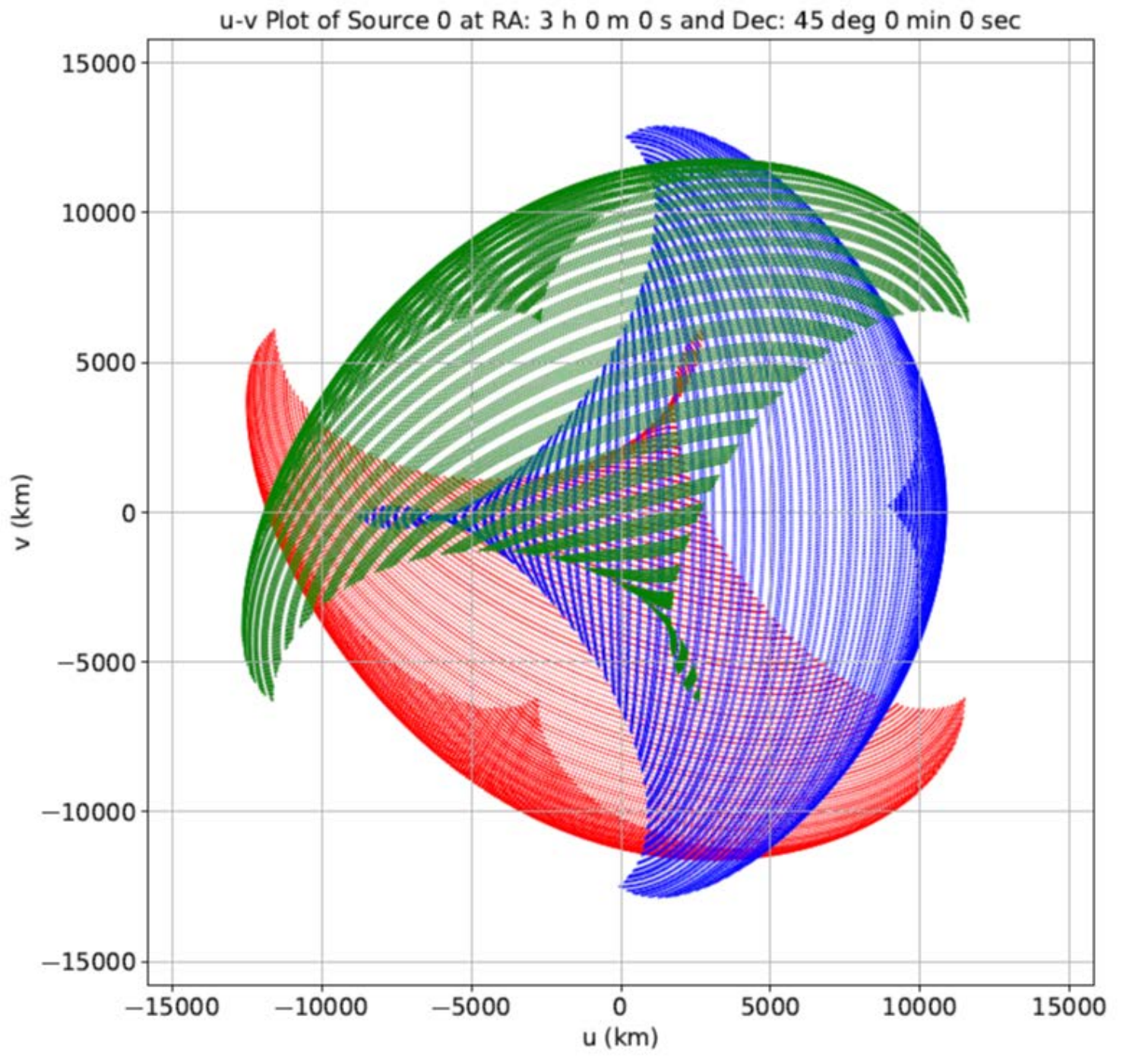}
			\includegraphics[width=0.33\linewidth]{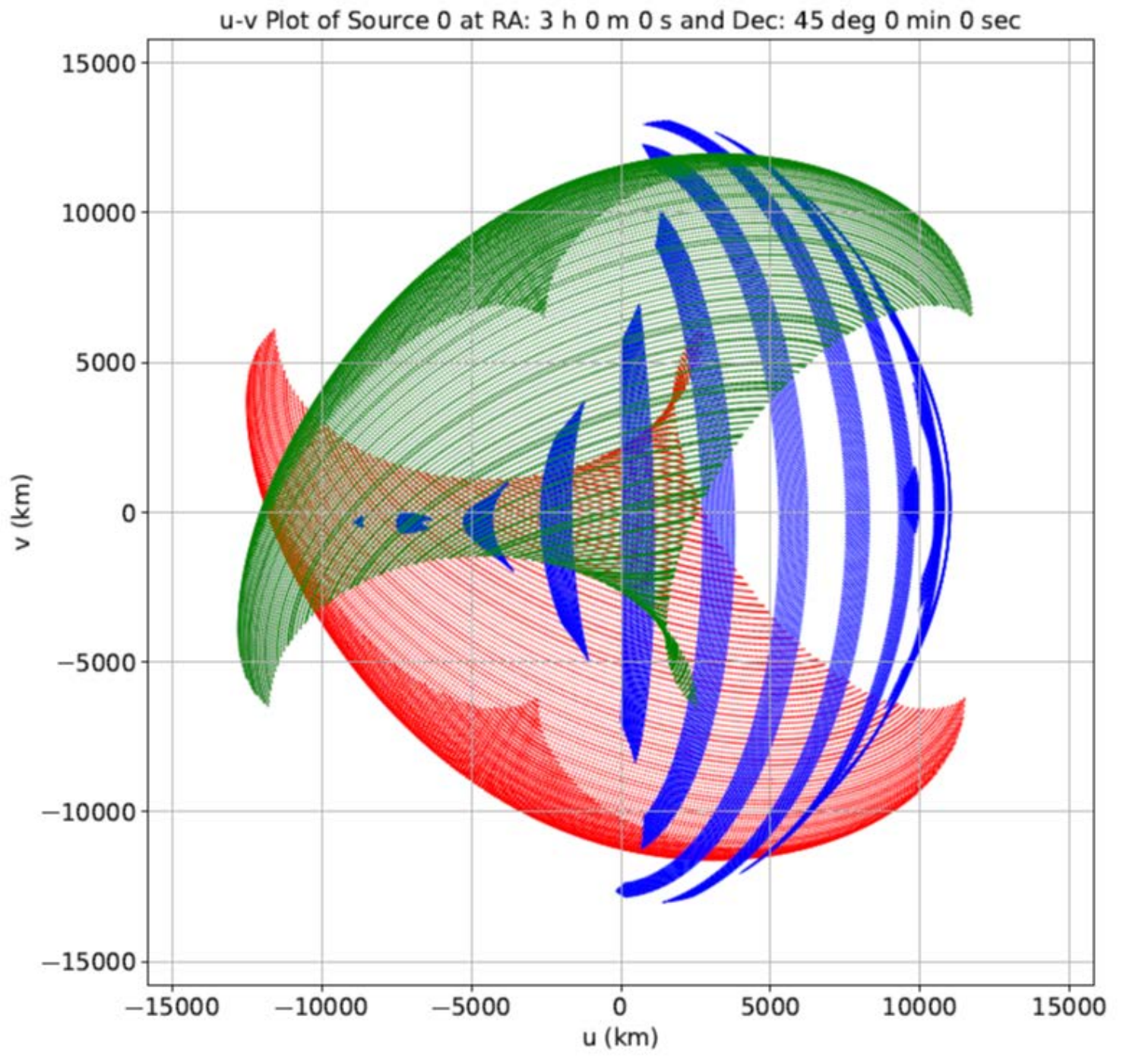}
		\end{subfigure}
		
		\caption{The $(u,v)$ coverage obtainable in a 16 days time span for 
			the source direction A is shown. For clarity, the symmetric counterpart of the
			coverage (implied by the Hermitian symmetric nature of the visibilities) is
			not displayed. The u and v axes are marked in km.
			A total of 5 configurations are shown for assessing sensitivity 
			to orbital periods (corresponding to assumed heights).
			The height of the orbit of the Satellite 1 is kept constant at 770 km,
			while coverages in the top and bottom rows assume the height 
			of the Satellite 2 orbit to be 1085 and 1185 km, respectively.
			In usual order, the plots correspond to the Satellite 3 height 
			of 1300, 1400, 1500 km in the top panels, 
			and 1400, 1600 km in the lower panels.
			From among these cases, the coverage with the fourth 
			configuration (lower-left panel,
			with the set of orbit heights 770, 1185 and 1400 km)
			appears the most uniform.} \label{figure4}
		
	\end{figure*}


	\section{Simulation Results and Discussion} \label{secrad}
	In this section, we examine and discuss the results of our simulation,
	in terms of the achievable $(u,v)$ coverage, and dependencies on
	model parameters/configurations, assessed for a set of source directions.
	
	During our simulations, we follow the locations of the satellites, 
	to estimate the $(u,v)$ spacings, at time interval of 10 seconds, and
	note the sampled spacing in the $(u,v)$ plane spanning $\pm$15750 km. 
	Although in such a time interval the baselines can change significantly,
	sometimes by as much as 100 km, the behaviour at finer time intervals,
	when required, can be found out to the desired accuracy directly, or even
	with suitable interpolation.
	This choice of time interval certainly makes the simulations
	computationally efficient, while still retaining the essential
	details of the $(u,v)$ coverage for the purpose of illustration.
	
	In order to simplify assessment of the maximum and minimum possible coverages 
	achievable with our model system for any source direction, we define 
	three special source directions (referred to as the three 
	exclusive directions as in the Figure~\ref{figure2}) for which we 
	compute and examine the $(u,v)$ coverage. 
	These three special source directions, namely A, B, and C, are defined as follows:
	
	\begin{enumerate}
		\item Direction A: RA 03:00:00; Dec +45 deg, inclined equally to all 
		three satellite orbit axes (equivalent situation would be
		encountered for RA of 9, 15 and 21 Hrs, 
		and/or declination of -45 deg).  
		\item Direction B: RA 03:00:00; Dec 0.0 deg, representing a set of specific directions
		that are perpendicular to 
		one of the satellite orbit axes (or in the plane of that orbit) 
		and inclined equally with the other two;  
		\item Direction C: RA 00:00:00; Dec 0.0 deg; Perpendicular to any two of the 
		satellite orbit axes and parallel to the third axis. 
	\end{enumerate}

	Figure~\ref{figure3} illustrates how the $(u,v)$ coverage differs 
	for each of three special source directions, 
	estimated over a default span of 16 days 
	\footnote{The evolution of the $(u,v)$ coverage 
	during the 16 days (corresponding
    to each of the maps in Figure 3), can be viewed at:  
    \url{https://bit.ly/SpaceInterferometer}}.
	The situation for any other choice of source direction would 
	typically be in between the cases equivalently represented by these
	three special source directions, and 
	the $(u,v)$ coverage would hence be also in between the range of 
	baseline-wise coverages seen for the special source directions 
	(a quantitative assessment of the coverages is provided 
	in the subsection~\ref{subsecqmc}). 
	The $u$ and $v$ axes in these figures are shown in km rather 
	than the usual wavelength ($\lambda$) units.
	Alternatively, the numerical values may be treated as referring to 
	spatial frequencies, in units of the wavelength, at the radio 
	frequency of 0.3 MHz,
	but can be scaled trivially to any other radio frequency.
	
	These results show that the system of three satellites 
	renders sufficiently large coverage for all of the 
	special source directions probed, and  
	can be expected to provide comparable coverage for 
	other sky directions as well. If the primary field of view
	were to be narrower than assumed presently, the
	coverage would be correspondingly less extended,
	with little change in the detailed sampling of remaining region.
	A major or rather a dramatic adverse impact on the coverage 
	should of course be expected when the narrowness of 
	the field of view hampers any simultaneous 
	observation of a given direction of interest
	by even one of the three satellites, with loss of two baselines.
	The directions which may fall outside those sampled 
	by two satellites yield no coverage at all.
	Thus, the importance of ensuring that the equipment on each satellite
	is capable of observing over a sufficiently wide spread of directions 
	(ideally $\pm 90^{\circ}$)
	about their respective central (radial) directions cannot be overstated.
	This would require either a suitable multi-beam arrangement that allows
	a wide-field coverage collectively, or 
	a single beam which can at least be steered to a chosen direction within
	the mentioned range, desirably without any reduction in the effective 
	aperture/collecting area, facilitating a corresponding 
	narrow-field measurement.
	
	We have also assessed if the attained coverage for a randomly 
    chosen source direction would be qualitatively similar to the 
    3 cases presented. The related simulations indeed show
    a similar coverage span and fineness of sampling in $(u,v)$, differing
    only in the overall orientation of the $(u,v)$ tracks and the patch
    covered by the three interferometers, depending on the chosen direction.
     
    Although for the presently discussed configurations, 
    much of the potentially obtainable $(u,v)$ coverage may be 
    {\it spanned} in typically a few tens of days (e.g., 16 days), 
    the considerations relevant for obtaining desired 
    image quality are not limited to the span and fineness in 
    $(u,v)$ coverage, but necessarily include the various consistency 
    checks between repeated observations, achievable sensitivity 
    on long integrations, RFI detection and excision, 
    calibratibility for various system parameters (in angular and 
    spectral domains), understanding and accounting for 
    systematics, etc.  
    Ensuring image fidelity requires even more demanding 
    considerations, which are beyond the scope of present paper, 
    but are pivotal to reaching the sensitivity required 
    for reliably detecting very weak signals, such as those 
    related to the Epoch of Reionisation (see, for example, the review by \cite{liu2020data}).

	\begin{figure*}[ht]
		\centering
		
		\begin{subfigure}{\linewidth}
			\centering
			\includegraphics[width=0.33\linewidth]{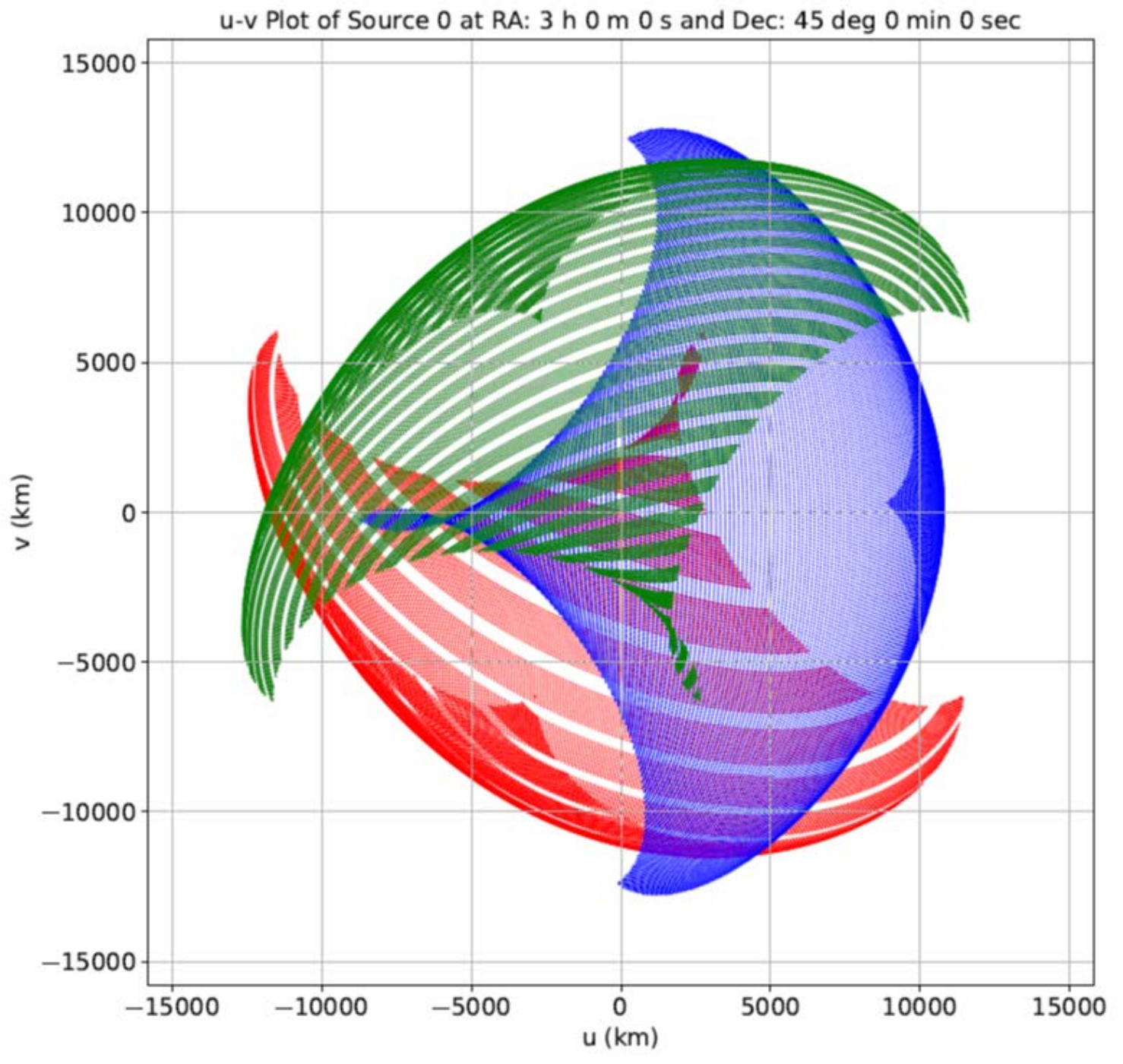}
			\includegraphics[width=0.33\linewidth]{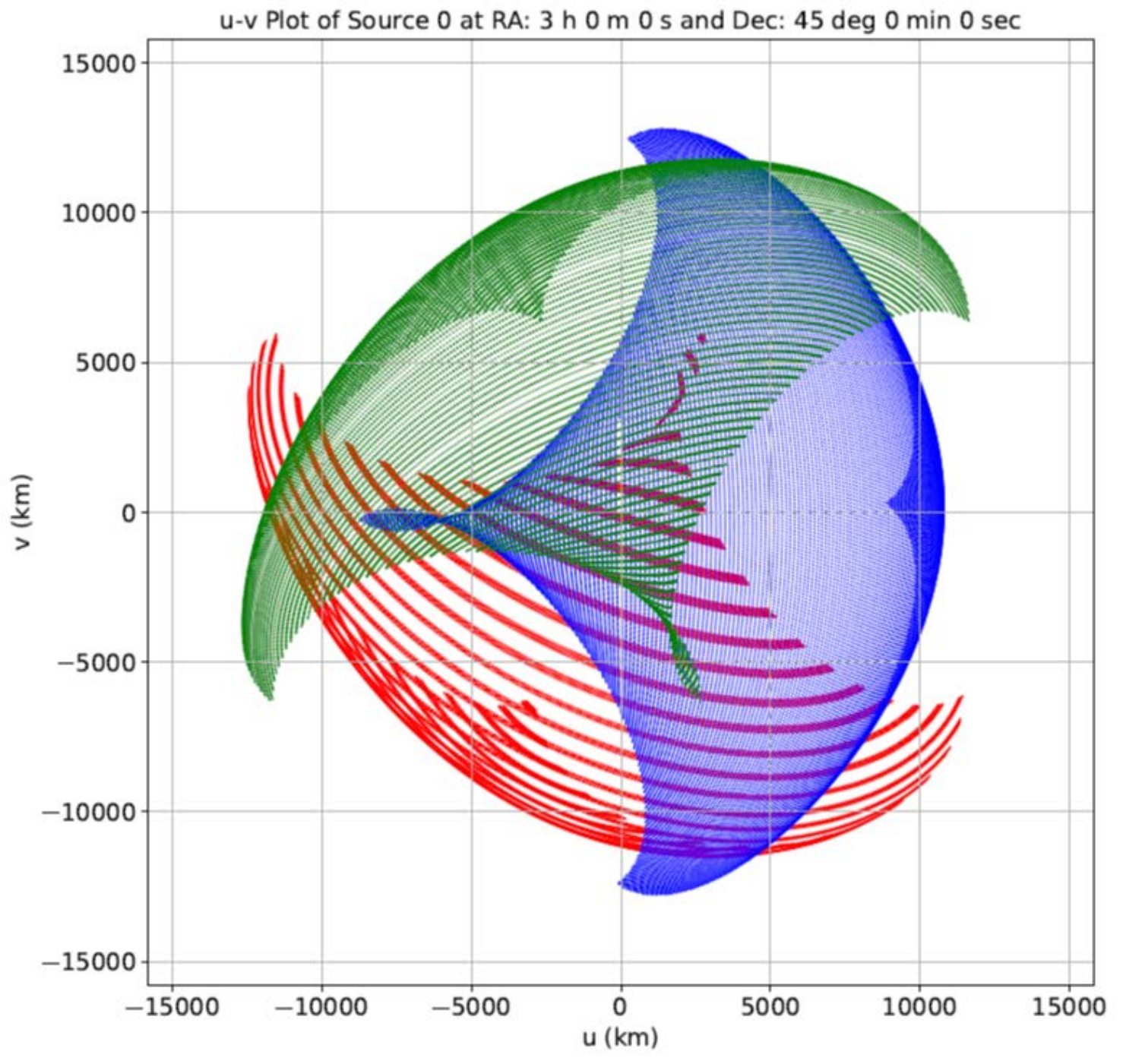}
		\end{subfigure}\par\medskip
		
		\begin{subfigure}{\linewidth}
			\centering
			\includegraphics[width=0.33\linewidth]{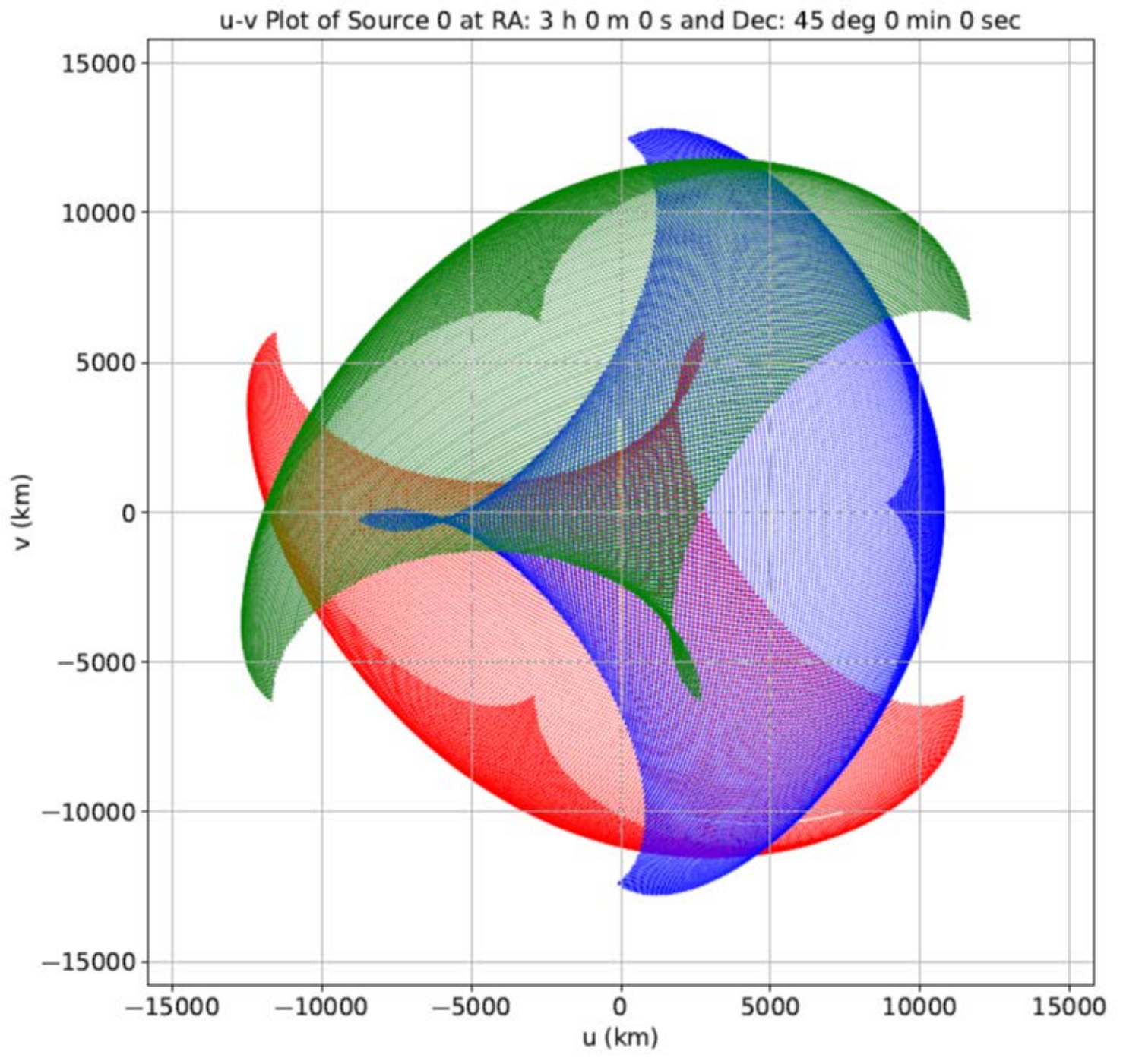}
			\includegraphics[width=0.33\linewidth]{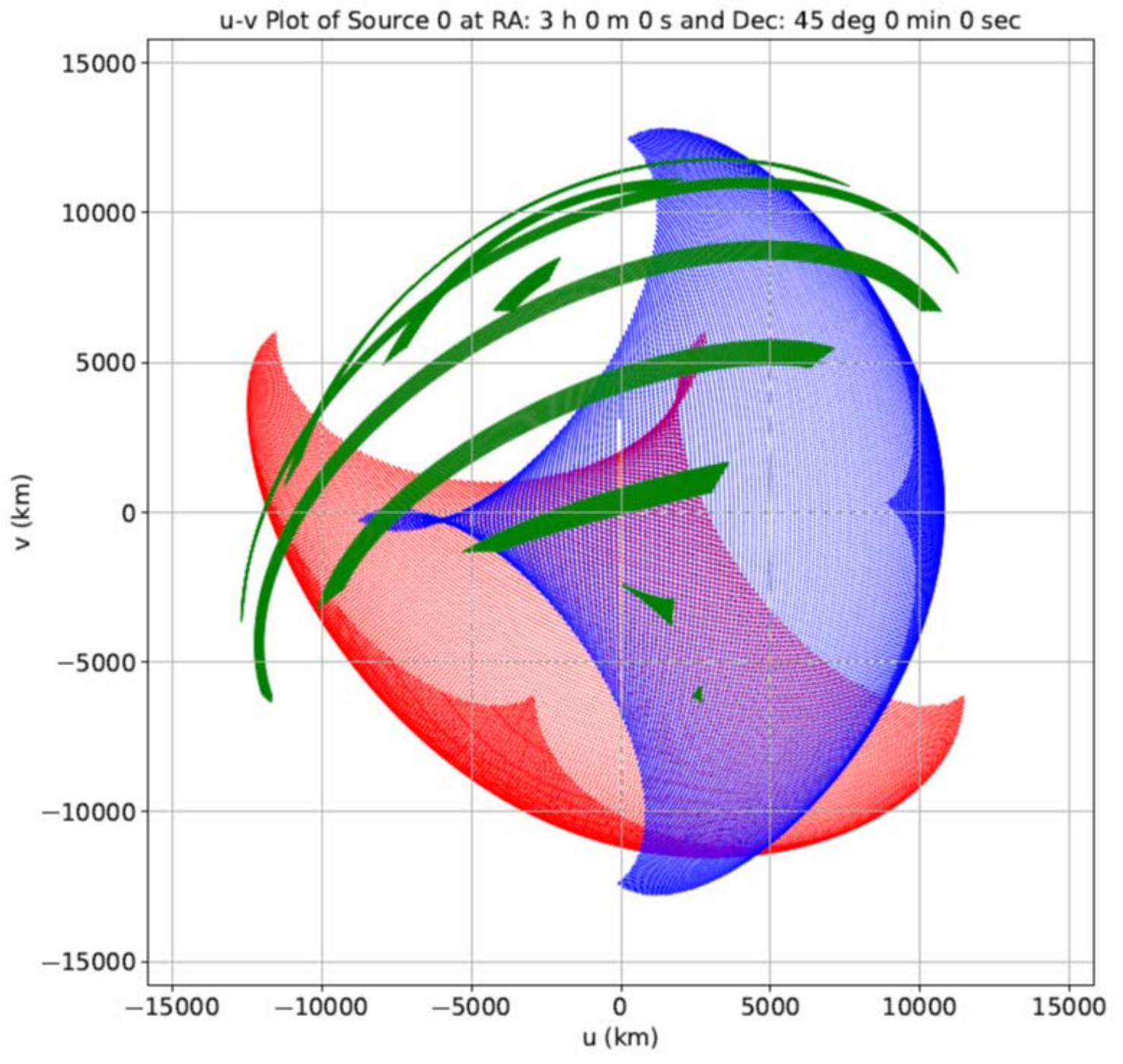}
		\end{subfigure}
		
		\caption{The $(u,v)$ coverage obtainable in a 16 days time span for 
			the source direction A is shown, excluding the symmetric counterparts, 
      and the u and v axes are in km. 
			In usual order, the panels show 4 cases corresponding to
			the following initial orbital phase 
			for Satellite 1:- 0$^{\circ}$, 10$^{\circ}$, 30$^{\circ}$, 45$^{\circ}$ || 
			The significantly higher uniformity in the $(u,v)$ coverage 
			is readily evident for the case with 30$^{\circ}$ phase.
		} \label{figure5}
	\end{figure*}
	
	\subsection{Sensitivity of the $(u,v)$ Coverage to the Orbital Parameters} \label{subsecscop}
	The sensitivity of coverage to the two key parameters, namely the
	orbital period and orbital phase,
	is assessed separately.  
	For these assessments, we probe the coverage for all of the three exclusive 
    source directions but present only a few illustrative samples, 
    all of them for source direction A, of the large ensemble 
	of cases/combinations probed.
	
	The sensitivity of coverage is assessed first by varying the orbital period 
	of any one satellite at a time, keeping the orbital periods of the other two
	unchanged, and then repeating the procedure with the 
	other two satellites. Since the orbital period of the satellite 
	is directly related to its distance above the Earth, 
	so we chose to vary these distances instead, and those values have 
	been mentioned here henceforth. 
	The corresponding orbital period can readily be 
	computed using Equation (\ref{timeperiod}). 
	
	Figure~\ref{figure4}
	shows a few examples of the $(u,v)$ coverage attainable in 16 days duration.
	Our results of these extensive assessment suggest that from
	among the different 
	orbital distance combinations, the configuration in which 
	the satellites 1, 2 and 3 are at 770 km, 1185 km and 1400 km 
	above the surface of the Earth, respectively, provides the maximum
	as well as most 
	uniform spatial frequency coverage. 
	
	\begin{table*}[h]
		\centering
		\caption{Pair-wise and total $(u,v)$ coverage with 3 satellites for 
        different chosen directions observed for a span of 16 days and 8 days 
        respectively.}\label{table1a}
		\begin{tabular*}{\textwidth}{@{}c\x c\x c\x c\x c\x c\x c@{}}
			\toprule
			\textit{Baseline or} & \multicolumn{3}{c}{Percentage Coverage for} & \multicolumn{3}{c}{Percentage Coverage for} \\
			\textit{Satellite Pair} & \multicolumn{3}{c}{16 days in Direction of} & \multicolumn{3}{c}{8 days in Direction of} \\
			\cmidrule(lr){2-4}\cmidrule(lr){5-7}
			~ & Source A & Source B & Source C & Source A & Source B & Source C \\ \midrule
			\textit{ 1-2} & 39.5 & 34.5 & 37   & 23   & 20.5 & 20  \\
			\textit{ 2-3} & 46   & 34   & 38.5 & 30.5 & 26   & 22.5 \\
			\textit{ 3-1} & 39.5 & 35.5 & 21.5 & 24   & 22   & 17.5  \\ \midrule
			\textit{All 3 Combined} & 64 & 55.5 & 62 & 52 & 47.5 & 43.5 \\ \bottomrule
		\end{tabular*}
	\end{table*}
	
	The sensitivity of coverage is also assessed by considering 
	different combinations of the relative initial phases of the satellite orbits. 
	Figure~\ref{figure5} illustrates results of some of these
	combinations, showing coverage over a duration of 16 days.
	From among the different combinations of
	initial orbital phases that were probed, the one in which the 
	satellites 1 has an initial orbital phase of 30$^{\circ}$ 
	is seen to provide the most uniform and highest density $(u,v)$ coverage.
	
	The final values of the orbital parameters
	(as mentioned in Table~\ref{table1}) were chosen after 
	probing numerous cases of coverage and obtaining the
    best possible uniformity, and hence the highest possible density, 
	of coverage, attainable over a duration of 16 days, for all 
	the three special source directions. 
	
	It is important to emphasize that the duration of 16 days, 
    as a specific time-span referred to here and later, is picked 
    merely as an indicating interval over which 85\% or greater of 
    the potential maximum coverage that a given baseline offers 
    is obtained. Beyond this time, the approach to the respective 
    maximum coverage is relatively slow.  It is also worth 
    remembering that the so-called coverage is assessed in 
    the $(u,v)$ plane which, for simplicity, has a rather coarse 
    grid (100 km) presently (more on this in subsection~\ref{subsecqmc} 
    and the corresponding footnote). Assessment on a finer grid 
    in $(u,v)$ plane would correspondingly stretch the time spans 
    required for similar fractional coverages. However, it 
    is important to note that much of the {\it extent} of 
    the potential coverage in $(u,v)$ is spanned in the mentioned 
    duration of about 16 days or even quicker, and 
    the {\it gaps} get filled progressively with increasing time.
	
	It should also be emphasized that while some 
    specific combinations of the 
	orbital parameters might appear to provide noticeably
	better coverage than others, its significance is to be appreciated
	more in terms of the speed of coverage. When assessed over
	suitably longer durations, most combinations of orbital parameters
	relevant to LEO would provide $(u,v)$ coverages that compare well with the best
	possible for the 3-satellite system.
	Thus, if higher coverage is a priority, then the sources can be observed 
	for a suitably longer duration even if the orbital heights (or periods) and 
	initial orbital phases of the satellites are not optimal.
	
	In general, optimizing one parameter at a time 
    sequentially may not yield optimal combination.
    However, in the present case, the altitude is merely a proxy for
    the orbital period (or frequency), which dictates the rate of change
    of orbital phase. Hence, it is not at all surprising that both,
    the altitude and the initial phase, have
    similar effect as they together decide the 
    orbital phase, and hence the location of a given satellite. 
    As a consequence, the baseline
    vector defined by relative locations for the pair of satellites would
    also respond to the combination of orbital phases (each in turn
    depends only on the combination of period and initial phase).
    
    Now we discuss the implications of the fact that, in practice,
    exact match to a specified orbit, such as in our model, may not be assured.
    There will, of course, be finite sensitivity to the parameters,
    including inclination and Right Ascension of the Ascending Node (RAAN). 
    Fortunately, the absolute RAANs or inclinations, and therefore 
    {\it absolute} orientations of the orbital axes,
    are not important in the present astronomy context,
    as long as the relevant differences ensure mutual orthogonality.
    With three mutually orthogonal orbits, one indeed ensures that
    each octant has similar set of coverages.
    Slight violation of mutual orthogonality of the orbits will be
    the essential outcome of any moderate non-idealities in the {\it relative}
    orientations of three orbital axes.
    The {\it pseudo} octants in such a case will differ from each other, and
    hence will the set of associated coverages. The most relevant change in
    the coverage will be for the outermost baselines. A slightly shrunken
    {\it pseudo} octant will have reduction in the respective baseline lengths,
    but only by a (rather slowly varying) factor of
    ($\cos\frac{\phi}{2} - \sin\frac{\phi}{2}$), where $\phi$ is the 
    reduction (or deviation) in angle from the desired orthogonality.
    The sources in opposite octant will see increase in corresponding baseline
    extent by a factor of ($\cos\frac{\phi}{2} + \sin\frac{\phi}{2}$).
    Thus, even a 5$^{\circ}$ deviation from orthogonality will change
    the corresponding baseline extent only by less than 5\%, 
    much less than the source direction dependent variation.
	
	\begin{figure*}[]
		\centering
		
		\begin{subfigure}{\linewidth}
			\includegraphics[width=0.33\linewidth]{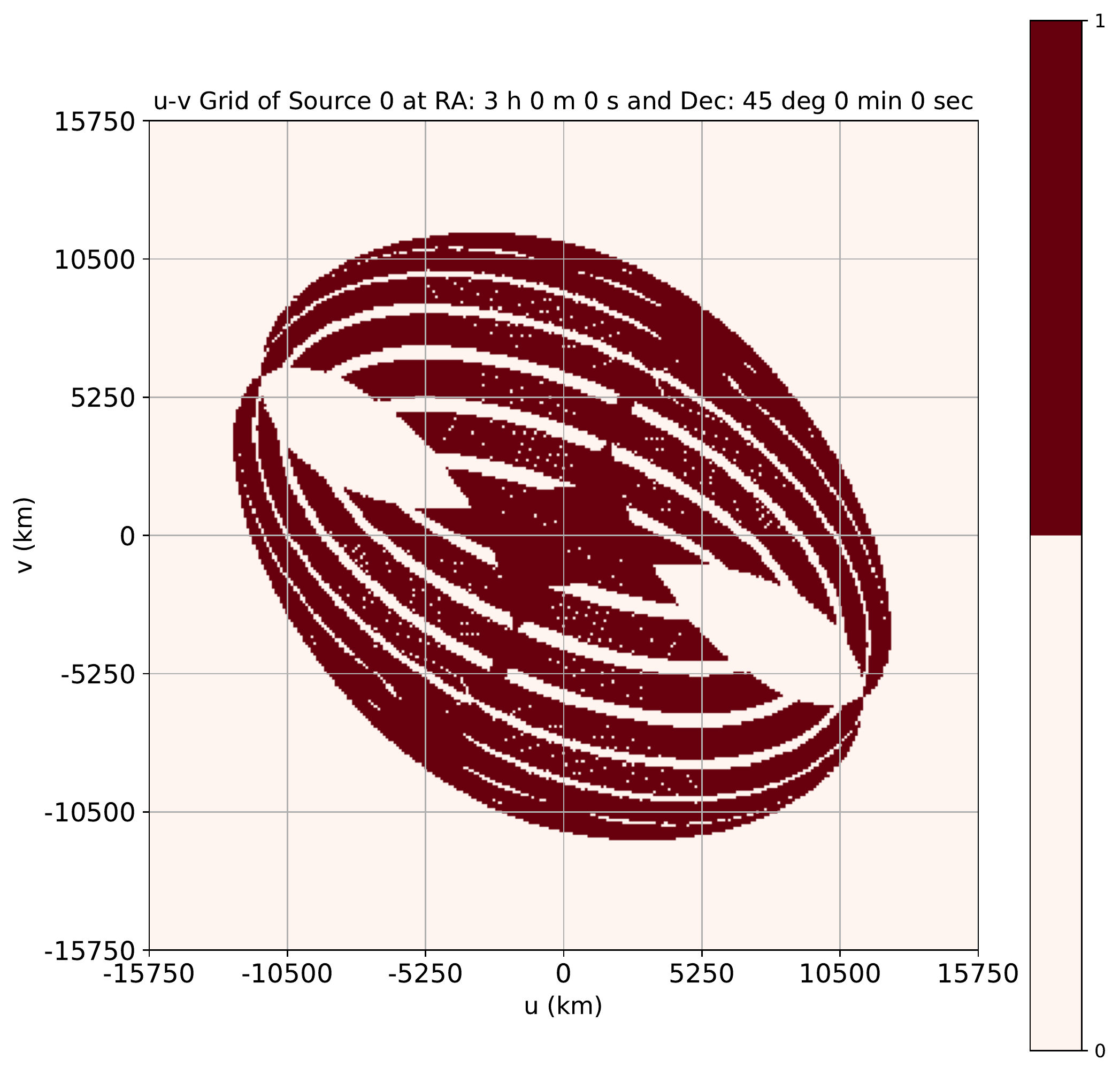}\hfill
			\includegraphics[width=0.33\linewidth]{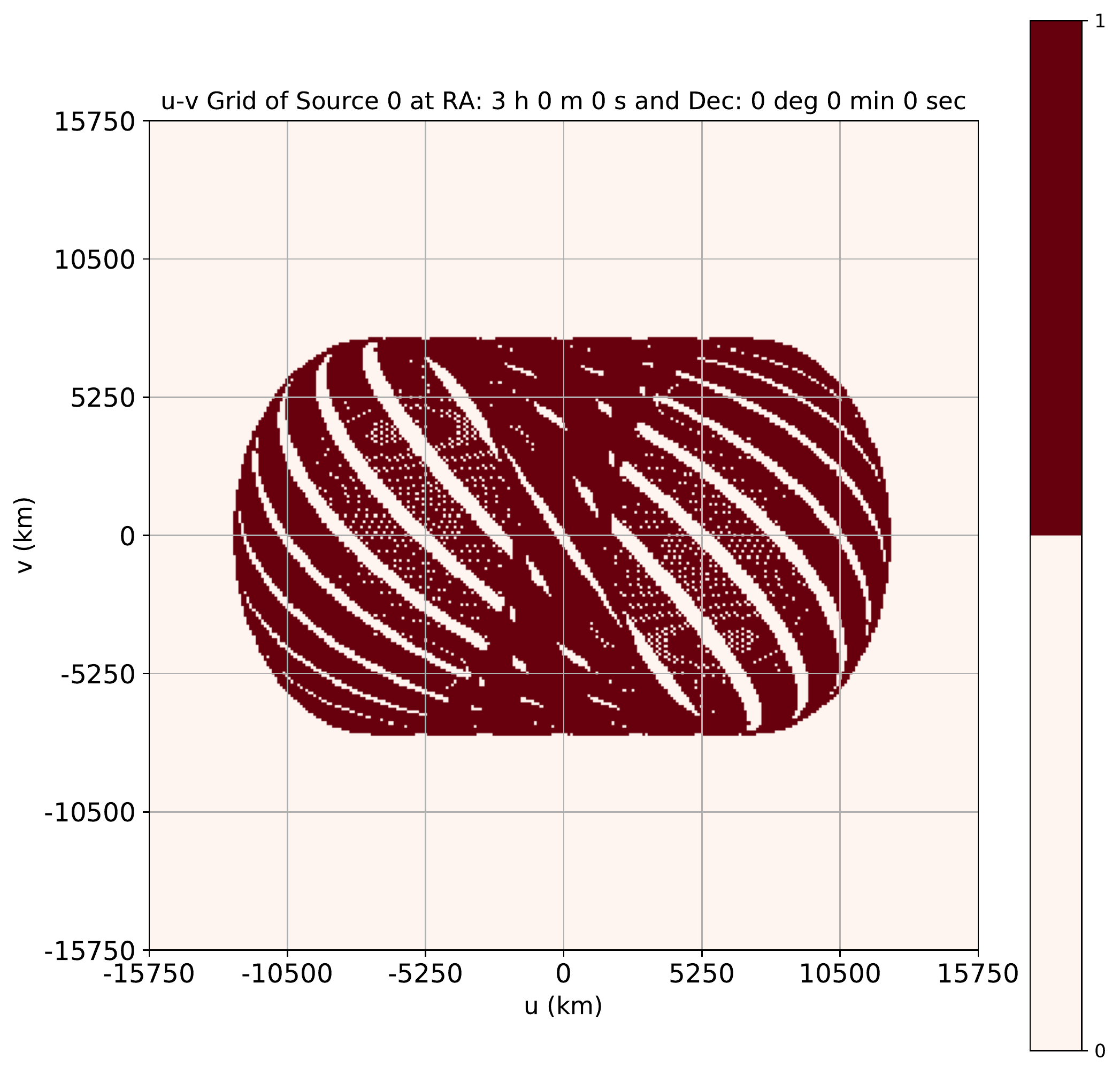}\hfill
			\includegraphics[width=0.33\linewidth]{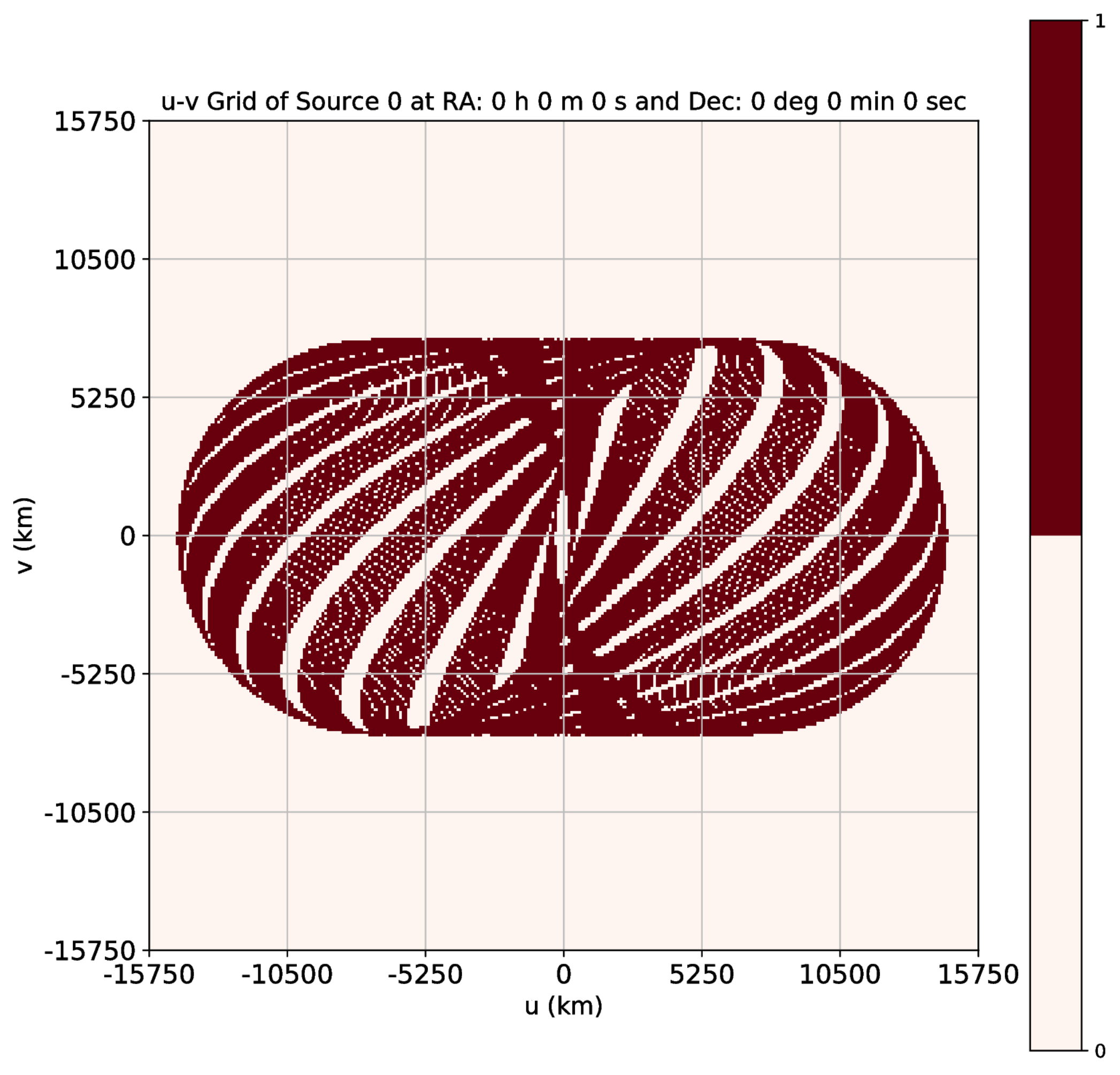}
		\end{subfigure}\par\medskip
		
		\begin{subfigure}{\linewidth}
			\includegraphics[width=0.33\linewidth]{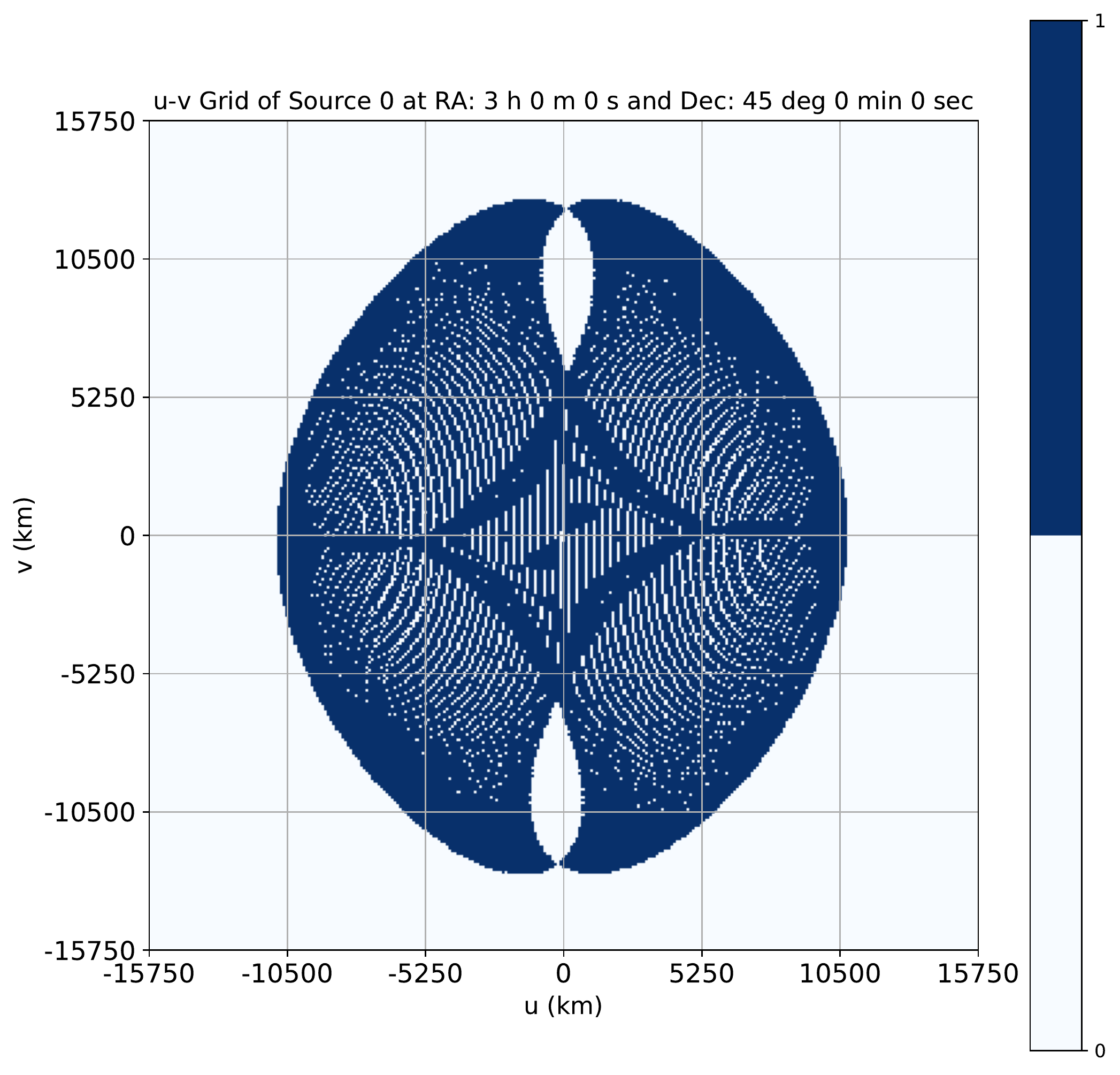}\hfill
			\includegraphics[width=0.33\linewidth]{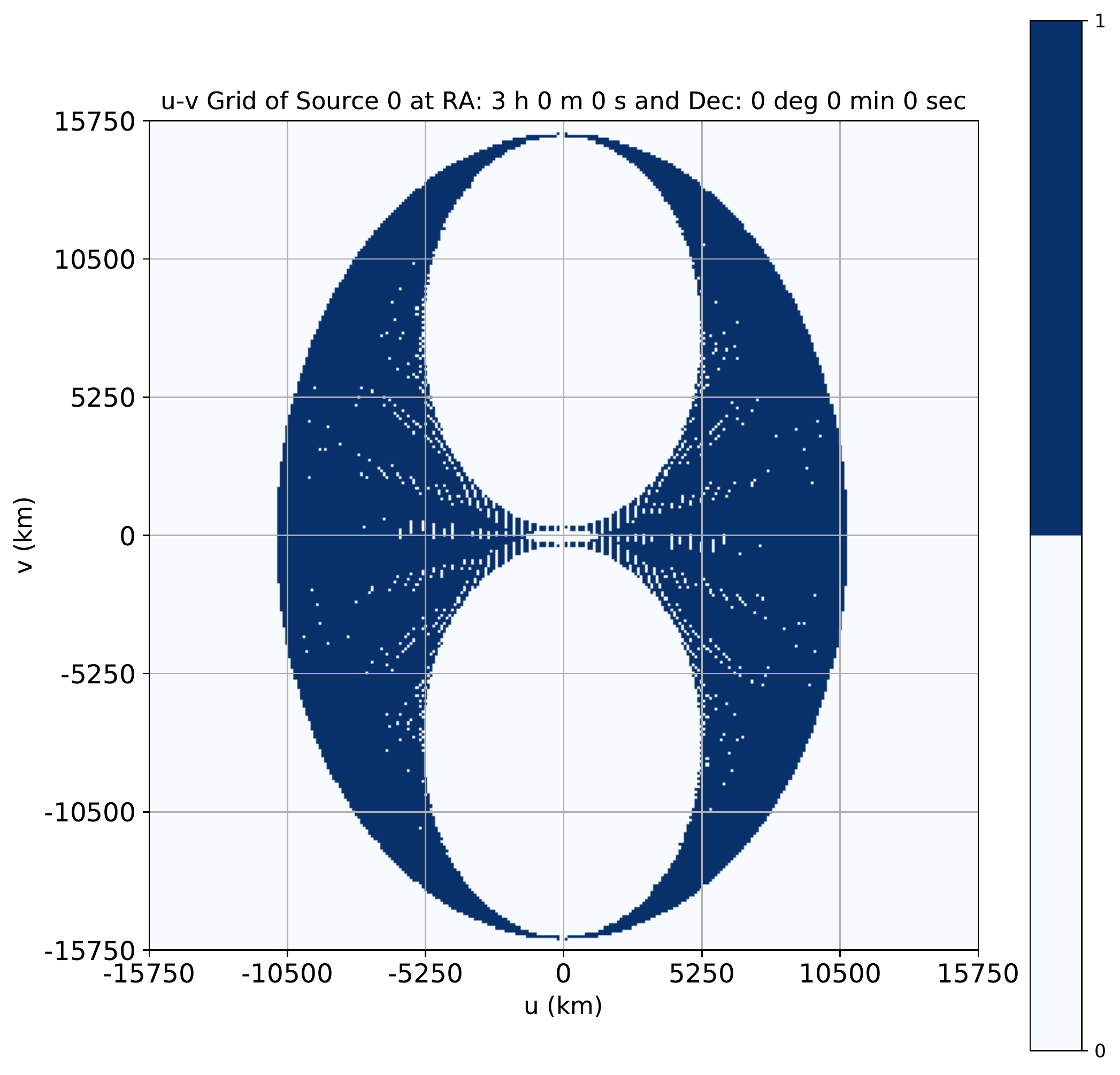}\hfill
			\includegraphics[width=0.33\linewidth]{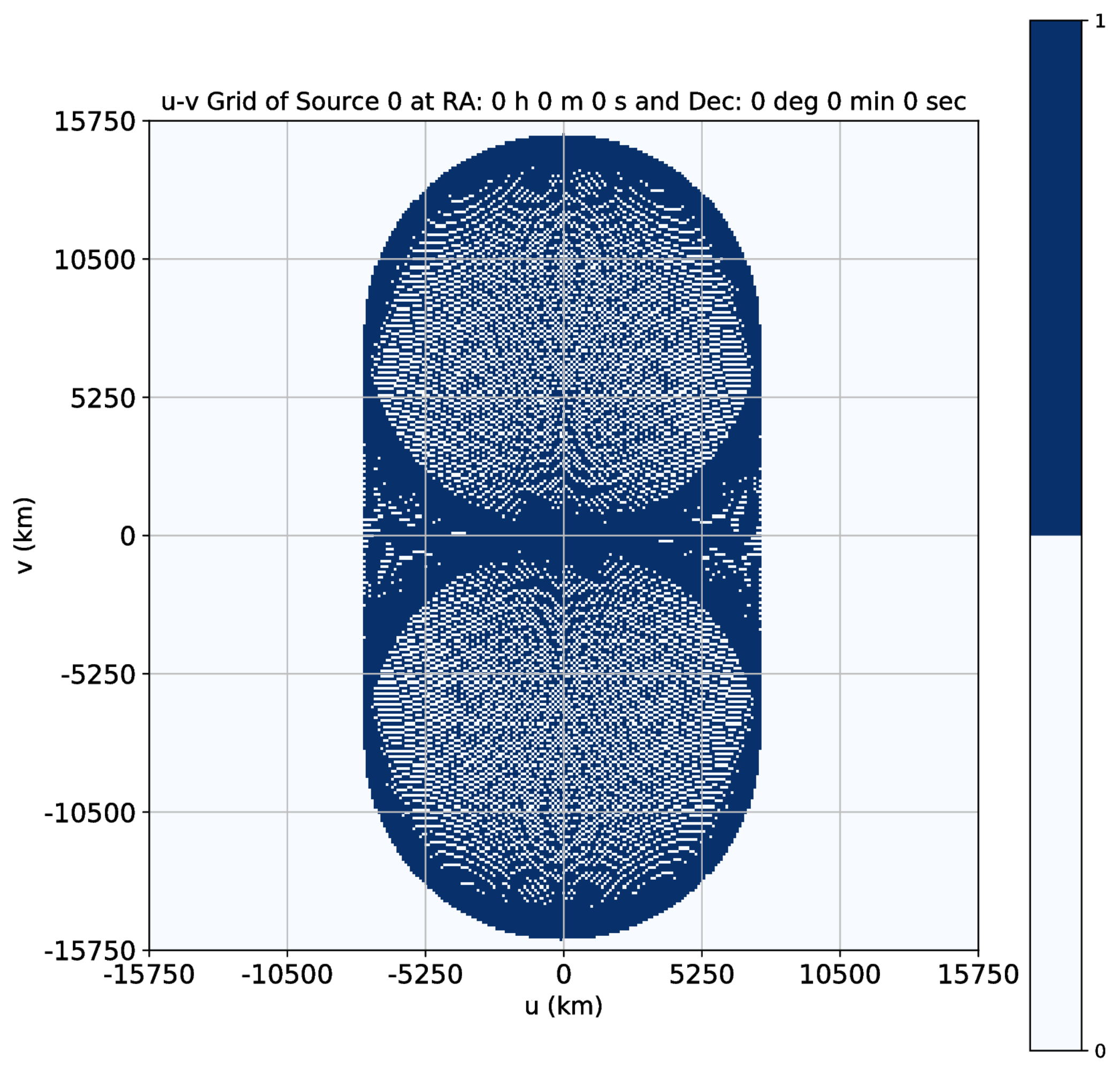}
		\end{subfigure}\par\medskip
		
		\begin{subfigure}{\linewidth}
			\includegraphics[width=0.33\linewidth]{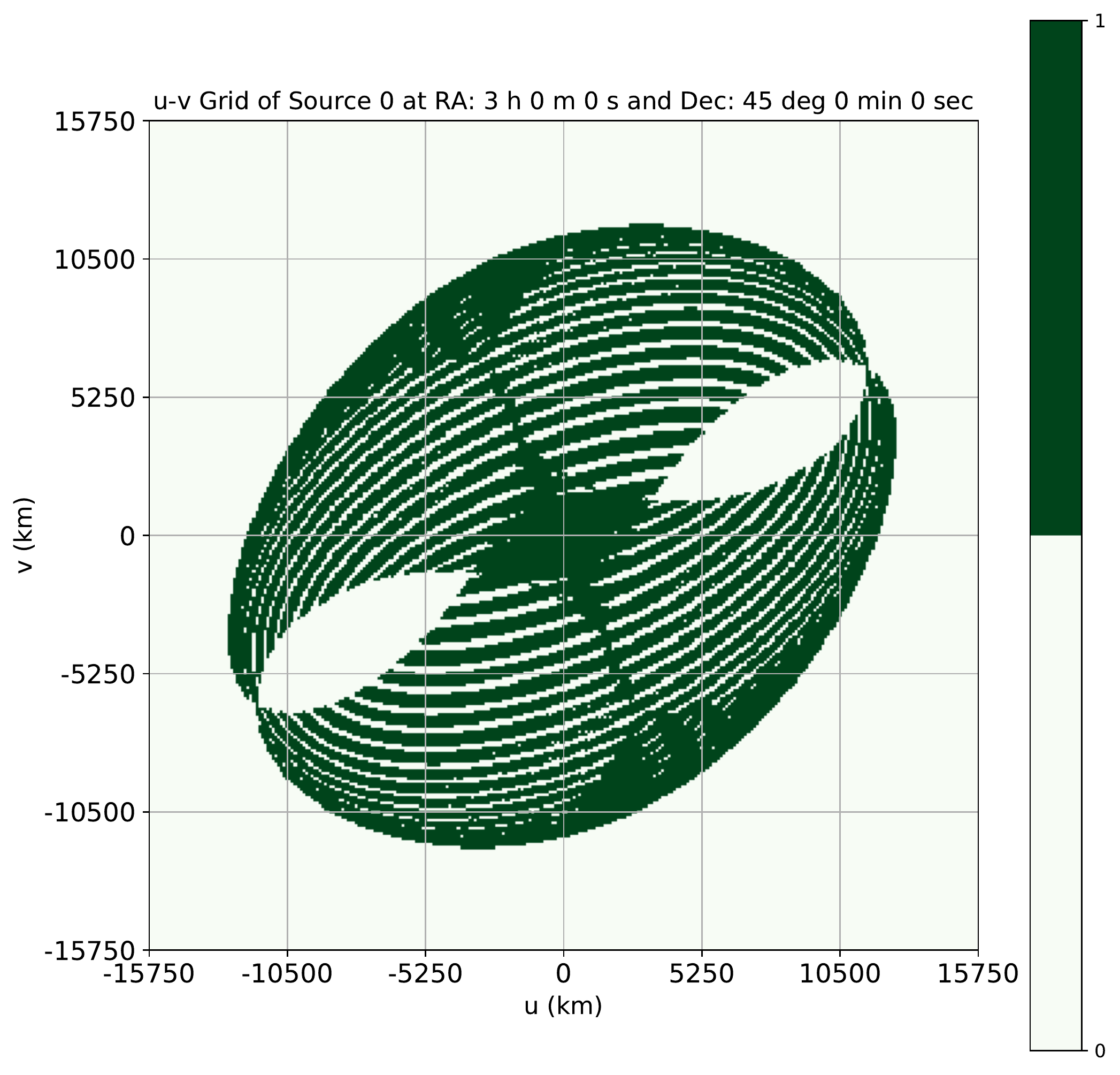}\hfill
			\includegraphics[width=0.33\linewidth]{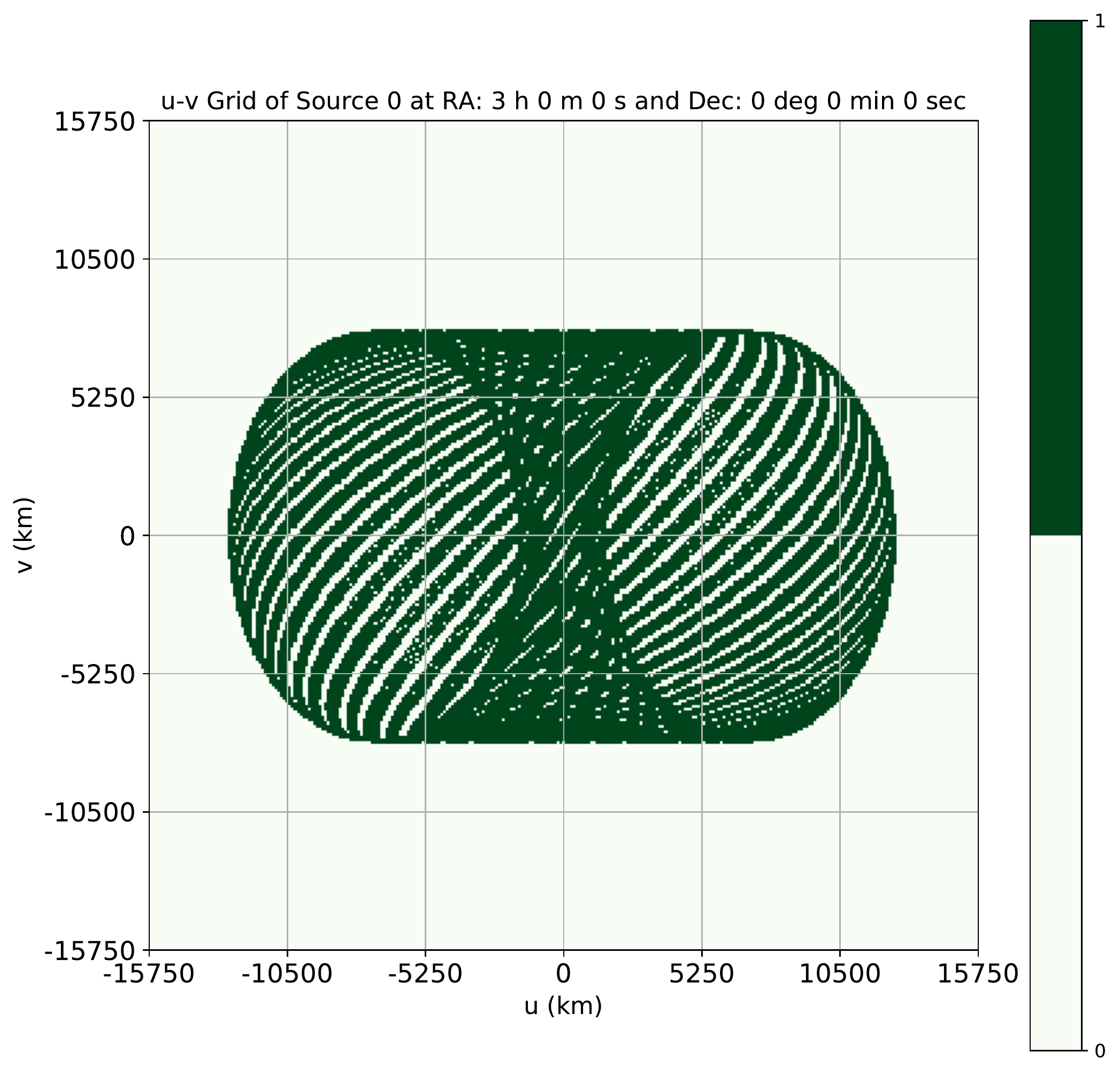}\hfill
			\includegraphics[width=0.33\linewidth]{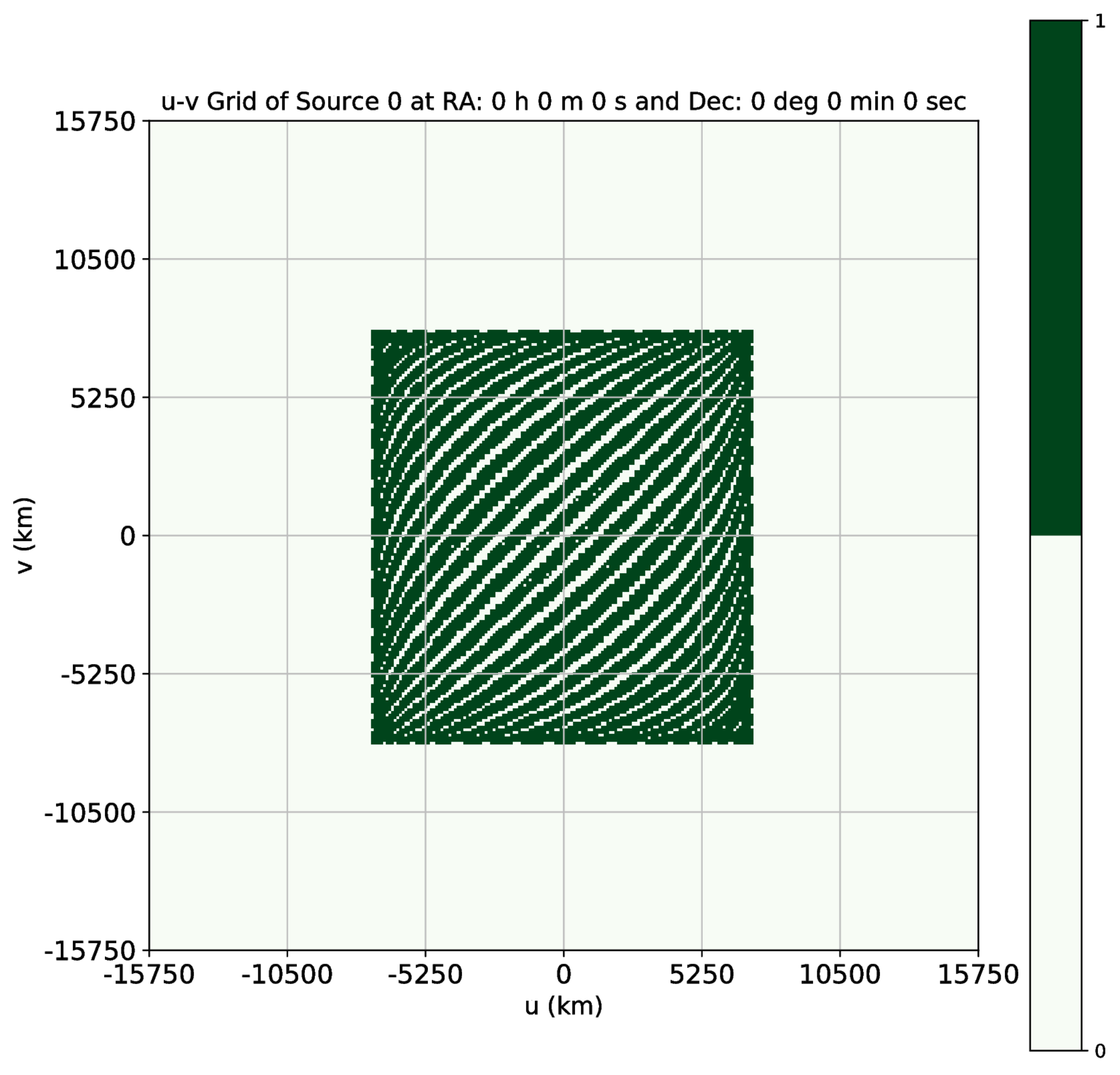}
		\end{subfigure}	
		
		\caption{The coarsely gridded $(u,v)$ coverages using the 3-satellite 
			system over a time span of 16 days are shown, with u and v axes marked in km.
			The three columns correspond to the three special source directions,
			namely the source A, B and C, in that order (see main text for definition 
            of these special directions).
			The different rows refer to the coverage corresponding to the
			different baselines, namely 1-2, 2-3 and 3-1, in that order.
		}\label{figure6}
		
	\end{figure*}
	
	\subsection{Quantitative Assessment of the $(u,v)$ Coverage} \label{subsecqmc}
	Having so far assessed the coverages more qualitatively,
	we proceed to present indicative quantitative measures of the
	coverage in the $(u,v)$ plane, 
	for all the three special source directions defined in the beginning
	of this section.
	
	For this purpose, we need to assert the gridding 
    scheme in the spatial frequency,
	i.e. the $(u,v)$ plane. Ideally, for widefield imaging configuration,
	the gridding might need to be as fine as
	the size of the aperture (in units of wavelength)
	associated with each of the interferometric elements, or a few times finer than
	that implied by the inverse of the angular size 
	of the field we wish to image at a time.
	For convenience in computation, although at the risk of our quantitative results 
	appearing somewhat misleading in their absolute measures, we have chosen 
	to grid the $(u,v)$ plane significantly coarsely. 
	Although the $(u,v)$ cell size can be
	varied in our simulations, the quantitative results here are based on 
	a cell size of 100 km x 100 km, as mentioned even earlier,
	which suffices for the relative assessment\footnote{Also, to relate this choice 
	of gridding to other relevant contexts, we note that the cell crossing 
    time would typically be 10 seconds
	or so, given the typical speeds of the satellites, 
	thus, defining the upper limit for time integration 
	of visibilities before cell migration.
	Another relevant parameter would be the spectral resolution, the field-of-view limit
	it would imply from the consideration of bandwidth decorrelation, and the scale of
	coherence it (or the so-called "delay beam") would in turn suggest for visibilities 
	across the $(u,v)$ plane. 
	For example, even 1-kHz for spectral resolution, would imply a delay beam with 
    (peak-to-null) width of about a degree. The coherence scale of visibilities when 
    observing at, say, 0.3 MHz
	would then be a large fraction of the cell size, making it not appear too coarse.}.
	By considering the cell 
	as filled if one or more $(u,v)$ points fall within that cell, 
	we estimate the so-called sampling function, $S(u,v)$,
	providing the description of the spatial frequency coverage,
	and then percentage coverage was 
	calculated by comparing it with the limiting $(u,v)$ span assumed as 
	a circle of radius 15750 km (maximum baseline length). 
	For completeness, we do include the coverage symmetry in $(u,v)$ plane, 
	resulting from the Hermitian symmetric nature of 
	the visibilities. 
	
	\begin{figure*}[t]
		\centering
		
		\begin{subfigure}{\linewidth}
			\includegraphics[width=0.33\linewidth]{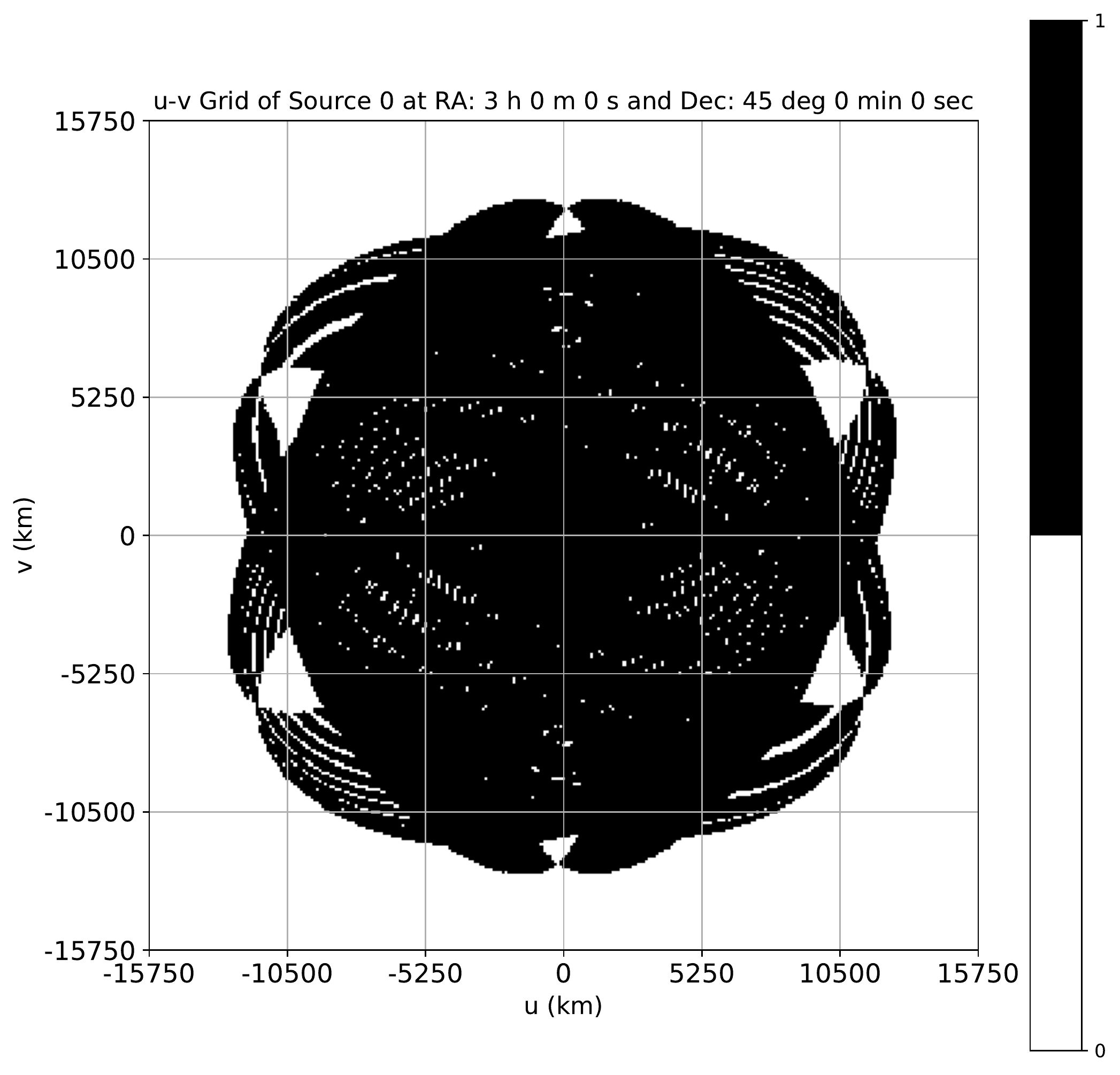}\hfill
			\includegraphics[width=0.33\linewidth]{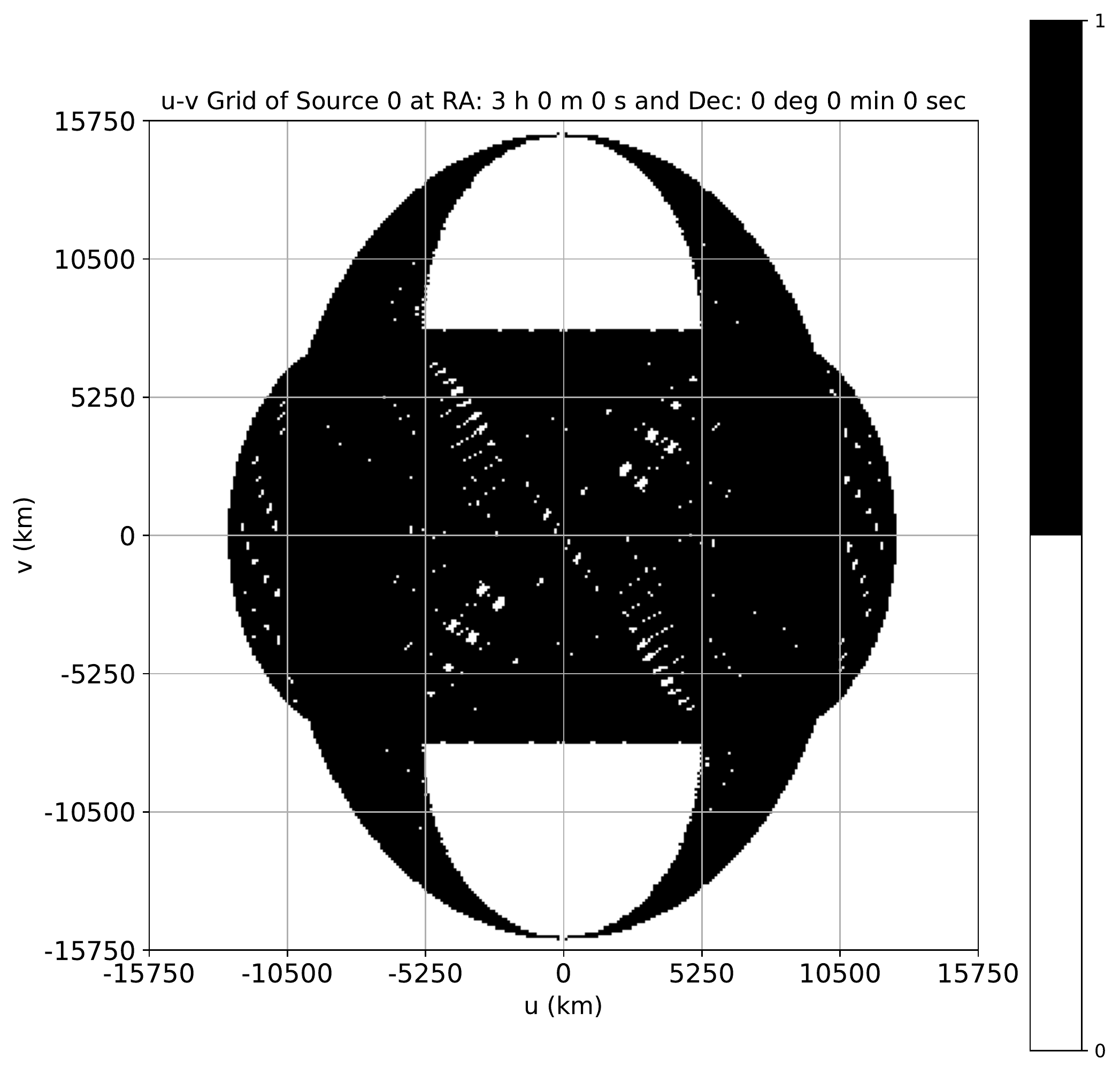}\hfill
			\includegraphics[width=0.33\linewidth]{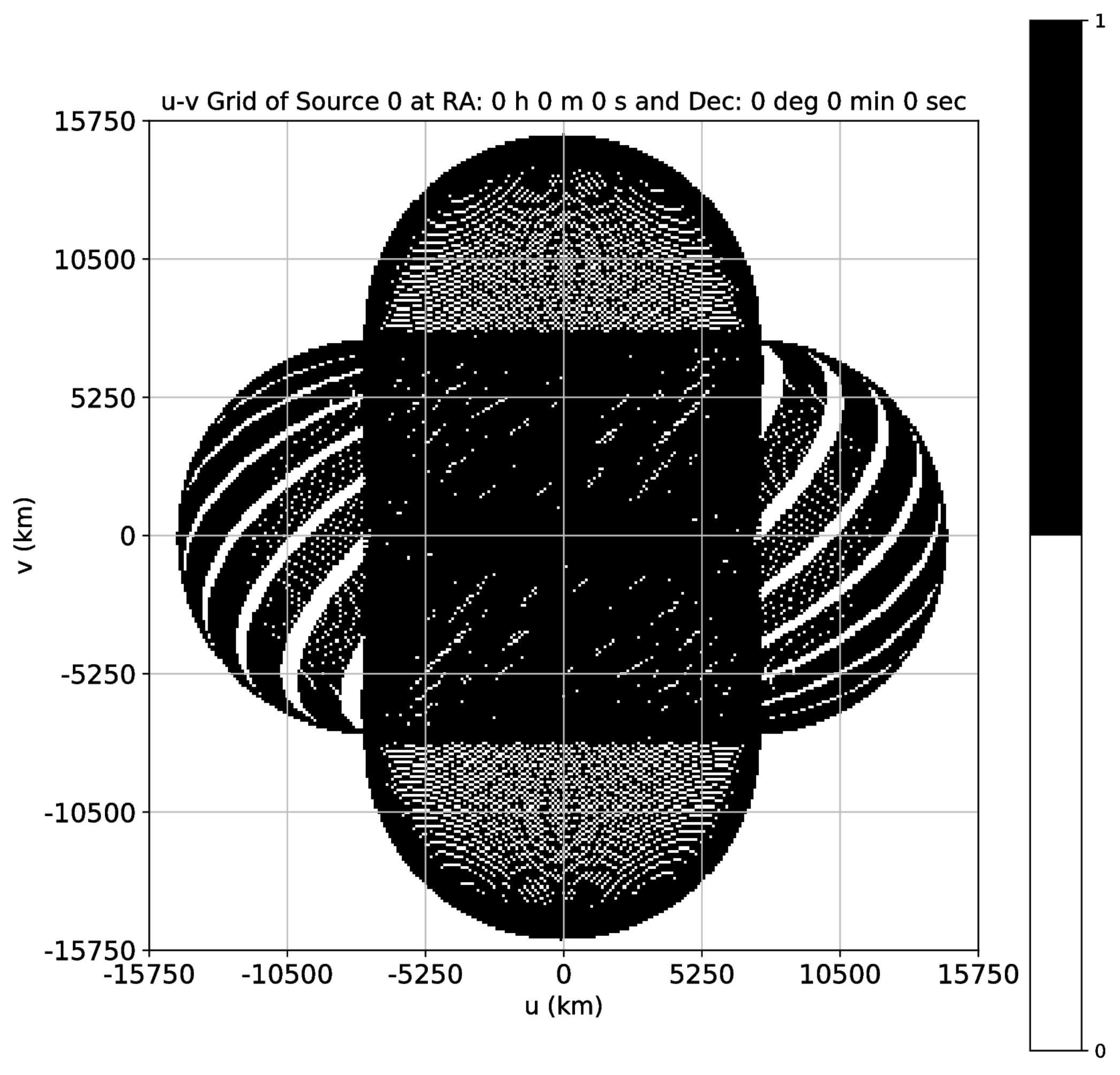}
		\end{subfigure}\par\medskip
	
		\begin{subfigure}{\linewidth}
			\includegraphics[width=0.33\linewidth]{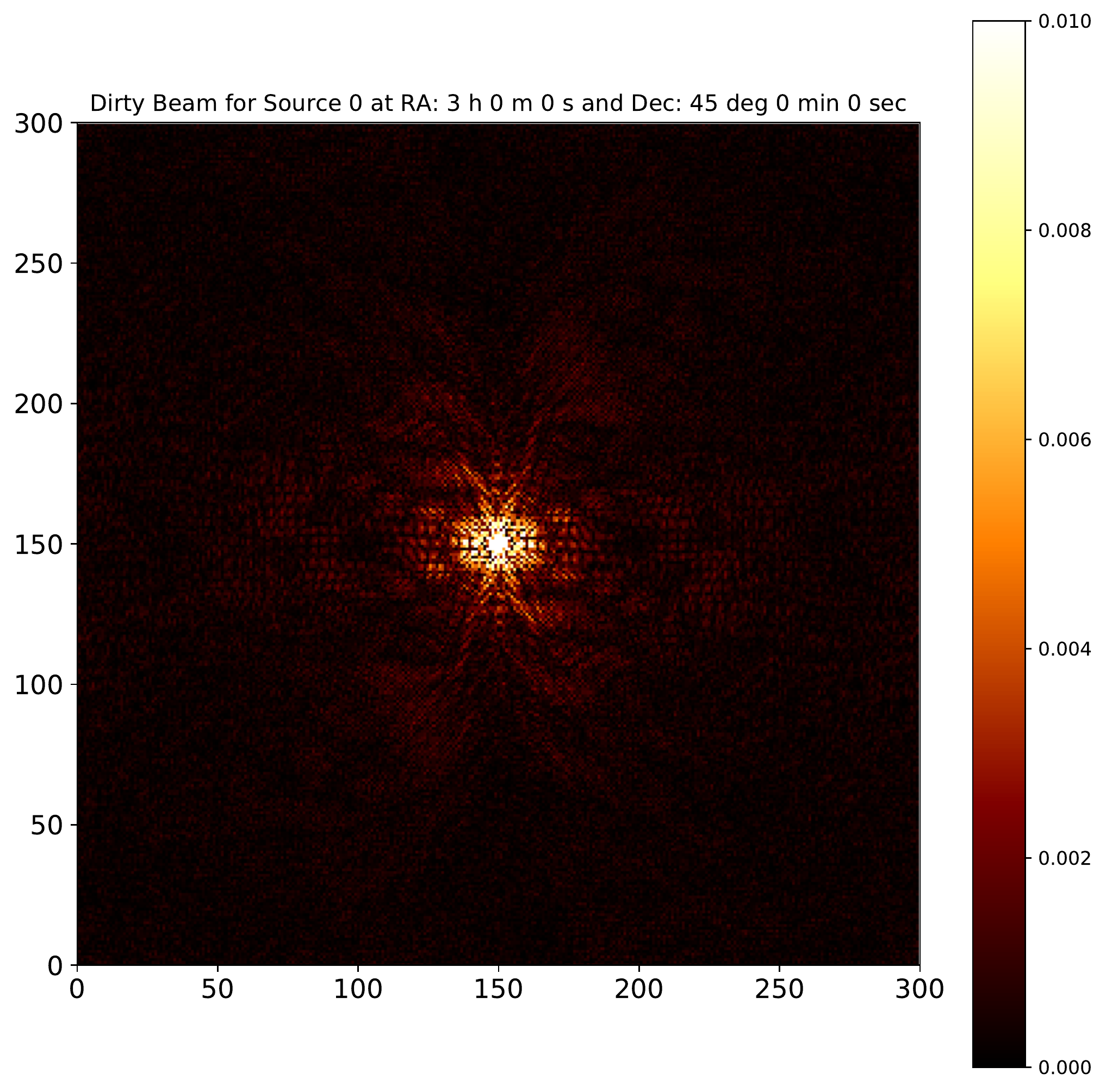}\hfill
			\includegraphics[width=0.33\linewidth]{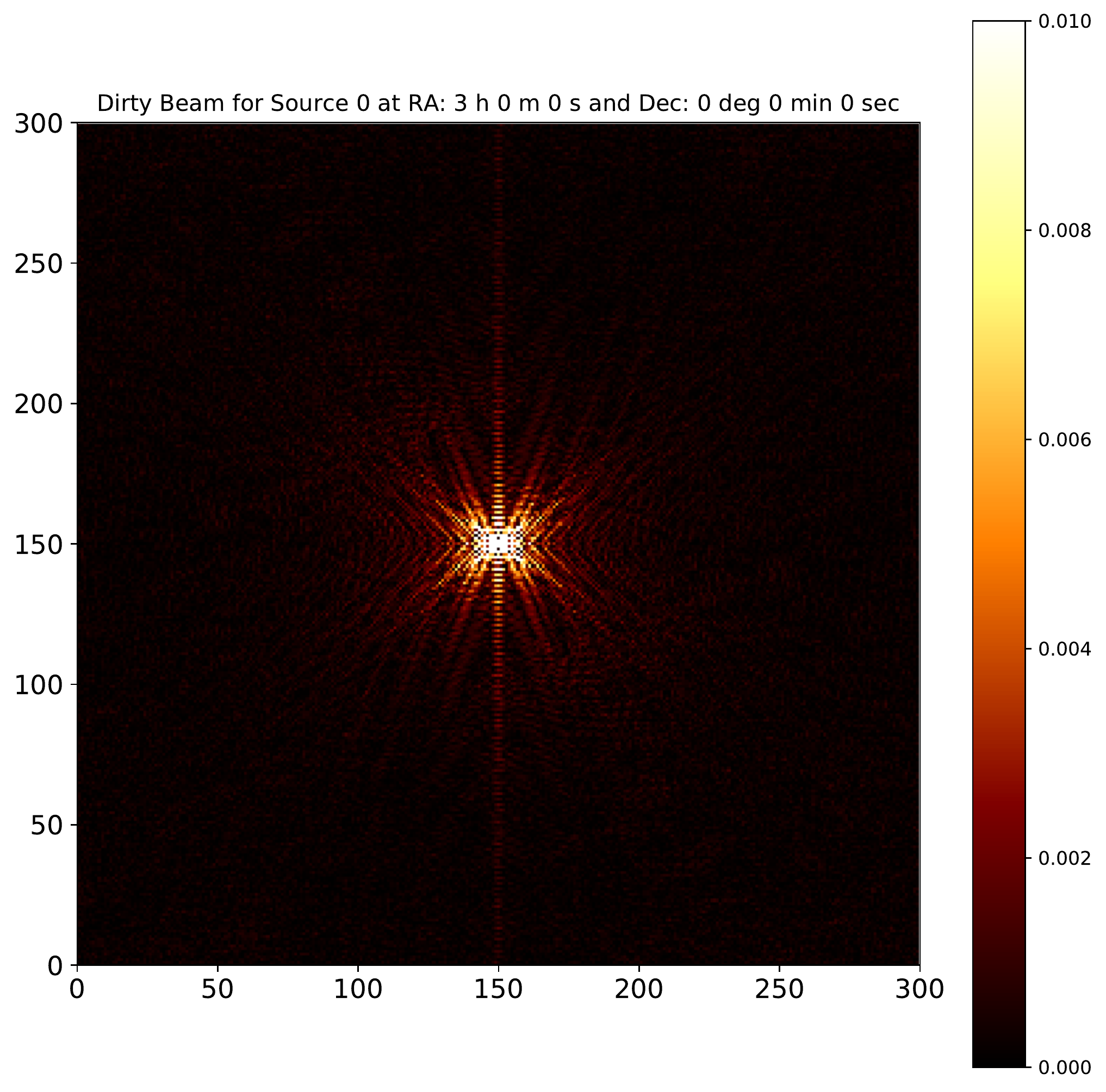}\hfill
			\includegraphics[width=0.33\linewidth]{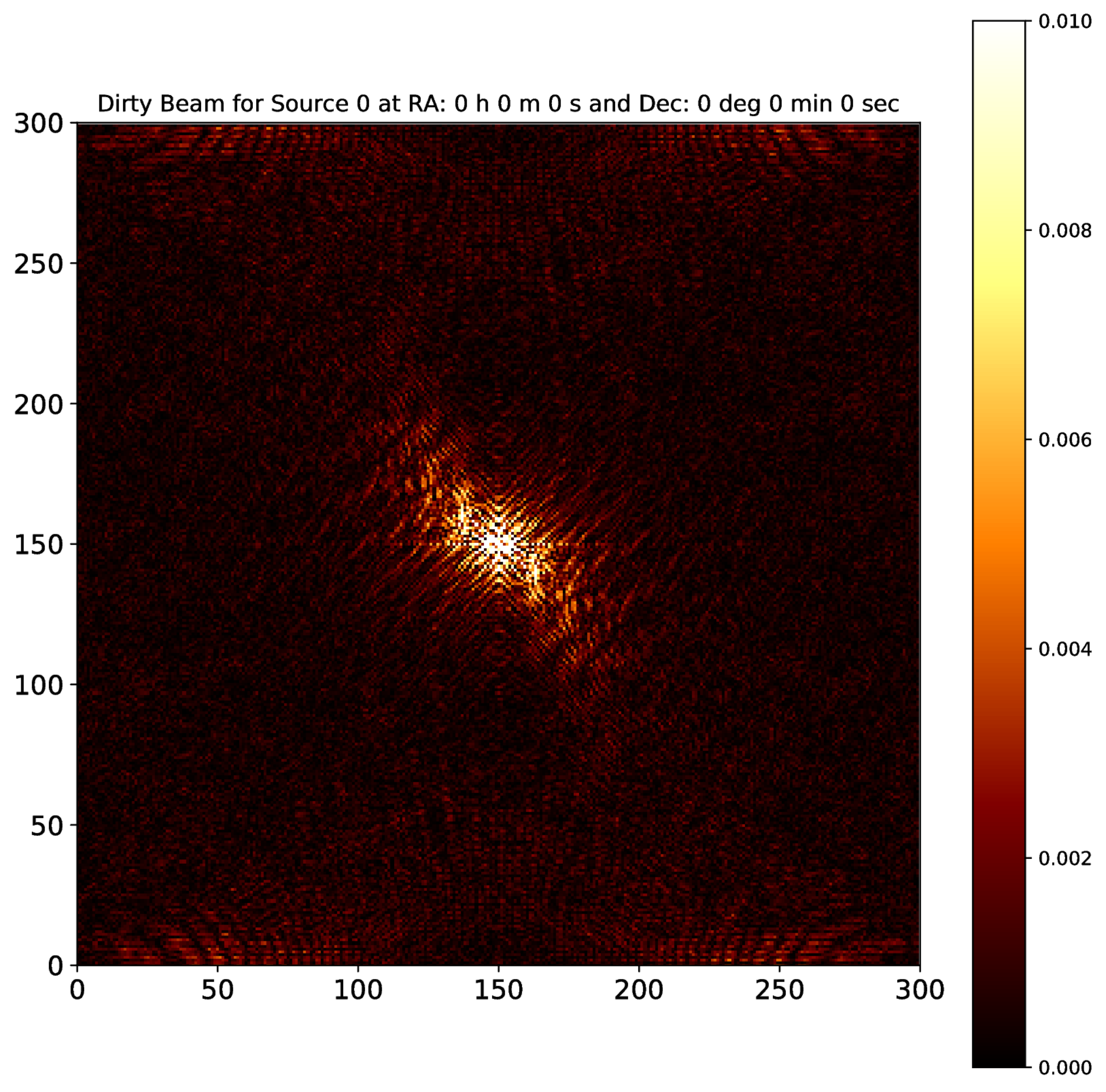}
		\end{subfigure}
		
		\caption{The top row panels show the combined $(u,v)$ coverage,  
			obtainable using the three-satellite configuration, 
			for the three special source directions (source A, B and C),
			and the corresponding dirty beams in the bottom row,
			assuming uniform weighting.
		} \label{figure7}
	\end{figure*}
	
	When the source direction A is observed for a duration of 16 days, 
	the percentage coverages from the
	baseline due to satellite pairs 1-2,2-3 and 3-1 are
	about 39.5, 46 and 39.5 \%, respectively, with the total
	coverage being about 64\%.
	Refer to Figure~\ref{figure6} for further details and results.
	
	These results (summarized in Table \ref{table1a})
	show that the total coverage for any source observed 
	for a time duration of 16 days would, thus, lie between 64\% and 
    55.5\%. 
	The table also has results corresponding to observations done for time durations of
	8 days for the same three chosen source directions. These values will be referred 
    to in subsection~\ref{subsec4sat}.  
	Also, the coverages due to our system are much greater than what 
	we usually get with the ground-based counterparts when compared 
	for the same time duration. This means that while having greater baselines, 
	our model system is also capable of giving significantly better coverage 
	for any source, which would result in very fine angular resolutions 
	at low radio frequencies.
	
	\subsection{Dirty Beams or the Point Spread Functions} \label{subsecdb}
	The dirty beam, or the point-spread-function, associated
	with the total spatial frequency (or $(u,v)$) coverage obtainable,
	for each of 
	the three special source directions, is estimated separately, by performing 
	the 2D (Inverse) Fourier Transform of the respective $(u,v)$ coverage data. 
	
	The dirty beam, $b(l,m)$ is given by: 
	
	\begin{equation} \label{dirtybeam}
	\hspace{1cm} b(l,m) = \mathcal{F}^{-1} [S(u,v)]
	\end{equation}
	
	where, $S(u,v)$ is the sampling function for the visibility measurements,
	describing the effective coverage in the spatial frequency plane, while  
	the dirty beam is a function of direction defined by $l$ and $m$, which
	are the direction cosines of angles in the planes 
	containing $u$ and $v$ respectively.  
	The sampling function $S(u,v)$ depends on actual (natural) sampling 
	described by $s(u,v)$
	and a user-specified weighting function $W(u,v)$, in the following way:
	
	\begin{equation} \label{weightingfunc}
	\hspace{1cm} S(u,v) = s(u,v)W(u,v)
	\end{equation}
	
	\begin{figure*}[ht]
		\centering
		
		\begin{subfigure}{\linewidth}
			\centering
			\includegraphics[width=0.33\linewidth]{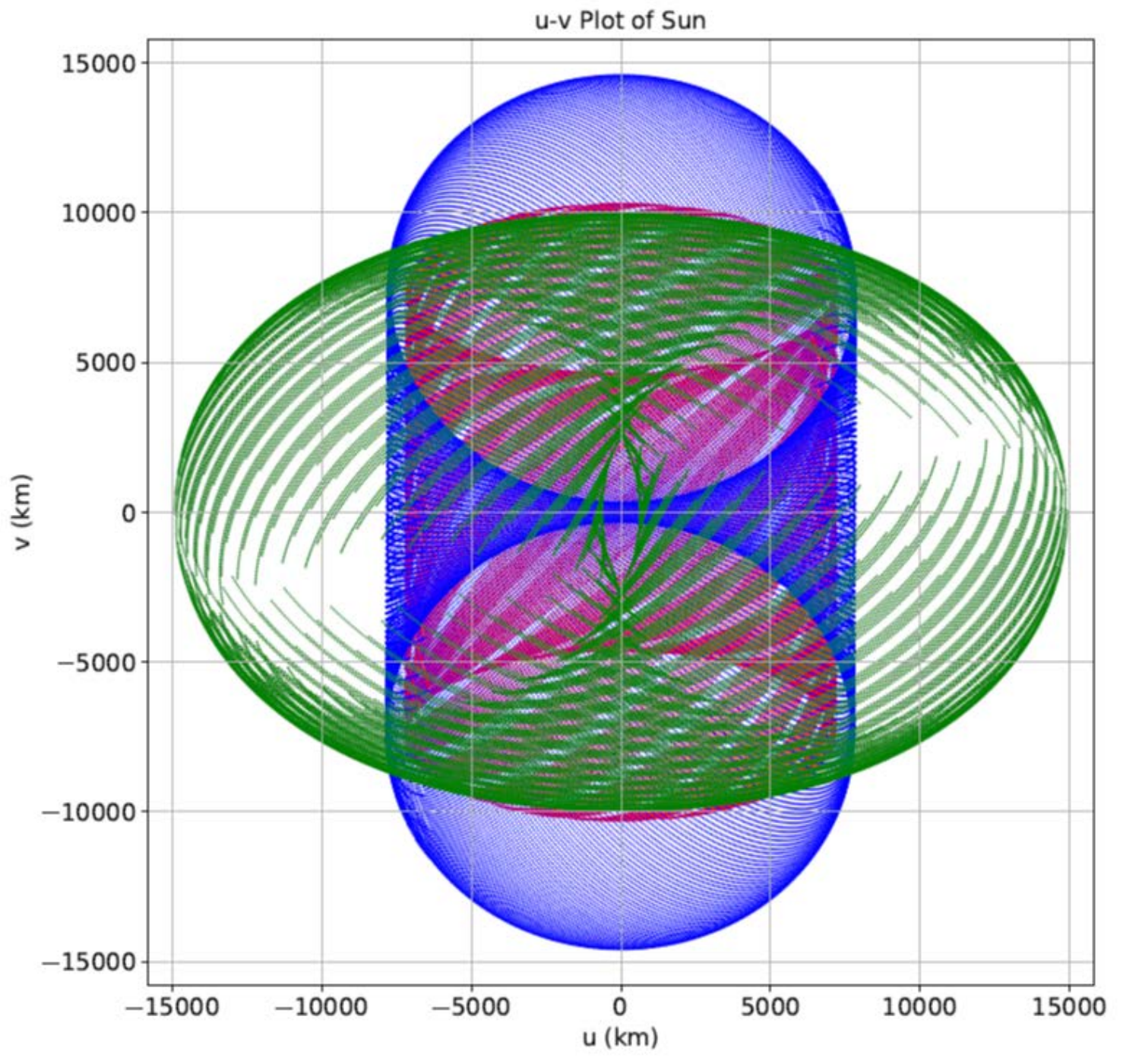}
			\includegraphics[width=0.33\linewidth]{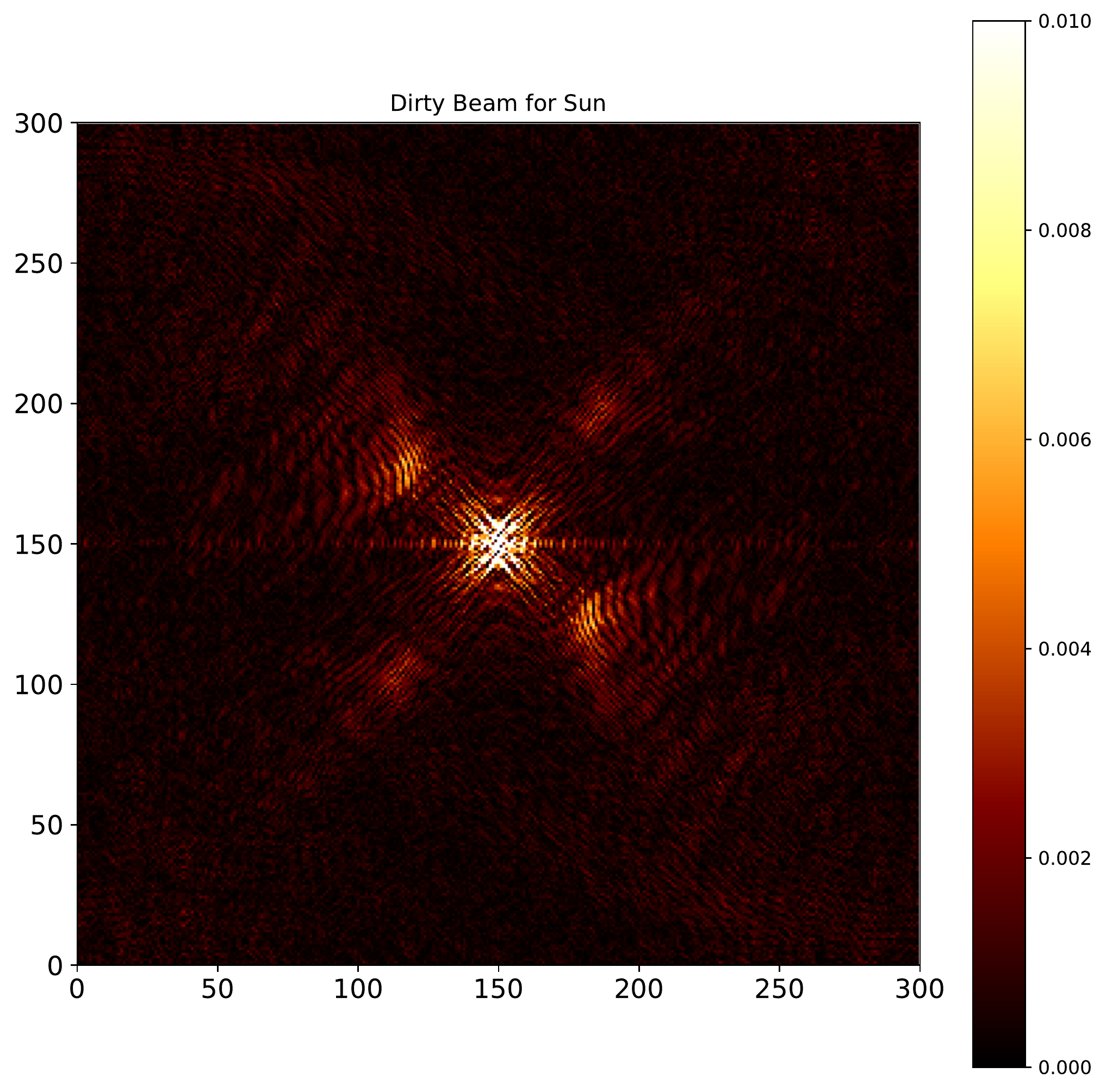}
		\end{subfigure}
		
		\caption{The total $(u,v)$ coverage plot for the Sun observation 
			for a duration of 16 days and its corresponding dirty beam. 
		} \label{figure8}
	\end{figure*}
	
	In our simulation, we have assumed uniform weighting, 
	in which $W(u,v)$ is inversely proportional to the local density 
	of $(u,v)$ points in $s(u,v)$. 
	The sum of weights in a $(u,v)$ cell is constant if the cell is filled 
	and zero if the cell is empty. This implementation ensures that 
	the $(u,v)$ plane is filled more uniformly and the dirty beam sidelobes 
	are minimum. Such an implementation enhances angular resolution 
	as it gives more weight to longer baselines, at the expense of 
	point source sensitivity. Since our model system has widefield 
	antennas with very large baselines, uniform weighing is 
	more appropriate (see for more details, \cite{wilner2010imaging}),
	and is employed in the estimation of the dirty beams shown in
	Figure~\ref{figure7}. It should be noted that the x and y axes 
	of the dirty beam plots are RA offset and Dec offset respectively, 
	but are not shown in the figures as we have not assumed any spatial frequencies 
	in the $(u,v)$ graphs. For example, when assuming a spatial frequency of 0.3 MHz, 
	the extent of both RA offset and Dec offset in the dirty beam plots would be from 
	about -3 arcsec to +3 arcsec. These values can be trivially scaled for any 
	other radio frequency in a similar manner. 
	
	\subsection{Sun Observation} \label{subsecdso}
	Here, we examine a special case for observation
	of the Sun using our space interferometer setup. 
	The Sun, as a source at a finite distance of 1 AU,
	represents indeed an additional special scenario, 
	as the Earth orbits around it, making its
	apparent direction (i.e., RA, Dec) change systematically.
	With appropriate modifications to take these aspects into
	account, the attainable $(u,v)$ coverage is estimated, assuming
	time duration of 16 days.
	The combined percentage coverage in this case is found to be about 63\%,
	reassuringly in the range of coverage noted in subsection~\ref{subsecqmc}.
	Figure~\ref{figure8} shows the relevant details of the $(u,v)$ coverage,  
	over a duration of 16 days, and the corresponding dirty beam.
	
	Admittedly, the Sun is a very broad source and more importantly
	highly variable on a range of time-scales, due to a wide variety of
	reasons. Naturally therefore,
	the synthesis imaging combining data over several days
	in such a case implies a very challenging situation, if not
	an ill-posed case. However, we have still included this case here merely
	for completeness, mainly to illustrate the potential $(u,v)$ coverage
	offered by the proposed configuration.

	\begin{table*}[ht]
		\centering
		\caption{Defined parameters of the 4-satellite model.}\label{table2}
		\begin{tabular*}{0.7\textwidth}{@{}c\x c\x c\x c@{}}
			\toprule
			\textit{Index} & Orbital Height & Orbital Velocity & Time Period \\
			~ & Above Earth's Surface & ~ & ~  \\ \midrule
			\textit{Satellite 1:} & 770 km & 7.48 km/s & 100.01 min \\  
			\textit{Satellite 2:} & 980 km & 7.37 km/s & 104.45 min \\ 
			\textit{Satellite 3:} & 1190 km & 7.27 km/s & 108.96 min \\  
			\textit{Satellite 4:} & 1400 km & 7.17 km/s & 113.53 min \\ \bottomrule
		\end{tabular*}
	\end{table*}

	\subsection{Comparison with a Four Satellite System} \label{subsec4sat}
	Although we have argued and demonstrated that the minimal 
    configuration of three satellites
	offers largely the desired level of $(u,v)$ coverage, it is important
	to ask if a 4-satellite system would do significantly better.
	To assess this, we indeed simulated a four satellite system,  
	to examine how much of an advantage adding a fourth satellite 
	would be, how that would be reflected in terms of the improvement 
	in $(u,v)$ coverage and the time taken to obtain desired coverage.
	
	\begin{figure}[t]
		\centering
		\includegraphics[width = 1\linewidth]{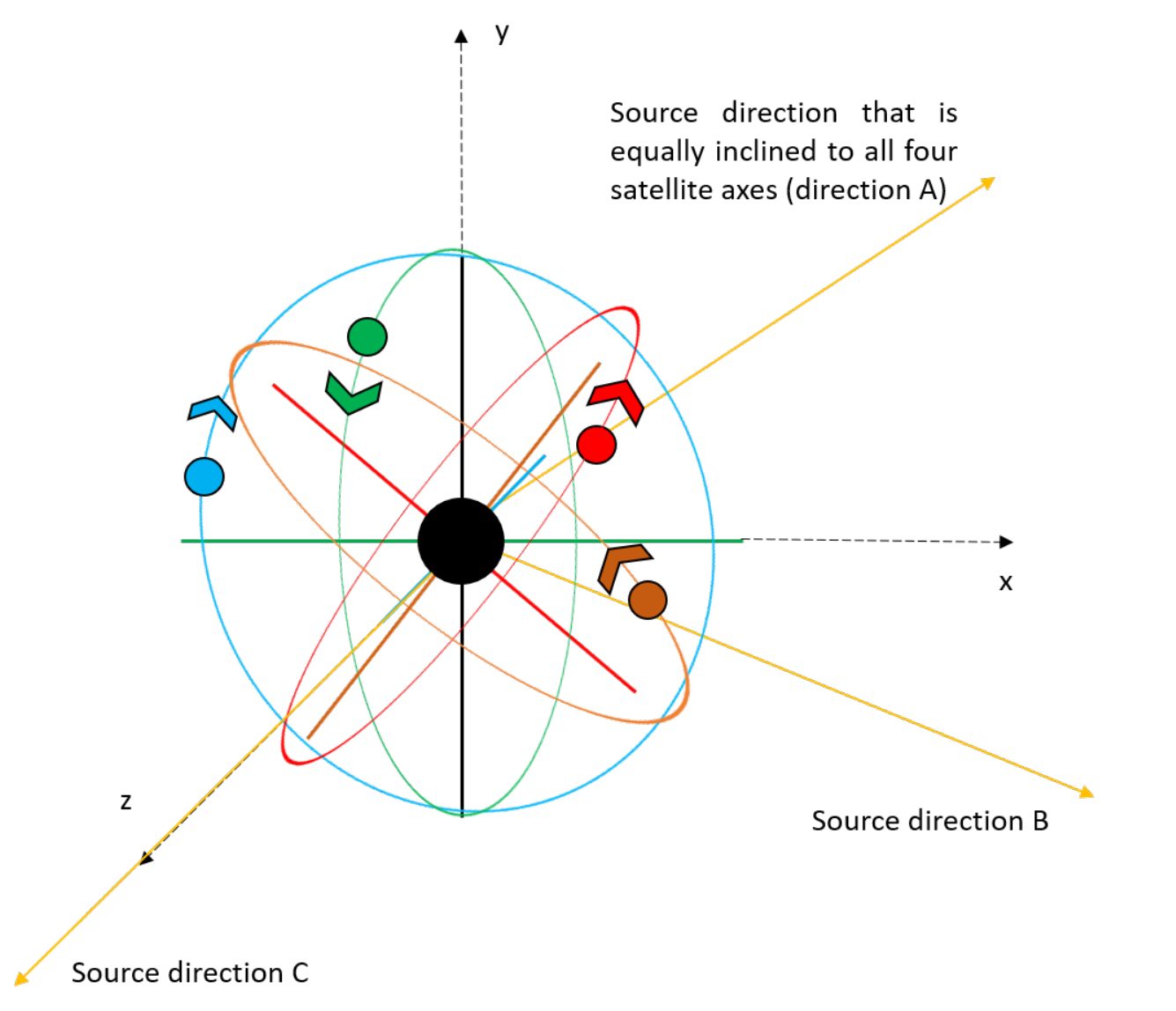}
		
		\caption{A line diagram depicting the model with four satellites. 
    The central black sphere represents the Earth, with the red, orange, 
    blue and green spheres representing the satellites 1, 2, 3 and 4 respectively. 
    Correspondingly, the red, orange, blue and green axes passing through the 
    Earth’s center represent the axes of revolution of the satellites 
    1, 2, 3 and 4 respectively. The yellow rays represent the direction 
    of the sources A, B and C. The details in the figure are not to scale.} \label{figure9}
	\end{figure}
	
	\begin{figure*}[ht]
		\centering
			
			\begin{subfigure}{\linewidth}
				\includegraphics[width=0.33\linewidth]{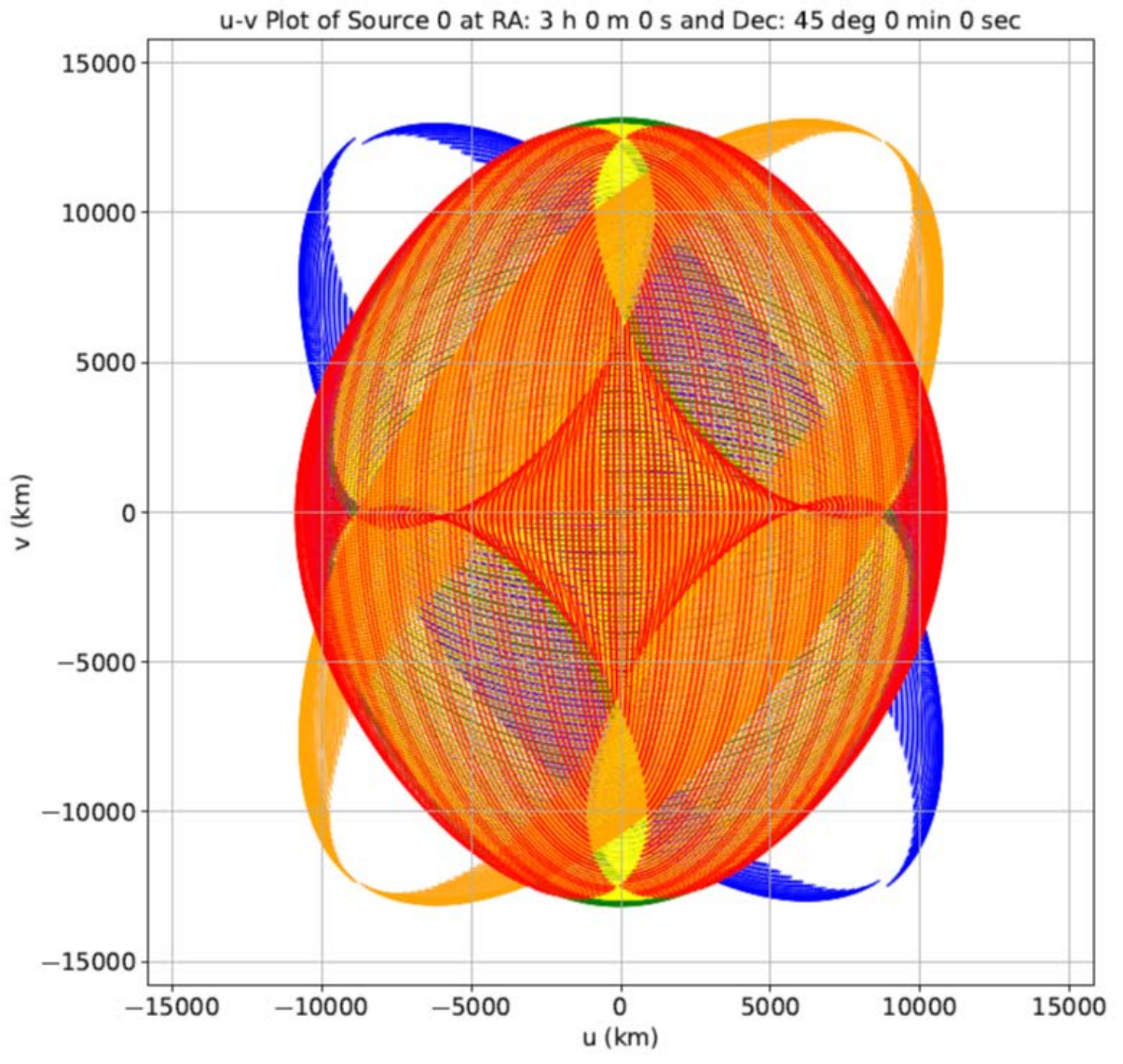}
				\includegraphics[width=0.33\linewidth]{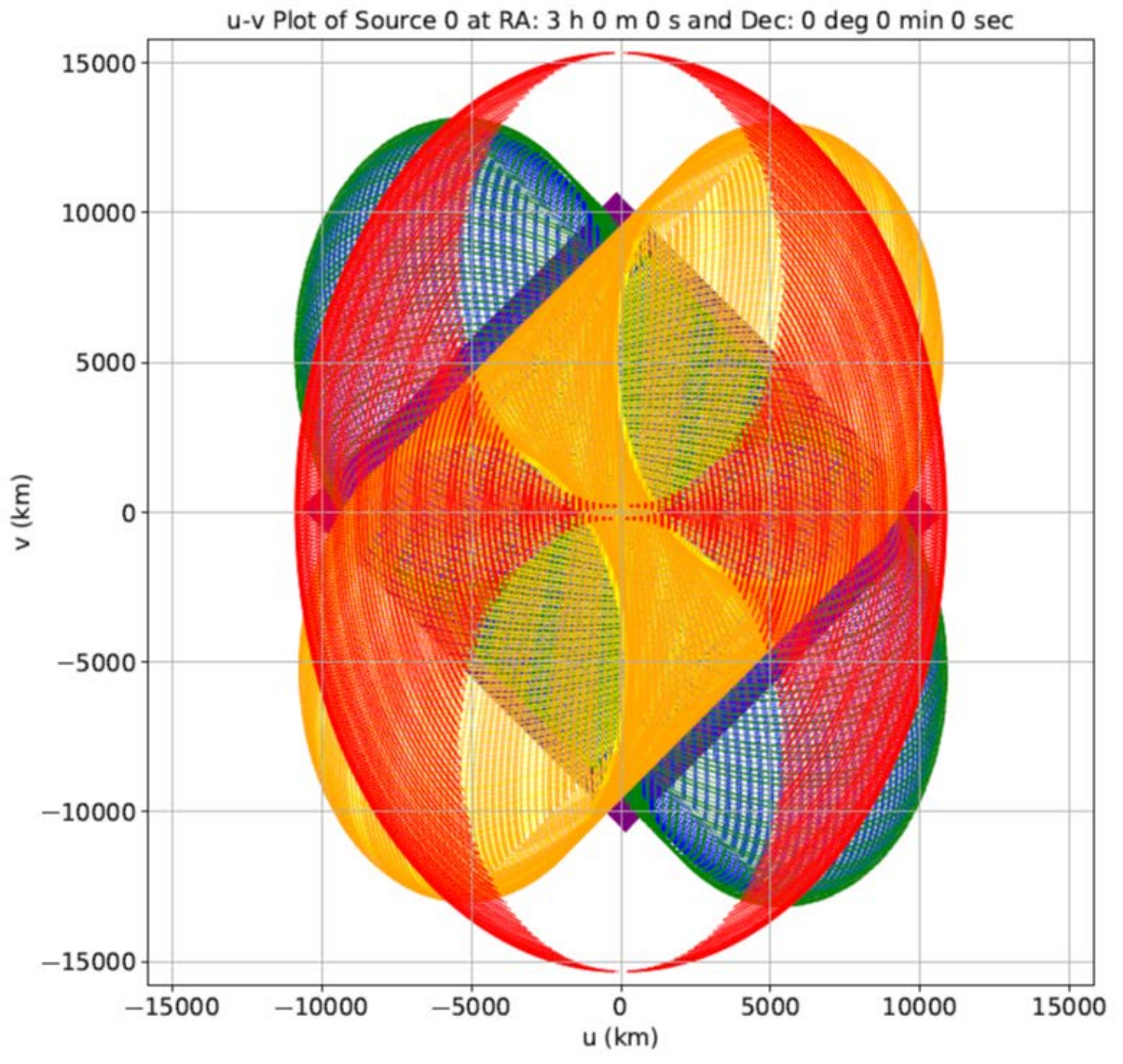}
				\includegraphics[width=0.33\linewidth]{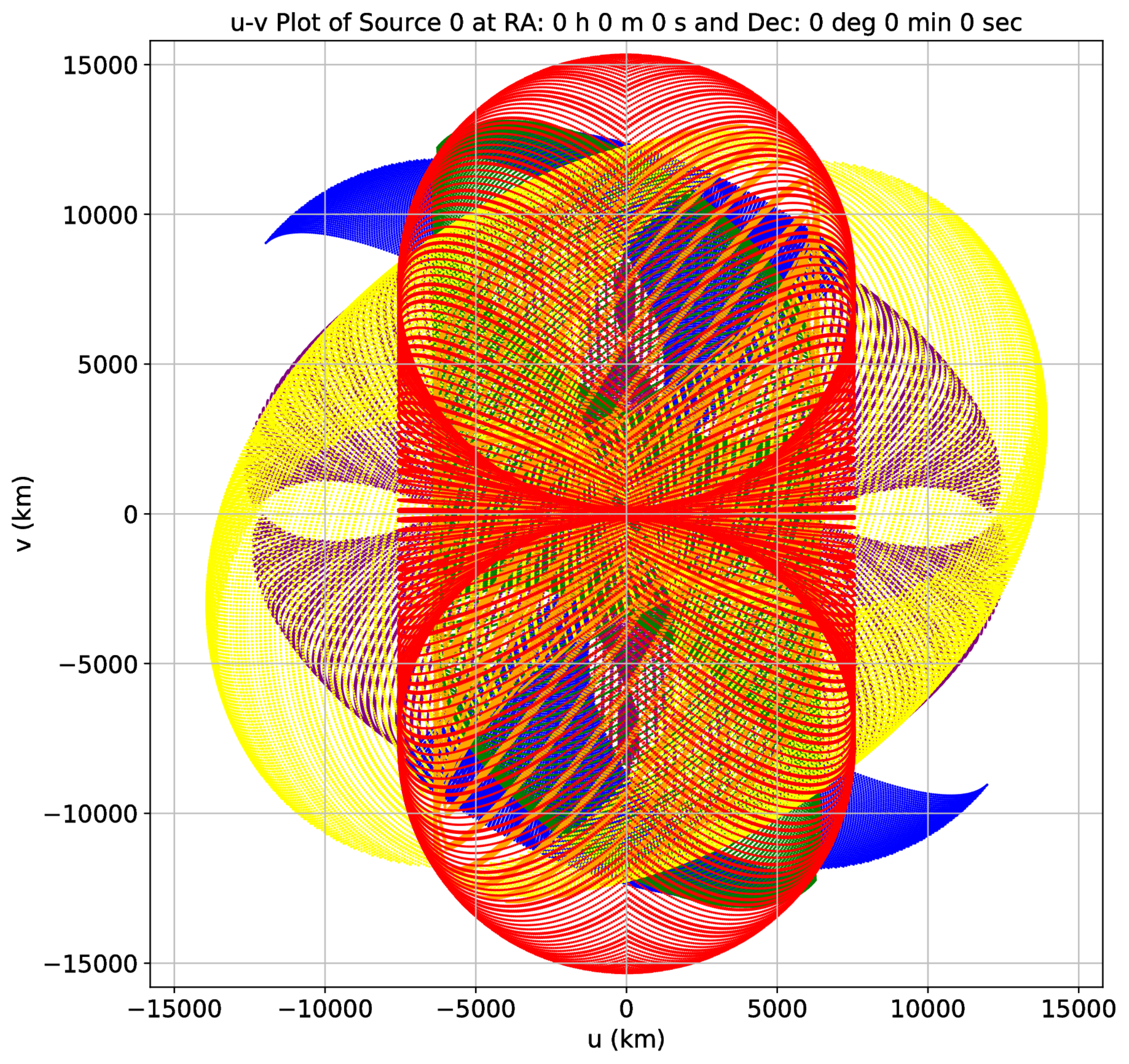}
			\end{subfigure}
			
			\caption{The $(u,v)$ coverages possible with a four satellite system,
				over a duration of 16 days, for the source directions A, B and C 
                respectively, are shown. 
				The purple, blue, green, yellow, orange and 
				red tracks correspond to the six baselines formed by 
				satellites 1-2, 1-3, 1-4, 2-3, 2-4, and 3-4, respectively. 
				The u, v axes are marked in km.
			} \label{figure10}
	\end{figure*}
	
	In this modified model, satellites 3 and 4 are in perpendicular 
	polar orbits similar to those of satellites 2 and 3 in the original model 
	(refer to Section~\ref{secm}) while satellites 1 and 2 are 
	in orbits perpendicular to each other with their axes inclined 
	to the Earth's rotation axis at an angle of 45$^{\circ}$ and 
	also equally inclined to the axes of satellites 3 and 4 respectively. 
	A simple line diagram in Figure~\ref{figure9} illustrates this 
	4-satellite configuration. 
	
	The orbital heights of the satellites above the surface of the 
	Earth are redefined but the maximum and minimum values are kept 
	the same as those in Table~\ref{table1}. The redefined parameters 
	are given in Table~\ref{table2}. The orbits of all satellites 
	are assumed to be nearly circular and follow the same equations as defined 
	in Section~\ref{secm}. The number of baselines for the 
	four satellite system is six (see Equation \ref{baselines}). 
	Therefore, this configuration offers twice the number of baselines as 
	that in the three-satellite system. 
	
	A special source direction is chosen in such a way that it is equally 
	inclined to all four satellite orbit axes, so as to correspond to the
	best case, offering maximum $(u,v)$ 
	coverage. This special direction happens to be the same as source direction A 
	(i.e. RA of 3 hrs and Dec of 45 deg).
	When a duration of 16 days is considered, 
	as has been standard in the majority of our simulations, 
	the percentage coverages for the source A are 41.5, 23 and 37.5
	for the baselines 1-2, 1-3, and 1-4 respectively.
	The other 3 baselines, namely 2-3, 2-4 and 3-4 offer
	percentage coverage of 41.5, 24.5 and 41, respectively.
	The combined coverage is 61.5\%. For the sake of completeness, 
    we also observed for source directions B and C
	(refer to Section~\ref{secrad}) with the 4-satellite system. 
    These results are summarized in Table~\ref{table2a}.
	
	These results indicate that even with a system of four satellites 
	and six baselines, when observing the source A, the total percentage coverage of 61.5\% 
	is essentially similar (and slightly lower) to that obtained with three satellites 
	when observed for the same time duration of 16 days. 
	So, when considering the maximum coverage possible, 
	there is no significant improvement even if we employ four satellites.
	Figure~\ref{figure10} shows the coverage with 
	four satellites for all the three special source directions, 
    assuming a duration of 16 days.
	
	However, when we observe the same sources for a duration of 8 days, 
	there is a noticeable improvement in the percentage coverage of 
	the four satellite system when compared with the three satellite 
    system for the same duration. 
	The percentage coverages when observing source A corresponding
	to the individual baselines
	1-2,1-3,1-4,2-3,2-4 and 3-4 are 27, 19, 26.5, 23.5, 22 and 30, respectively,
	while the combined coverage is about 59.5\%.
	In comparison, the percentage coverage for the same source direction when 
	observed with the three satellite setup for 8 days is,
	23, 30.5, and 24 for the baselines 1-2,2-3 and 3-1, respectively,
	while their combined coverage is close to 52\%. The results for all the three 
	source directions in case of 3-satellite and 4-satellite systems are summarized in 
	Tables~\ref{table1a} and \ref{table2a} respectively.
	
	\begin{table*}[h]
		\centering
		\caption{Pair-wise and total $(u,v)$ coverage with 4 satellites for different chosen directions observed for a span of 16 days and 8 days respectively.}\label{table2a}
		\begin{tabular*}{\textwidth}{@{}c\x c\x c\x c\x c\x c\x c@{}}
			\toprule
			\textit{Baseline or} & \multicolumn{3}{c}{Percentage Coverage for} & \multicolumn{3}{c}{Percentage Coverage for} \\
			\textit{Satellite Pair} & \multicolumn{3}{c}{16 days in Direction of} & \multicolumn{3}{c}{8 days in Direction of} \\
			\cmidrule(lr){2-4}\cmidrule(lr){5-7}
			~ & Source A & Source B & Source C & Source A & Source B & Source C \\ \midrule
			\textit{ 1-2} & 41.5 & 29.5 & 34.5 & 27   & 23   & 24.5 \\
			\textit{ 1-3} & 23   & 34   & 17   & 19   & 24.5 & 4.5  \\
			\textit{ 1-4} & 37.5 & 30   & 29   & 26.5 & 21   & 20   \\
			\textit{ 2-3} & 41.5 & 33.5 & 41.5 & 23.5 & 20   & 31.5 \\
			\textit{ 2-4} & 24.5 & 36   & 16   & 22   & 26   & 11.5 \\
			\textit{ 3-4} & 41   & 32   & 33.5 & 30   & 25   & 22.5 \\ \midrule
			\textit{All 4 Combined} & 61.5 & 66 & 75 & 59.5 & 62 & 65.5 \\ \bottomrule
		\end{tabular*}
	\end{table*}
	
	These results show that having a four satellite system can be of great 
	advantage when observing for shorter intervals of time. 
	As already mentioned, when observed for 8 days, the four satellite 
	system offers an additional 7.5\% 
	coverage than that with a three satellite system, when observing the source A. This 
	difference in coverage increases even more for sources B (by about 14.5\%) 
    and C (by about 22\%).
	But if observed long enough, the minimal configuration with
	just three LEO satellites is seen to be sufficient. 
	
	In assessing if any 4-satellite system does better, it is important to
    see if the betterment is significant and commensurate with the fractional
    increase (of about 30\%) in the resource employed.
    Our choice of orbits in the 4-satellite system is prompted by
    the requirement that coverage, assessed in the three source directions,
    is as uniform as possible across all octants.
    Any additional orbit, keeping the 3 original orbits, will not provide
    the desired uniformity across octants, even if significant benefits
    in coverage may be seen in certain directions or octants. 
	

	\section{Discussion and Conclusion} \label{seccfs}
	
	In this last section, we first discuss some of the assumptions
    and considerations, along with relevant justification and implications.

    The encouraging outcome of our present focused exploration, 
    that is, our identification of a novel minimal configuration 
    for apertures in LEO to facilitate 
    high-resolution synthesis imaging at low radio frequencies,
    paves way to proceed to the next step. This
    more challenging phase of designing a fuller system, taking 
    into account a range of practical considerations relevant 
    to eventual implementation of this idea, is beyond the 
    scope of present paper.
    Nonetheless, later in this section, we do allude to some
    of these aspects relevant to interferometry, 
    such as synchronization and effect of RFI, and briefly 
    discuss the advantages and challenges our minimal 
    configuration implies for operations at low radio frequencies.
    
    \bigskip\medskip
	
	In our model, we have assumed the beam of the antenna to be 
	hemispherical with a beam angle of exactly 180$^{\circ}$ 
	(as shown in Figure~\ref{figure1}) but in practice, realizing such 
	a beam response for a single element antenna is not possible.
	Given the science goals 
	of our system, a simple dipole or 
	a tripole antenna would suffice, which however 
	would have a non-uniform directivity. 
	Nevertheless, a suitably arranged array of antenna elements
	may serve the purpose for catering to the wide-angle coverage.
	
	\bigskip\medskip
	
	While estimating the dirty beams, we have employed 
	the uniform weighting function, to maximize angular resolution
	at the expense of point-source sensitivity
	(for example, \cite{wilner2010imaging}). 
	At such low radio frequencies, the interstellar scattering would 
	be expected to be severe, and the consequent angular broadening
	(see for example, \cite{goodman1985slow}, and references therein)
	should be expected to smoothen the apparent sky brightness distribution
	by an arcsecond to an arcminute scale, of course depending largely on the
	frequency and also the sky direction. 
	In such a case, the very high angular resolution offered by our proposed
	space interferometer would not be as useful, and hence 
	tapering of visibilities at high spatial frequencies
	might appear more profitable, to reduce the side-lobes as well as
	to improve sensitivity, now at the expense of angular resolution. 
	
	\bigskip\medskip
	
	While defining the parameters of our model, we assumed only 
	the gravitational pulls due to the Earth and the Sun on the satellites. 
	Although, this assumption would be valid for the most part, 
	in order to design a practical system, we need to consider all the forces 
	that might affect the motion of the satellite over longer periods. 
	Effects due to the gravitational pulls of the moon and other planets, 
	the precession and nutation of the Earth and the apparent forces 
	acting on the satellites cannot be ignored while considering 
	the evolution of the satellite orbits over very long periods of time. 
	
	It is also possible to deliberately use inclined orbits, which will
    precess at a predictable rate\footnote{We thank one of our anonymous referees
    for drawing our attention to this interesting and relevant possibility.}.
    More specifically, such slightly inclined orbits, for the two presently
    ideal polar orbits in our model, could be arranged to precess with
    periods many times the nominal 16-day cycles. A 45$^{\circ}$ precession
    of the plane of the orbit about the pole in about 3 months (0.5$^{\circ}$/day)
    would need about 4$^{\circ}$ to 5$^{\circ}$ inclination, given the altitudes of
    our polar orbits (see an early discussion by \cite{searle1958perturbations}). 
    While the inclination will not adversely 
    impact the baselines in any additional manner than due to the effect of
    non-orthogonality of the orbits 
    (as already assessed/discussed in subsection~\ref{subsecscop}),
    over the precession cycle, all sources at a given declination will 
    benefit from same coverage, making the total coverage independent of RA
    (for example, the coverage for all sources 
    at zero declination will be closely described by a combination of coverages 
    shown for present source directions B and C).
     
    \bigskip\medskip
	
	As can be noted from \cite{stankov2003new}, the Total Electron Content (TEC) 
    (and thus, the plasma content) of the ionosphere peaks at about 400 km above 
    the surface of the Earth and then drastically falls to a minimum after 700 km. 
    Therefore, in order to avoid the ionospheric plasma and allow the most amount 
    of incident radiation onto our antennas, we chose the 
	orbital heights of the satellites in our model to be well above 700 km. 
    For more details on the attenuation 
	of radio waves due to the ionospheric plasma, see \cite{chengalur2007low}.
	
	\bigskip\medskip
	
	Intrinsic variability timescales for radio sources, as apparent from flares 
    originating from objects ranging from flare stars to supermassive 
    black holes in AGNs (excluding pulsars and FRBs, but including
    events like SNe and GRBs), is known to vary from minutes to years
    (for details see, \cite{pietka2015variability}).
    Variations induced by the effect of intervening medium, such as scintillations, 
    are seen on timescales ranging from seconds to years, the longer timescale 
    being related to refractive effects (see, \cite{goodman1985slow}). 

    Each sky direction will be in view of a single aperture (with $\pm 90^{\circ}$ FoV) 
    in LEO for at least half of its orbital period. The considered configuration
    implies 12-15 orbits in a day for each of the satellites, providing sampling 
    of a comparable number of tracks in the $(u,v)$ plane, per baseline. 
    The revisit time for a given part of sky, as viewed by one of the 
    interferometers, would nominally be half of the longest orbital period 
    (say, an hour), while continuous monitoring is possible over duration of 
    a $(u,v)$ track. If $T_{1}$ and $T_{2}$ are the orbital periods of two LEO
    apertures, then the revisits, offering similar set of sampling 
    tracks in $(u,v)$ by the interferometer, are to be expected on time 
    scales decided by the beat between the orbital cycles 
    (i.e. $(\frac{1}{T_1}-\frac{1}{T_2})^{-1}$), which in our case, ranges 
    between 0.6 to 1.2 days.
    Over days, the gaps between the already sampled tracks (extending across
    the total span) are filled.

    As already mentioned in subsection~\ref{subsecscop}, the 16-day span is 
    merely an indicator of the duration 
    over which a major fraction (over 85\%) of the potential sampling in $(u,v)$ 
    plane is attained, and does not imply any lower limit for the time-scales on
    which variability can be studied. Although the implicit assumption in 
    synthesis imaging, that the sky distribution is unchanging, is rendered 
    invalid in instances of source variability, the variations themselves 
    are of immense interest to astronomers and any effect of the variability 
    on imaging quality can be desirably mitigated by measuring 
    the variation in adequate detail, and duly accounting for its effect
    on the image.
     
    \bigskip\medskip
	
	Terrestrial man-made RFI continues to be an important and unavoidable 
    issue even in Earth Orbits, and more so at low radio frequencies 
    (see \cite{bentum2016rfi}, and references therein).  
    As already mentioned, our consideration of avoiding ionospheric 
    attenuation of astronomical signals prompts even the closest orbit to 
    be outside the ionosphere. Thus, the severe attenuation of the very 
    low radio frequency signals by the ionosphere, owing to the large 
    electron densities over significant pathlengths, would 
    naturally shield radio sky measurements from contamination 
    due to the terrestrial sources of RFI to certain extent.
    In addition, our suggested confinement of the field of view
    implicitly excludes, in principle, any significant antenna 
    response in the directions of the terrestrial sources of RFI, 
    and in practice, any efforts to ensure highly attenuated backlobe 
    responses would not only be desired, but would be richly rewarding. 
    This advantage is not easy to gain for the system using a swarm 
    of satellites, unless such blind-to-Earth mode is 
    explicitly incorporated.

    Even if a finite amount of RFI contamination makes its way through 
    the ionosphere to our interferometric elements, on long baselines 
    the individual elements are unlikely to "see" the same sources 
    of RFI (given separated footprints), and hence the picked RFI 
    can be expected to be mutually uncorrelated. Similar argument 
    for uncorrelatedness is applicable to any EMI/RFI from the respective 
    crafts. In instances of contamination from any common source of 
    terrestrial RFI, the large (relative) delays introduced by the 
    ionosphere (particularly at low frequencies and due to the high TEC), 
    as seen on a long baseline, would imply too small decorrelation 
    bandwidths for the RFI to contaminate the visibility measurement 
    in any significant manner. These considerations would be of reducing
    benefit when the footprints of a pair of satellites overlap, 
    making the paths through the ionosphere less oblique, in addition 
    to seeing the correlated RFI. Hence, the visibility measurements 
    on relatively shorter baselines (<3000 km) may get affected to the extent  
    that the backlobe response may fail to adequately attenuate 
    signals from the Earth, and their level of surviving phase coherence.

    Independent of these considerations, as is widely appreciated, it is
    essential to ensure that at least the first amplifiers possess high 
    dynamic range, to avoid creation of any intermodulation products, 
    so that at the first opportunity in the following receiver chain 
    it becomes possible to employ suitable spectral filtering, where relevant.
    As long as RFI is kept to within their respective native bands, 
    suitable detection and excision techniques can be employed in the 
    post processing (e.g., \cite{deshpande2005correlations}).
  
    \bigskip\medskip
    
    On the other side, 
    the ionospheric attenuation in our band would not be 
    as significant as will be for the terrestrial RFI, and 
    hence any RFI from the possible numerous satellites, in the space above 
    our LEO orbits, would be "seen" within the wide field-of-view 
    of one or more of our satellites. 
    Any formal radio downlink 
    transmission from such satellites would nonetheless be at 
    frequencies much higher than the band of our interest.

    However, any out-of-band (OOB) or spurious signal radiation
    from other satellites, amounting their lack of 
    Electromagnetic compatibility (EMC), can potentially 
    contaminate our band of interest.  If such RFI is 
    narrow-band, then usual detection and excision procedures, 
    applied separately at each element of our interferometers, 
    would suffice, and the impact can be expected to be limited,
    in terms of some amount of data loss (usually a small fraction)
    and corresponding reduction in the system sensitivity. 
    Any broadband RFI from other satellite systems 
    would however need a different strategy, noting the associated
    challenge and the opportunity. It is easy to see that 
    any broadband RFI from a well defined direction would be 
    strongly correlated on our baselines, with delay 
    corresponding to relative path difference 
    to our satellites, and these delays would change 
    predictably but distinctly differently from those expected  
    for signals from sky. Thus, sensitive detection of such 
    broadband RFI from each of the other earth satellites 
    would be possible by identifying and monitoring the 
    associated peaks and their tracks in the dynamic 
    delay spectrum (that is Fourier transform of the 
    cross-correlation spectrum). Appropriate 
    filtering based on delay-rate would effect desired excision
    of any significant broadband RFI from such differently moving sources. 
    As an attractive opportunity, such
    data on the relative delays and their 
    rates of change for all identifiable moving sources of 
    broadband RFI can be used beneficially
    towards refined monitoring of the changing 
    positions of {\it our} satellites, as well as to obtain 
    instructive sampling of delays associated with the
    upper ionosphere.
    
    \bigskip\medskip
    
    The high data rates implied by the proposed wide-band wide-field
    observations present additional challenges.
    For our band up to 20 MHz, in dual polarization, the 
    raw voltage sampling would amount to 80 mega-samples per second,
    accumulating typically to about 10 GB (for 8-bit samples), 
    at each of our satellites over the duration of  
    respective orbital periods. The downlink channel capacity, thus,
    needs to be adequately high to transfer $10^{10}$ samples per primary beam
    or antenna element, from each satellite  within a fraction of 
    their respective orbital period whenever ground station access becomes possible. 
    Partial processing performed
    locally at each satellite, to compute fine resolution spectra 
    and to employ primary detection and excision of any dominant RFI,
    would help in reducing the dynamic range requirement, and thus,
    justify any reduction in bit-length per sample, if required.
    Fortunately, suitable optical/millimeter-wave downlinks 
    can provide attractive bandwidths catering to high rates of 
    data transfers. Similar data exchanges can be considered 
    also between our satellites 
    (e.g., \cite{bitragunta2020best}, and references therein), 
    which would facilitate on-board computations,
    including cross-correlation, enabling allowed level of
    time integration of visibilities to reduce data size, 
    and thus, easing the data rate requirements for 
    the downlinks. While this would facilitate desired immediate 
    local checks on quality of interferometric data, the entire 
    raw voltage data download would remain the preferred mode
    to enable detailed and sophisticated processing and 
    refinements. In case of stringent constraints from capacity
    of downlinks, the effective bandwidth to be downloaded 
    can be reduced accordingly, either as a narrower contiguous 
    band or with picket-fence sampling across our entire spectral 
    span.
      
    \bigskip\medskip
  
    Two of the most important and integral parts of any interferometry
    setup are a) the aspect of time synchronization, and b) knowledge of
    locations of the interferometer elements. Operating at frequencies
    below 20 MHz relaxes some of the otherwise demanding requirements
    encountered at shorter wavelengths. Thus, in our case, a sub-meter 
    accuracy in positions of phase centres of our apertures would suffice.
    Similarly, native synchronization at the level of a few nanoseconds would
    ensure retention of any coherence intrinsic to the sky signal, 
    while assessing correlation even with the entire bandwidth (say, up to 20 MHz). 
    However, for catering to the wide-field imaging with long baselines, 
    use of fine spectral resolution is essential as already discussed, 
    and hence, the decorrelation delays would be correspondingly large. 
    Even discounting this possible relaxation in time synchronization requirements,
    the required accuracies are modest and can be easily met by use of a suitable
    local frequency reference that has high stability on short-term, 
    and phase-locking it to a common standard that has stability on long 
    timescales. In our experience, a Rubidium oscillator disciplined using
    1 PPS (one pulse per second) signal from the Global Positioning System (GPS) 
    receiver (see, for example, \cite{maan2013rri}) readily provides the 
    presently required accuracies, and is a reliable combination
    for setup on-board each of our satellites. A 1 PPS pulse
    produced by most Rubidium oscillator systems provides a time stamping
    reference with 1 ns resolution, and a 10 MHz output which forms the
    reference for phase-locking all local clocks and oscillators.
    GPS-based orbit determination (see, for example,
    \cite{montenbruck2004proceedings}), and verification through calibration
    observations on bright astronomical sources, will form the
    basis for estimation of position information with the desired accuracy.
    A more detailed discussion on these and related aspects, though
    important, is beyond the scope of present paper.

	\bigskip\medskip
	
	To summarize, recognizing  
	the need of a fully space-based low frequency radio observation setup,
	we have proposed and studied in detail a minimal configuration of 
	only three apertures, 
	each aboard a LEO satellite, which would be sufficient to map the entire sky, 
	while giving baselines greater than 15000 km and resolutions finer 
	than 10 arcsec for frequencies under 20 MHz. We have shown 
	that the percentage coverage of this system is also greater 
	than its Earth-based and Earth-and-space-based counterparts 
	and it is able to achieve this in a shorter time span. 
	We have also discussed the special case of the Sun,
	which does not have fixed RA and Dec.
	Our assessment of the 4-satellite system suggests that
	adding a fourth satellite would not be of any significant advantage 
	and having just three satellites would be 
	optimum scientifically, 
	technologically and economically. 
	Although motivated by the requirement of an optimum setup at low radio
	frequencies, the various aspects discussed here, including the minimal
	configuration, are relevant for space interferometry in other wavebands as well.


	\begin{acknowledgements}
	We thank our two anonymous referees for their valuable comments,
    which have helped towards significant improvement of our manuscript.
	Akhil Jaini (AJ) would like to thank Raman Research Institute, Bangalore 
	for providing Visiting Studentship and support to facilitate this study, 
	and Birla Institute of Technology and Science-Pilani, Pilani campus, for providing 
	the opportunity to do this project at the host institute.
	AJ would also like to thank Hrishikesh Shetgaonkar, Pavan Uttarkar, Pratik Kumar and 
	Aleena Baby for helpful discussions and insightful comments.
	\end{acknowledgements}
	
	
	\bibliographystyle{pasa-mnras}
	\bibliography{space_interferometer_revised_bib}

\begin{thebibliography}{}
\makeatletter
\relax
\def\mn@urlcharsother{\let\do\@makeother \do\$\do\&\do\#\do\^\do\_\do\%\do\~}
\definecolor{darkblue}{rgb}{0,0,0.597656}
\def\mndoi{\begingroup\mn@urlcharsother \@ifnextchar [ {\mndoi@} {\mndoi@[]}}
\def\mndoi@[#1]#2{\def\@tempa{#1}\ifx\@tempa\@empty \href
  {http://dx.doi.org/#2} {\textcolor{darkblue}{doi:#2}}\else \href
  {http://dx.doi.org/#2} {\textcolor{darkblue}{#1}}\fi \endgroup}
\def\mn@eprint#1#2{\mn@eprint@#1:#2::\@nil}
\def\mn@eprint@arXiv#1{\href {http://arxiv.org/abs/#1} {{\tt arXiv:#1}}}
\def\mn@eprint@dblp#1{\href {http://dblp.uni-trier.de/rec/bibtex/#1.xml}
  {dblp:#1}}
\def\mn@eprint@#1:#2:#3:#4\@nil{\def\@tempa {#1}\def\@tempb {#2}\def\@tempc
  {#3}\ifx \@tempc \@empty \let \@tempc \@tempb \let \@tempb \@tempa \fi \ifx
  \@tempb \@empty \def\@tempb {arXiv}\fi \@ifundefined
  {mn@eprint@\@tempb}{\@tempb:\@tempc}{\expandafter \expandafter \csname
  mn@eprint@\@tempb\endcsname \expandafter{\@tempc}}}

\bibitem[\protect\citeauthoryear{An et~al.,}{An et~al.}{2020}]{an2020space}
An T.,  et~al., 2020, \mndoi [Advances in Space Research]
  {https://doi.org/10.1016/j.asr.2019.03.030}, 65, 850

\bibitem[\protect\citeauthoryear{Apoorva, Bitragunta  \& Nitundil}{Apoorva
  et~al.}{2020}]{bitragunta2020best}
Apoorva .,  Bitragunta S.,   Nitundil S.,  2020, \mndoi [IET Communications]
  {https://doi.org/10.1049/iet-com.2020.0515}, 14, 3350

\bibitem[\protect\citeauthoryear{Baan}{Baan}{2013}]{baan2012suro}
Baan W.,  2013, in Proceedings of the meeting from Antikythera to the Square
  Kilometre Array: Lessons from the Ancients (Antikythera \& SKA). Kerastari,
  Greece. p.~045, \mndoi{10.22323/1.170.0045}

\bibitem[\protect\citeauthoryear{Balmino}{Balmino}{1974}]{balmino1974coriolis}
Balmino G.,  1974, \mndoi [Celestial mechanics] {doi.org/10.1007/BF01229119},
  10, 423

\bibitem[\protect\citeauthoryear{Bentum \& Boonstra}{Bentum \&
  Boonstra}{2016}]{bentum2016rfi}
Bentum M.,  Boonstra A.-J.,  2016, in 2016 Radio Frequency Interference (RFI).
  pp~1--6, \mndoi{10.1109/RFINT.2016.7833521}

\bibitem[\protect\citeauthoryear{Bentum et~al.,}{Bentum
  et~al.}{2020}]{bentum2020roadmap}
Bentum M.,  et~al., 2020, \mndoi [Advances in Space Research]
  {https://doi.org/10.1016/j.asr.2019.09.007}, 65, 856

\bibitem[\protect\citeauthoryear{Deshpande}{Deshpande}{2005}]{deshpande2005correlations}
Deshpande A.~A.,  2005, \mndoi [Radio Science] {10.1029/2004RS003156}, 40, 1

\bibitem[\protect\citeauthoryear{Engelen, Verhoeven  \& Bentum}{Engelen
  et~al.}{2010}]{engelen2010olfar}
Engelen S.,  Verhoeven C. J.~M.,   Bentum M.~J.,  2010, in USU Conference on
  Small Satellites, Logan, UT. \url
  {https://digitalcommons.usu.edu/smallsat/2010/all2010/20}

\bibitem[\protect\citeauthoryear{Goodman \& Narayan}{Goodman \&
  Narayan}{1985}]{goodman1985slow}
Goodman J.,  Narayan R.,  1985, \mndoi [Monthly Notices of the Royal
  Astronomical Society] {10.1093/mnras/214.4.519}, 214, 519

\bibitem[\protect\citeauthoryear{Gurvits}{Gurvits}{2018}]{gurvits2018radio}
Gurvits L.~I.,  2018, Radio Interferometers Larger than Earth: Lessons Learned
  and Forward Look of Space VLBI (\mn@eprint {arXiv} {1810.01230})

\bibitem[\protect\citeauthoryear{Hirabayashi et~al.,}{Hirabayashi
  et~al.}{1998}]{hirabayashi1998overview}
Hirabayashi H.,  et~al., 1998, \mndoi [Science]
  {10.1126/science.281.5384.1825}, 281, 1825

\bibitem[\protect\citeauthoryear{Kardashev, Kovalev  \& Kellermann}{Kardashev
  et~al.}{2012}]{kardashev2012radioastron}
Kardashev N.~S.,  Kovalev Y.~Y.,   Kellermann K.~I.,  2012, \mndoi [URSI Radio
  Science Bulletin] {10.23919/URSIRSB.2012.7910147}, 2012, 22

\bibitem[\protect\citeauthoryear{Liu \& Shaw}{Liu \& Shaw}{2020}]{liu2020data}
Liu A.,  Shaw J.~R.,  2020, \mndoi [Publications of the Astronomical Society of
  the Pacific] {10.1088/1538-3873/ab5bfd}, 132, 062001

\bibitem[\protect\citeauthoryear{Maan et~al.,}{Maan et~al.}{2013}]{maan2013rri}
Maan Y.,  et~al., 2013, \mndoi [The Astrophysical Journal Supplement Series]
  {10.1088/0067-0049/204/1/12}, 204, 12

\bibitem[\protect\citeauthoryear{Montenbruck}{Montenbruck}{2004}]{montenbruck2004proceedings}
Montenbruck O.,  2004, Proceedings of the 18th International Symposium on Space
  Flight Dynamics, \url {https://elib.dlr.de/21412/}

\bibitem[\protect\citeauthoryear{Pietka, Fender  \& Keane}{Pietka
  et~al.}{2014}]{pietka2015variability}
Pietka M.,  Fender R.~P.,   Keane E.~F.,  2014, \mndoi [Monthly Notices of the
  Royal Astronomical Society] {10.1093/mnras/stu2335}, 446, 3687

\bibitem[\protect\citeauthoryear{Rao}{Rao}{2007}]{chengalur2007low}
Rao A.~P.,  2007, in Chengalur J.~N.,  Gupta Y.,   Dwarkanath K.~S.,  eds, ,
  Low frequency radio astronomy 3rd edition.
NCRA-TIFR, Chapt.~16, \url
  {http://www.ncra.tifr.res.in/ncra/gmrt/gmrt-users/low-frequency-radio-astronomy}

\bibitem[\protect\citeauthoryear{Searle}{Searle}{1958}]{searle1958perturbations}
Searle L.,  1958, Journal of the Royal Astronomical Society of Canada, \href
  {https://ui.adsabs.harvard.edu/abs/1958JRASC..52...65S} {52, 65}

\bibitem[\protect\citeauthoryear{Stankov, Jakowski, Heise, Muhtarov, Kutiev  \&
  Warnant}{Stankov et~al.}{2003}]{stankov2003new}
Stankov S.~M.,  Jakowski N.,  Heise S.,  Muhtarov P.,  Kutiev I.,   Warnant R.,
   2003, \mndoi [Journal of Geophysical Research: Space Physics]
  {https://doi.org/10.1029/2002JA009570}, 108

\bibitem[\protect\citeauthoryear{Teles, Samii  \& Doll}{Teles
  et~al.}{1995}]{teles1995overview}
Teles J.,  Samii M.~V.,   Doll C.~E.,  1995, \mndoi [Advances in Space
  Research] {https://doi.org/10.1016/0273-1177(95)98783-K}, 16, 67

\bibitem[\protect\citeauthoryear{Wilner}{Wilner}{2010}]{wilner2010imaging}
Wilner D.~J.,  2010, in NRAO 12th Synthesis Imaging workshop. \url {http://www.
  aoc. nrao. edu/events/synthesis/2010/lectures10.html}

\makeatother
\end{thebibliography}
	
\end{document}